\documentclass [11pt]{article}

\usepackage{jheppub}

\usepackage[utf8]{inputenc}

\usepackage{amsfonts}

\usepackage{amsmath,amssymb}

\usepackage{mathtools}

\usepackage{graphicx,subfigure}
\usepackage[space]{grffile}

\usepackage{simplewick}

\usepackage{array}

\usepackage{float}

\usepackage{color}

\usepackage{dsdshorthand}

\usepackage{tikz}
\usetikzlibrary{trees}
\usetikzlibrary{decorations.pathmorphing}
\usetikzlibrary{decorations.markings}
\usetikzlibrary{shapes.misc}

\tikzset{
	threept/.style={
		circle,
		draw,
		inner sep=2pt,
	},
	twopt/.style={
		circle,
		draw,
		fill=black,
		inner sep=1pt,
		minimum size=1pt
	},
	cross/.style={
		cross out,
		draw=black, 
		minimum size=7pt, 
		inner sep=0pt,
		outer sep=0pt
	},
	scalar/.style={
		thick,
		dashed,
		postaction={
			decorate,
			decoration={
				markings,
				mark=at position 0.5 with {\arrow{>}}
			}
		}
	},
	spinning/.style={
		thick,
		postaction={
			decorate,
			decoration={
				markings,
				mark=at position 0.5 with {\arrow{>}}
			}
		}
	},
	spinning no arrow/.style={
		thick,
	},
	finite with arrow/.style={
		decoration={
			snake,
			amplitude=1pt,
			segment length=6pt,
			post length=2pt
		},
		decorate,
		thick,->
	},
	finite/.style={
		decoration={
			snake,
			amplitude=1pt,
			segment length=6pt,
		},
		decorate,
		thick
	}
}

\def\nn{\nonumber}

\newcommand{\point}{{\bf p}}

\newcommand{\ii}{\mathbf{i}}

\DeclareMathOperator{\tr}{tr}

\newcommand*{\uniq}{\raisebox{-0.7ex}{\scalebox{1.8}{$\cdot$}}}

\newcommand{\beq}{\begin{equation}}
\newcommand{\eeq}{\end{equation}}
\newcommand{\eref}[1]{(\ref{#1})}
\newcommand{\sref}[1]{section~\ref{#1}}
\newcommand{\Figref}[1]{fig.~\ref{#1}}

\newcommand{\structgeneral}{
	\left[
	\begin{matrix}
	q_1 & q_2 & q_3 & q_4 \\
	\bar q_1 & \bar q_2 &\bar q_3 &\bar q_4	
	\end{matrix}
	\right]
}
\newcommand{\struct}[8]{
	\left[
	\begin{matrix}
	#1 & #2 & #3 & #4 \\
	#5 & #6 & #7 & #8	
	\end{matrix}
	\right]
}

\newcommand{\II}{\mathbb{I}}
\newcommand{\JJ}{\mathbb{J}}
\newcommand{\KK}{\mathbb{K}}

\usepackage{marginnote}
\usepackage[normalem]{ulem}

\def\bea{\begin{eqnarray}} \def\eea{\end{eqnarray}}

\title{\boldmath Fermion Conformal Bootstrap in 4d}

\author[a]{Denis Karateev,}
\author[b]{Petr Kravchuk,}
\author[c,d]{Marco Serone}
\author[a]{and Alessandro Vichi}

\affiliation[a]{Institute of Physics, EPFL, CH-1015 Lausanne, Switzerland}
\affiliation[b]{School of Natural Sciences, Institute for Advanced Study, Princeton, New Jersey 08540, USA}
\affiliation[c]{SISSA and INFN, Via Bonomea 265, I-34136 Trieste, Italy}
\affiliation[d]{ICTP, Strada Costiera 11, I-34151 Trieste, Italy}

\abstract{
	We apply numerical conformal bootstrap techniques to the four-point function of a Weyl spinor in 4d non-supersymmetric CFTs. We find universal bounds on operator dimensions and OPE coefficients, including bounds on operators in mixed symmetry representations of the Lorentz group, which were inaccessible in previous bootstrap studies. We find discontinuities in some of the bounds on operator dimensions, and we show that they arise due to a generic yet previously unobserved ``fake primary'' effect, which is related to the existence of poles in conformal blocks. We show that this effect is also responsible for \mbox{similar} discontinuities found in four-fermion bootstrap in 3d, as well as in the mixed-correlator \mbox{analysis} of the 3d Ising CFT. As an important byproduct of our work, we develop a practical technology for numerical approximation of general 4d conformal blocks.}

\begin{document}

\maketitle

\section{Introduction and summary of results}
\label{sec:intro}

Considerable progress has been achieved during the last ten years in the understanding of conformal field theories (CFTs) in $d\geq 3$ space-time dimensions. This was triggered by the pioneering work~\cite{Rattazzi:2008pe} where it was shown how to efficiently apply the conformal bootstrap program~\cite{Ferrara:1973yt,Polyakov:1974gs} using numerical methods. Invoking first principles only, such as crossing symmetry, operator product expansion (OPE) and unitarity, rigorous and general bounds can be put on the space of CFTs in 
various number of dimensions. See~\cite{Poland:2018epd} for a review and a comprehensive list of references on what is now a well-developed field of research.

Previous works have shown that certain theories, such as the 2d and 3d Ising models and the 3d $\mathrm{O}(N)$ vector models \cite{Rychkov:2009ij,ElShowk:2012ht,El-Showk:2014dwa,Kos:2013tga}, sit at the boundary between the allowed and forbidden regions of the parameter space, in points that appear to have a kink-like discontinuity. Using as heuristic guiding principle the idea that discontinuities of this kind are hints of the presence of consistent CFTs, the numerical conformal bootstrap allows to discover new theories and compute their CFT data by using extremal functional methods \cite{Poland:2011ey,ElShowk:2012hu,El-Showk:2016mxr,Simmons-Duffin:2016wlq,Mazac:2016qev,Mazac:2018mdx,Mazac:2018ycv}.

Unfortunately, in 4d non-supersymmetric CFTs the boundary between the allowed and the forbidden regions in the parameter space is rather smooth and
no kink-like discontinuities have been found~\cite{Rattazzi:2008pe,Rychkov:2009ij,Caracciolo:2009bx,Rattazzi:2010gj,Poland:2010wg,Rattazzi:2010yc,Vichi:2011ux,Poland:2011ey,Caracciolo:2014cxa,Iha:2016ppj,Nakayama:2016knq}.\footnote{Supersymmetric theories are on a different footing. For instance a kink (already noticed in~\cite{Poland:2011ey}) has been conjectured to be associated to a minimal ${\cal N}=1$ supersymmetric CFT \cite{Poland:2015mta}.
From now on, we leave implicit that in this paper, unless explicitly stated, we consider non-supersymmetric CFTs only.} However these studies were all based on four-point functions with external scalar operators only. 
The study of non-scalar correlators has been hindered for some time by the need of knowing the 4d conformal blocks associated to correlators involving spin.\footnote{On the other hand, several numerical bootstrap studies with spin correlators in 3d CFTs have already been made \cite{Iliesiu:2015qra,Iliesiu:2017nrv,Dymarsky:2017xzb,Dymarsky:2017yzx}.}
Due to recent results~\cite{Costa:2011mg,Costa:2011dw,SimmonsDuffin:2012uy,Echeverri:2015rwa,Penedones:2015aga,Iliesiu:2015akf,Echeverri:2016dun,Costa:2016hju,Kravchuk:2016qvl,Cuomo:2017wme,Karateev:2017jgd,Kravchuk:2017dzd} this is no longer an issue and
it is then natural to address numerically various non-scalar correlators in the hope of finding hints of new CFTs that were not present in scalar setups.

The aim of this paper is to continue exploring 4d non-supersymmetric CFTs,  by con\-sidering a four-fermion correlator.
In particular we study the constraints coming from unitarity and crossing symmetry of the four-point function
\be
    \label{eq:the_fermion_four_point_function}
	\<\psi^\dagger_{\dot\a_1}(x_1)\psi_{\a_2}(x_2)\psi^\dagger_{\dot\a_3}(x_3)\psi_{\a_4}(x_4)\>,
\ee
which consists of two identical $(1,0)$ Weyl fermions $\psi_{\alpha}$\footnote{In our convention a vector transforms in the representation $(1,1)$, and a Dirac spinor in $(1,0)\oplus(0,1)$. 
Our Weyl spinor $\psi_\a$ can be a chiral component of a Dirac spinor.}  
with scaling dimension $\De_\psi$ and their hermitian conjugates $\psi^\dag_{\dot\a}$. 
We assume the existence of a $\mathrm{U}(1)$ ``baryon"  global symmetry in the CFT under which $\psi$ and  $\psi^\dag$ carry $q=+1$ and  $q=-1$ charges respectively.\footnote{A unit charge under a $\Z_n$ group for $n>2$ (for example under the discrete remnant of an axial symmetry broken by the ABJ anomaly) leads to the same analysis, except in the bounds where we consider the $\mathrm{U}(1)$ current. We will also make some comments that apply to the case $q=0$.}

There are two different types of operator product expansion (OPE) 
that one can take in~\eqref{eq:the_fermion_four_point_function}. We refer to the OPE $\psi^\dagger \psi$ as the neutral channel and to the OPE $\psi\psi$ (and its hermitian conjugate) as the charged channel. Both channels contain traceless symmetric  (TS) tensors  in the $(\ell, \ell)$ and non-traceless symmetric (NTS) tensors  in the $(\ell+2, \ell)$ and  $(\ell, \ell+2)$ spin representations \cite{Mack:1969rr,Elkhidir:2014woa}.
The correlator~\eqref{eq:the_fermion_four_point_function} allows us to access operators with non-trivial ``baryon'' charge and, at the same time, NTS operators for the first time.

We determine all the four-fermion conformal blocks by using differential operators~\cite{Echeverri:2015rwa,Karateev:2017jgd} that relate them to the known seed conformal blocks~\cite{Echeverri:2016dun}. To efficiently construct their rational approximations needed for~\texttt{SDPB}~\cite{Simmons-Duffin:2015qma} we implement the following strategy. First, we generate rational approximations of scalar blocks using the Dolan-Osborn closed form expressions~\cite{DO1,DO2}, in order to bypass subtleties associated with the double poles in the traditional  Zamolodchikov-like recursion relations~\cite{Zamolodchikov:1987,Kos:2013tga,Kos:2014bka,Penedones:2015aga}.\footnote{Closer to the completion of this work we have implemented a Zamolodchikov-like recursion algorithm for scalar conformal blocks in $d=4$, taking care of the double poles. This algorithm was used only in the lower bound on $C_T$ in section~\ref{sec:numerical_results}, and will be described elsewhere.} Second, we apply the recursion relations of~\cite{Karateev:2017jgd} to obtain the rational approximation of the relevant seed blocks. Finally, we derive the rational approximation for the four-fermion blocks using their expression in terms of the seed blocks.

We determine the CFT data associated to~\eqref{eq:the_fermion_four_point_function} when $\psi$ is a generalized free fermion using the algebraic expressions for the four-fermion conformal blocks. This generalized free theory (GFT) provides a consistency check for our setup and a useful reference point in the numerical analysis that follows.

We construct numerically various bounds: bounds on scaling dimensions of charged and neutral operators (TS and NTS), bounds on the central charges $C_T$ and $C_J$, associated to the energy momentum tensor and the $\mathrm{U}(1)$ conserved current respectively and bounds on the OPE coefficients between two Weyl fermions and a scalar (charged and neutral).

Interestingly enough, we find jump-like discontinuities in the upper bounds for all TS operators in the charged channel (see figures~\ref{fig:charged00} and~\ref{fig:charged11-22})
and all NTS operators with $\ell >0$ in the neutral channel (see figure~\ref{fig:neutral13-24}). The jumps occur when the upper bound under consideration on the operator with $(\ell,\bar\ell)$ spin crosses an integer value
\begin{equation}
\label{eq:jump_value}
\Delta_\text{jump} = 4 + \frac{\ell+\bar\ell}{2}.
\end{equation}
These discontinuities, however, appear to be associated not to new CFTs, but to a general mechanism which we refer to as the \emph{fake primary effect}. In a nutshell, the fake primary effect works in the following way. Given an operator $\cO$ which contributes to the four-point function, the associated conformal block generically has a pole in $\De$ at the unitarity bound. The residue of this pole is the contribution of a particular descendant of $\cO$, together with the conformal multiplet generated from it. Strictly at the unitarity bound this descendant becomes a primary, and the residue is thus again a conformal block~\cite{Zamolodchikov:1987,Kos:2013tga, Penedones:2015aga}. As the normalization of conformal blocks is ambiguous, we can say that at the unitarity bound the conformal block of $\cO$ is simply equal to that of the descendant. Since the descendant generically transforms in a different spin representation than $\cO$, the descendant conformal block effectively fakes the presence of a primary operator with some new spin and dimension.\footnote{In most cases the dimension of the fake primary is $\De_\cO+1$ and its spin is $j_\cO-1$, where $\De_\cO$ is at the unitarity bound. In our 4d case this implies that the spin and the dimension of the fake primaries are connected by~\eqref{eq:jump_value}.}

This connection between different parts of the spectrum forces us to reinterpret our numerical bounds. As we argue in section~\ref{sec:topology_blocks} and verify numerically in section~\ref{sec:numerical_results}, the jumps in our bounds occur precisely due to the unexpected presence of fake primaries. We also classify in section~\ref{sec:topology_blocks} the cases when fake primaries occur more generally. Most notably, they never appear in single scalar correlator bounds, since in this case it happens that the descendant contribution discussed above vanishes.

Although it does not appear to have been previously understood, the fake primary effect has already manifested itself in several other works. 
For instance, jumps similar to ours were observed in the 3d fermion setup~\cite{Iliesiu:2015qra,Iliesiu:2017nrv} (see figures 1 and 3 respectively) and
in the 3d mixed scalar correlator setups~\cite{Nakayama:2016jhq} (see figures 2 and 6) and~\cite{Kos:2014bka} (see figure 1). We discuss some of these setups in section~\ref{sec:topology_blocks:other}. In section~\ref{sec:topology_blocks:ising} we show that the latter case~\cite{Kos:2014bka} is rather non-trivial since the jump-like feature is driven by the fake primary effect and the presence of a physical CFT at the same time (see figure~\ref{fig:isingsigmaprime}). 

It is likely that this phenomenon will affect future bootstrap studies, and it will be important to carefully take it into account and understand how to remove it. We discuss one possible way of doing so, by adding appropriate gaps above unitarity bounds (see figure~\ref{fig:removingJump}).

The conformal bootstrap approach is insensitive to the UV realizations of the CFT describing the IR theory, and is only characterised by the CFT data associated to 
the primary operators and their OPE coefficients.  Yet it might be useful to list some possible ways to interpret the external fermion operator $\psi_\alpha$ entering our correlator in terms of the UV degrees of freedom.
The main evidence we have for the existence of non-trivial non-supersymmetric 4d CFTs arises from UV Lagrangian descriptions based on gauge theories coupled to matter that
flow in the IR to a weakly coupled Caswell-Banks-Zaks (CBZ) fixed point \cite{Caswell:1974gg,Banks:1981nn}.\footnote{Lattice simulations provide also numerical evidence for CBZ fixed points not accessible in perturbation theory. See e.g.~\cite{DeGrand:2015zxa} for a review, in particular Table I and Table II for a summary of results for $\SU(3)$ gauge theories with 12 fundamental fermions and $\SU(2)$ theories with 2 (Dirac) adjoint fermions.} 
In the notable case of an $\SU(2n+1)$ gauge theory with elementary fermions $\chi_\alpha^a$ in the fundamental representation, $\psi_\a$ can be identified with the gauge-invariant baryon operator
\be
\label{eq:UV_composite_fermion_option1}
\psi_\a\propto \chi_\a^a\chi_{\b_1}^{b_1} \chi_{\s_1}^{c_1} \cdots \chi_{\b_n}^{b_n} \chi_{\s_n}^{c_n} \e_{ab_1c_1\ldots b_n c_n} \e^{\b_1\s_1}\cdots \e^{\b_n\s_n},\quad \Delta^{{\rm UV}}_\psi =
\frac 32 + 3n \,.
\ee
If we have elementary fermions $\chi_\alpha^a$ with $a$ in the adjoint representation, one can also consider
\be
\label{eq:UV_composite_fermion_option2}
	\psi_\a \propto \tr(\chi^\b F_{\a\b}), \quad  \Delta^{{\rm UV}}_\psi =\frac 72\,,
	\ee
where $F^a_{\alpha\beta}$ is the self-dual component of the gauge field strength. If elementary scalars in the appropriate representation $\f^a$  are present in the UV theory,\footnote{In presence of scalars UV asymptotic freedom becomes non-trivial because of quartic and Yukawa couplings. One has also to check that in the IR the theory does not undergo spontaneous symmetry breaking. See e.g.~\cite{Hansen:2017pwe} for a perturbative study of a class of gauge theories with fermion and scalar matter where consistent flows to a weakly coupled CFT have been found.} another option is provided by meson-like fermions of the form\footnote{In principle, one could also consider UV Lagrangians with fermion singlets $\chi_\alpha$ coupled to Yukawa couplings with other charged scalars and fermions, in which case we can simply identify the fermion singlet $\chi_\alpha$ with the CFT fermion $\psi_\a$ with $\Delta^{{\rm UV}}_\psi =3/2$.
As far as we know no perturbative gauge theory model with fermion singlets featuring a stable CBZ fixed-point has been constructed in 4d. We thus do not discuss further this possibility.}
\be
\label{eq:UV_composite_fermion_option3}
	\psi_\a \propto \f^{\dagger}_a\chi^a_\a,\quad   \Delta^{{\rm UV}}_\psi =\frac 52  \,.
\ee

We expect the CFTs originating from UV gauge theories to appear on various bounds as some features for the values of $\De_\psi$ in the vicinity of~\eqref{eq:UV_composite_fermion_option1}, \eqref{eq:UV_composite_fermion_option2} and \eqref{eq:UV_composite_fermion_option3}. Aside from the unphysical jumps driven by the fake primary effect, unfortunately we have not observed any such features. There are two main reasons for that. First, similarly to previous studies, the numerical bounds become weak rather quickly as $\De_\psi$ increases from its free field value $3/2$. Thus, theories of the type~\eqref{eq:UV_composite_fermion_option1} or \eqref{eq:UV_composite_fermion_option2} are always deep in the allowed region and cannot generate any kink-like features. However, theories of the type~\eqref{eq:UV_composite_fermion_option3} might still be reachable with stronger (theory specific) assumptions. Second, all the numerical bounds contain the fermion GFT line in the allowed region, however interesting theories might be sitting  below it and thus again cannot manifest themselves as features on the boundary. It turns out to be very difficult to find assumptions which robustly rule out the GFT, but not CBZ fixed points.

The structure of the paper is as follows. We start in section~\ref{sec:setup} by setting up  in detail the stage for our numerical study. 
Section~\ref{sec:conformal_blocks_approximations} is devoted to the computation and rational approximation of the four-fermion conformal blocks.
In section~\ref{sec:GFT} we study the generalized free fermion theory, determining completely its CFT data.  
In section~\ref{sec:topology_blocks} we discuss the fake primary effect in detail and show its impact on numerical bootstrap studies.
We finally present our numerical results in section~\ref{sec:numerical_results}.  We conclude in section~\ref{sec:conclusions} and discuss some further technical details in appendices. Various results of the paper are also summarized in attached Mathematica notebooks.

\section{Setup}
\label{sec:setup}

In this section we define and discuss in detail all the ingredients that are necessary to perform our numerical study.
We use the index-free conventions of~\cite{Cuomo:2017wme}, in which we write
\be
\label{eq:weyl_fermion}
	\psi(x,s)\equiv s^\a \psi_\a(x),\quad
	\bar\psi(x,\bar s) = \left(\psi(x,s)\right)^\dagger\,,
\ee
where $s_\alpha$ and $\bar s^{\dot \a}$ are auxiliary constant polarization spinors. 
We work in the Lorentzian signature, but take the points $x_i$ space-like separated, so that fermions anticommute with each other.
We consider the simplest setup containing a single Weyl fermion, thus the effect of non-abelian global symmetries cannot be addressed. We assume however the existence of a $\mathrm{U}(1)$ global symmetry in the CFT under which  $\psi$ and  $\psi^\dag$ carry $q=+1$ and $q=-1$ charges respectively. Charge conservation implies that the only non-vanishing four-fermion correlator is given by~\eqref{eq:the_fermion_four_point_function}.
Alternatively, we could have considered the case of neutral $\psi$ ($q=0$) (or CFTs with no global $\mathrm{U}(1)$ symmetry) which would require to study a set of four-fermion correlators involving all possible combinations of $\psi$'s and $\psi^\dagger$'s. The constraints imposed by~\eqref{eq:the_fermion_four_point_function} would still be valid but might not be optimal. Space parity or time reversal symmetry are not assumed.\footnote{We discuss implications of parity symmetry and $q=0$ case in appendix~\ref{app:parity}.}

Using a combined argument $\point\equiv(x,s,\bar s)$, the fermion four-point function~\eqref{eq:the_fermion_four_point_function} can compactly be written as
\be\label{eq:masterfourpt}
\<\bar\psi(\point_1)\psi(\point_2)\bar\psi(\point_3)\psi(\point_4)\>.
\ee
Associativity of the OPE requires that the following $s$-$t$ and $u$-$t$ crossing equations should hold:
\begin{align}
\label{eq:s-t-crossing_equations}
\contraction{\<}{\bar\psi}{(\point_1)}{\psi}
\contraction{\<\bar\psi(\point_1)\psi(\point_2)}{\bar\psi}{(\point_3)}{\psi}
\<\bar\psi(\point_1)\psi(\point_2)\bar\psi(\point_3)\psi(\point_4)\>
&=
\contraction{\<}{\bar\psi}{(\point_1)\psi(\point_2)\bar\psi(\point_3)}{\psi}
\contraction[2ex]{\<\bar\psi(\point_1)}{\psi}{(\point_2)}{\bar\psi}
\<\bar\psi(\point_1)\psi(\point_2)\bar\psi(\point_3)\psi(\point_4)\>,\\
\label{eq:u-t-crossing_equations}
\contraction{\<}{\bar\psi}{(\point_1)\psi(\point_2)}{\bar\psi}
\contraction[2ex]{\<\bar\psi(\point_1)}{\psi}{(\point_2)\bar\psi(\point_3)}{\psi}
\<\bar\psi(\point_1)\psi(\point_2)\bar\psi(\point_3)\psi(\point_4)\>
&=
\contraction{\<}{\bar\psi}{(\point_1)\psi(\point_2)\bar\psi(\point_3)}{\psi}
\contraction[2ex]{\<\bar\psi(\point_1)}{\psi}{(\point_2)}{\bar\psi}
\<\bar\psi(\point_1)\psi(\point_2)\bar\psi(\point_3)\psi(\point_4)\>,
\end{align}
where the lines connecting two operators denote their OPE.  

Before continuing the discussion, let us introduce our notation. We will denote operators appearing in the OPE expansions of \eref{eq:s-t-crossing_equations} and \eref{eq:u-t-crossing_equations} as
\begin{equation}
\label{eq:operator_naming}
\cO^{(\ell,\bar\ell)}_{\De,Q},
\end{equation}
Here $Q=0,\pm 2$ is the $\mathrm{U}(1)$ charge. In the rest of the paper we will write for simplicity $Q=0,\,\pm$ instead.\footnote{In principle there can be degeneracies in the spectrum, in which case this notation does not fully specify the primary operators. In such cases an additional label must be added, but for simplicity we will ignore it and discuss degeneracies only when they are important.}
In what follows we will also need the hermitian conjugate operators, which we denote as
\begin{equation}
\label{eq:conjugated_operator_naming}
\bar{\cO}^{(\ell,\bar\ell)}_{\De,Q}(\point) \equiv  \left( \cO^{(\bar \ell,\ell)}_{\De,-Q} (\point)\right)^\dagger\,. 
\end{equation}
Note that in the left-hand side of the above expression the labels refer to the complex conjugate operator. Given the definitions \eqref{eq:conjugated_operator_naming}, we observe that in the traceless symmetric ($\ell=\bar\ell$) case with vanishing charge ($Q=0$) the local operators can be chosen to be hermitian and operators with $\ell\neq\bar\ell$ can be grouped in hermitian-conjugate pairs, i.e.
\be
\label{eq:hermitian_conjugation_neutral}
    \cO^{(\ell,\ell)}_{\De,0}=\bar \cO^{(\ell,\ell)}_{\De,0},\quad
	\cO^{(\ell,\ell+2)}_{\De,0}=\bar \cO^{(\ell,\ell+2)}_{\De,0}.
\ee

With the above notation, the neutral channel OPE reads
\be
\label{eq:neutral_ope}
\contraction{}{\bar\psi}{(\point_1)}{\psi}
\bar\psi(\point_1)\psi(\point_2) =
\sum_{\cO}\sum_a
\lambda_{\<\bar\psi\psi\cO\>}^a C_{\<\bar\psi\psi\cO\>}^a(\point_1,\point_2,\partial_{x_1},\partial_{s},\partial_{\bar s})
\bar\cO^{(\bar\ell,\ell)}_{\De,0}(x_1,s,\bar s),
\ee
where we sum over the primary operators $\cO$ with $Q=0$. 
The operator $\bar\cO$ is the hermitian conjugate of $\cO$. The scaling dimensions of $\cO$ and $\bar\cO$ are denoted by $\De$ and their Lorentz representations by $(\ell,\bar\ell)$ and $(\bar\ell,\ell)$, respectively. As mentioned in the introduction, the allowed Lorentz representations for the exchanged operators $\cO$ are\footnote{In the Young diagram language these mixed-symmetry representations  are represented by a ``hook'' with $\ell+1$ boxes in the first and one box in the second row. The two-row Young diagrams can have (anti-)self-duality constraints in 4d, which is what distinguishes $(\ell+2,\ell)$ from $(\ell,\ell+2)$.} 
\be\label{eq:intermediatelorentz}
(\ell,\bar \ell)\in \left\{(\ell,\ell),\;(\ell+2,\ell),\;(\ell,\ell+2)\right\}.
\ee
It is convenient in the following  to introduce a parameter $p$ defined as
\begin{equation}
p\equiv |\ell - \bar \ell|\,,
\end{equation}
which for TS and NTS operators reads as $p=0$ and $p=2$ respectively. The two types of NTS operators appearing in~\eqref{eq:neutral_ope} are related by hermitian conjugation~\eqref{eq:hermitian_conjugation_neutral} or, equivalently, by the CPT-symmetry.
The $\lambda$'s in~\eqref{eq:neutral_ope} denote the OPE coefficients, while the $C$'s are functions completely fixed by the conformal symmetry that encode the contribution of all the descendant operators associated to $\bar\cO$. The OPE of non-scalar fields involves in general several OPE coefficients and functions $C$, which are taken into account by the index $a$. We use the subscript $\<\bar\psi\psi\cO\>$ as part of the naming in order to fully specify the objects belonging to this particular OPE channel.\footnote{The reason for the use of $\cO$ instead of $\bar\cO$ in the subscript will become clear in section~\ref{sec:three_point_functions}.}

The OPE in the charged channel reads as
\be
\label{eq:charged_ope}
\contraction{}{\psi}{(\point_1)}{\psi}
\psi(\point_1)\psi(\point_2) =
\sum_{\cO}\sum_a
\lambda_{\<\psi\psi\cO\>}^a
C_{\<\psi\psi\cO\>}^a(\point_1,\point_2,\partial_{x_1},\partial_{s},\partial_{\bar s})
\bar\cO^{(\bar\ell,\ell)}_{\De,+}(x_1,s,\bar s),
\ee
where the operators $\cO$ and $\bar\cO$ have $Q=-2$ and $Q=+2$ charges respectively. The spin representations $(\ell,\bar\ell)$  of the operator $\cO$ are the same as in~\eqref{eq:intermediatelorentz}. Since $\cO$ in this sum have non-zero charge $Q$, we cannot relate anymore the $(\ell,\ell+2)$ and $(\ell+2,\ell)$ NTS operators by hermitian conjugation.

Throughout this work we use conventions and notation of~\cite{Cuomo:2017wme}. We also use their ``CFTs4D'' package to perform all the algebraic computations below.

\subsection{Two- and three-point functions}
\label{sec:three_point_functions}

We choose our basis of operators to be orthogonal in the sense that non-vanishing two-point correlation functions only appear for conjugate pairs of operators~\eqref{eq:operator_naming} and~\eqref{eq:conjugated_operator_naming}
\begin{equation}
\label{eq:two-point functions}
\<\bar\cO^{(\bar\ell,\ell)}_{\De,-Q}(\point_1)\cO^{(\ell,\bar\ell)}_{\De,Q}(\point_2)\>=
i^{\ell-\bar\ell}\,
x_{12}^{-2\De-\ell-\bar\ell}
\left(\hat\II^{12}\right)^\ell\left(\hat\II^{21}\right)^{\bar\ell},
\end{equation}
where $\bar\ell=\ell$ or $\bar\ell=\ell+2$ and
\begin{equation}
x_{ij}\equiv |x_i^\mu - x_j^\mu|,\quad
\hat\II^{ij}\equiv x_{ij}^\mu (\bar s_i\bar\sigma_\mu s_j).
\end{equation}
In writing~\eqref{eq:two-point functions} we have used the freedom of changing the normalization of a primary operator to achieve a standard form for all two-point functions.\footnote{Our normalization of two-point function follows the conventions of~\cite{Cuomo:2017wme}.} As we discuss below, for conserved currents there is a different natural normalization, and~\eqref{eq:two-point functions} has to be modified.

The scaling dimensions of operators~\eqref{eq:operator_naming} and~\eqref{eq:conjugated_operator_naming} are subject to the unitarity bounds
\be
\label{eq:unitarity_bound}
\De\geq\De_\text{unitary}(\ell,\bar\ell)=&\begin{cases}
	2+\frac{\ell+\bar\ell}{2}, & \ell\bar \ell\neq 0,\\
	1+\frac{\ell+\bar\ell}{2},& \ell\bar\ell=0.
\end{cases}
\ee
An operator saturating these bounds for $\ell\bar\ell\neq 0$ is necessarily a conserved current.\footnote{The identity operator is the only special case for which the unitarity bounds~\eqref{eq:unitarity_bound} do not apply. Operators saturating the unitarity bounds for $\ell\bar\ell=0$ only exist in free theories~\cite{Weinberg:2012cd}.} Traceless symmetric spin-1 and spin-2 currents are just the familiar global symmetry currents and the stress tensor,
\begin{equation}
J\equiv \cO^{(1,1)}_{3,0},\quad
T\equiv \cO^{(2,2)}_{4,0}.
\end{equation}
Normalization of $J$ and $T$ is fixed by the Ward identities and there is no more freedom in choosing the overall scale of their two-point functions. Thus, instead of~\eqref{eq:two-point functions} their two point functions have the form
\begin{equation}
\label{eq:definitions:central_charges}
\<J(\point_1)J(\point_2)\>
=C_J \times x_{12}^{-8}\,\hat\II^{12}\hat\II^{21},\quad
\<T(\point_1)T(\point_2)\>
=C_T \times x_{12}^{-12}\,\left(\hat\II^{12}\hat\II^{21}\right)^2,
\end{equation}
where $C_J$ and $C_T$ are often called the central charges.

The information contained in the OPE~\eqref{eq:neutral_ope} and~\eqref{eq:charged_ope} is equivalent to the one contained in the three-point functions\footnote{\label{foot:C_and_tensor_structures}To see this one can multiply~\eqref{eq:neutral_ope} and~\eqref{eq:charged_ope} by $\cO$ respectively and take the vacuum expectation value. Given that the two-point function is uniquely determined one obtains a relation between the form of the three-point function and the functions $C$.}
\begin{equation}
\label{eq:three_point_functions}
\<\bar\psi(\point_1)\psi(\point_2)\cO^{(\ell,\bar\ell)}_{\De,0}(\point_3)\>,\quad
\<\psi(\point_1)\psi(\point_2)\cO^{(\ell,\bar\ell)}_{\De,-}(\point_3)\>,
\end{equation}
where the allowed representations $(\ell, \bar\ell)$ are listed in~\eqref{eq:intermediatelorentz}. The three-point functions~\eqref{eq:three_point_functions} have a simple dependence on scaling dimensions through the kinematic factor
\begin{equation} \label{eq:3dkinfactor}
K_3^{-1} \equiv
x_{12}^{2\De_\psi-\De-\ell-p/2+1}\;
x_{13}^{\De+\ell+p/2}\;
x_{23}^{\De+\ell+p/2}.
\end{equation}
Analogously to the case of two-point functions the spin dependence is encoded into tensor structures given by products of basic invariant objects. In case of three-point functions besides $\hat\II$ we get three more invariants $\hat\JJ,\, \hat\KK$ and $\hat{\bar\KK}$. We do not report their explicit form here and instead refer the reader to appendix D in~\cite{Cuomo:2017wme}. In the remainder of this section we analyze the three-point functions~\eqref{eq:three_point_functions} in detail.

\paragraph{Neutral channel} The first class of three-point functions in~\eqref{eq:three_point_functions} reads as
\begin{align}
\nn
\<\bar\psi(\point_1)\psi(\point_2)\cO^{(\ell,\ell)}_{\De,0}(\point_3)\>
&=
\lambda^{a=1}_{\<\bar\psi\psi\cO^{(\ell,\ell)}_{\De,0}\>} \times  K_3\, \hat\II^{12}\left(\hat\JJ_{12}^3\right)^\ell+
\lambda^{a=2}_{\<\bar\psi\psi\cO^{(\ell,\ell)}_{\De,0}\>}\times
K_3\hat\II^{13}\hat\II^{32}\left(\hat\JJ_{12}^3\right)^{\ell-1}
,\\
\nn
\<\bar\psi(\point_1)\psi(\point_2)\cO^{(\ell+2,\ell)}_{\De,0}(\point_3)\>
&= \lambda_{\<\bar\psi\psi\cO^{(\ell+2,\ell)}_{\De,0}\>} \times
K_3\; \hat\II^{13}\hat\KK^{23}_1\left(\hat\JJ_{12}^3\right)^\ell,\\
\label{eq:neutral_correlator_3}
\<\bar\psi(\point_1)\psi(\point_2){\cO}^{(\ell,\ell+2)}_{\De,0}(\point_3)\>
&= \lambda_{\<\bar\psi\psi{\cO}^{(\ell,\ell+2)}_{\De,0}\>}\times
K_3\; \hat\II^{32}\hat{\bar\KK}^{13}_2\left(\hat\JJ_{12}^3\right)^\ell.
\end{align}
Here $\lambda$'s are the OPE coefficients and the objects multiplying them are the tensor structures. For instance, the very first correlator in~\eqref{eq:neutral_correlator_3} has two tensor structures for $\ell\geq 1$. Following~\cite{Karateev:2017jgd} it is convenient to denote tensor structures by
\begin{equation}
\label{eq:tensor_structures_TS_eutral}
\<\bar\psi(\point_1)\psi(\point_2)\cO^{(\ell,\ell)}_{\De,0}(\point_3)\>^{(a)},
\end{equation}
where the superscript $(a)$ enumerates different structures and additionally indicates that it is not a physical correlator. So we have, for example,
\be
\<\bar\psi(\point_1)\psi(\point_2)\cO^{(\ell,\ell)}_{\De,0}(\point_3)\>^{(1)}\equiv& K_3\,\hat\II^{12}\left(\hat\JJ_{12}^3\right)^\ell,\nn\\
\<\bar\psi(\point_1)\psi(\point_2)\cO^{(\ell,\ell)}_{\De,0}(\point_3)\>^{(2)}\equiv &
K_3\,\hat\II^{13}\hat\II^{32}\left(\hat\JJ_{12}^3\right)^{\ell-1}.
\ee
When $\ell=0$ only the $(a)=(1)$ structure exists. When there is a unique tensor structure we often use the notation $(a)=(\uniq)$ to stress the uniqueness, e.g.
\begin{equation}
\label{eq:neutral_correlator_3l0}
\<\bar\psi(\point_1)\psi(\point_2)\cO^{(0,0)}_{\De,0}(\point_3)\>^{(\uniq)} \equiv
K_3\hat\II^{12}.
\end{equation}

Three-point functions in~\eqref{eq:neutral_correlator_3} are invariant under a $\pi_{12}$ permutation augmented by complex conjugation, where 
a general permutation $\pi_{ij}$ is defined by
\begin{equation}
\label{eq:definition_permutation}
\pi_{ij}:\quad\point_i\leftrightarrow\point_j.
\end{equation}
We work in equal time quantization in Lorentzian signature, and complex conjugation acts on a generic $n$-point correlator as follows\footnote{Note that complex conjugation does not act on coordinates of local operators.}
\be
\label{eq:CCdef}
\<\cO_1(\point_1)\cO_2(\point_2) \ldots \cO_n(\point_n)\>^*=\<\bar\cO_n(\point_n)\ldots \bar\cO_2(\point_2)\bar\cO_1(\point_1)\> \,.
\ee
This leads to the following properties of the OPE coefficients:
\begin{equation}
\label{eq:neutral_ope_properties_TS}
\lambda^{*a}_{\<\bar\psi\psi\cO^{(\ell,\ell)}_{\De,0}\>}=
(-1)^{\ell+1}
\lambda^{a}_{\<\bar\psi\psi\cO^{(\ell,\ell)}_{\De,0}\>},\quad
\lambda^*_{\<\bar\psi\psi\cO^{(\ell+2,\ell)}_{\De,0}\>}=
(-1)^{\ell}
\lambda_{\<\bar\psi\psi\bar\cO^{(\ell,\ell+2)}_{\De,0}\>}.
\end{equation}
From~\eqref{eq:neutral_ope_properties_TS} it is clear that the OPE coefficients of TS operators are purely imaginary for even $\ell$ and purely real for odd $\ell$. No similar statement can be made about the NTS operators.

When $\cO$ is a conserved NTS operator, its OPE coefficients in~\eqref{eq:neutral_correlator_3} must vanish to satisfy the conservation constraint.\footnote{This will lead to 
a fake primary effect when discussing upper bounds on scaling dimensions of NTS neutral operators, as we will see in sections~\ref{sec:topology_blocks} and \ref{sec:numerical_results}.}
On the contrary, conserved TS operators automatically satisfy the conservation condition. When $\cO$ is the conserved current $J$ or the stress tensor $T$ one can additionally use the Ward identities to relate the associated OPE coefficients to the $\mathrm{U}(1)$ and conformal charges of $\psi$. For our case this was done in~\cite{Elkhidir:2017iov}.\footnote{See formula (3.15) and appendix A of~\cite{Elkhidir:2017iov}. Note the different conventions between the three-point tensor structure (3.13) in~\cite{Elkhidir:2017iov} and ~\eqref{eq:neutral_set_2} below. There is a relative factor $-1$ in front of the second structure. There is an additional factor of $(-\sqrt{2})^\ell$ due to the difference in vector-spinor map, see appendix~\ref{app:vectorspinor} of this paper.} For the conserved current they find
\begin{equation}
\label{eq:J_ope_relation}
2\,\lambda^1_{\<\bar \psi  \psi J\>}+\lambda^2_{\<\bar \psi \psi J\>} =
\frac{q}{\sqrt{2}\pi^2},
\end{equation}
where $q=+1$.\footnote{As explained later, in our setup we are only sensitive to the ratio $q^2/C_J$, thus we can always reabsorb the charge in the definition of $C_J$. In supersymmetric CFTs one should be more careful: if $\psi$ is part of a chiral multiplet and $J$ is the R-charge, then $q$ is fixed by the superconformal algebra in terms of $\Delta_\psi$ and $C_J$ is fixed in terms of $C_T$.} It is convenient to parametrize the OPE coefficients of $J$ in such a way that the Ward identity~\eqref{eq:J_ope_relation} is manifest; we adopt the following option
\begin{equation}
\label{eq:theta_parametrization}
\lambda^1_{\<\bar \psi  \psi J\>} = \frac{1}{2\sqrt{2}\,\pi^2}\times \frac{\cos\theta}{\cos\theta+\sin\theta},\quad
\lambda^2_{\<\bar \psi  \psi J\>} = \frac{1}{\sqrt{2}\pi^2}\times \frac{\sin\theta}{\cos\theta+\sin\theta},\quad
\theta\in[-\frac{\pi}{4},\;\frac{3\pi}{4}].
\end{equation}
For the stress tensor the result of \cite{Elkhidir:2017iov} reads as
\begin{align}
\lambda^1_{\<\bar\psi\psi T\>}=-\frac{i}{3\pi^2}\times(\De_\psi-3/2),\quad
\lambda^2_{\<\bar\psi\psi T\>}=-\frac{i}{\pi^2}.
\label{eq:TmnOPEcoeff}
\end{align}
We provide a simple derivation of~\eqref{eq:J_ope_relation} and~\eqref{eq:TmnOPEcoeff} in appendix~\ref{app:ward_identities} using weight-shifting operators. Notice that only in the case of stress tensor all the OPE coefficients are fixed. For the conserved current only one linear combination of the OPE coefficients is constrained.

One can consider different ordering of operators in~\eqref{eq:neutral_correlator_3}, and for technical purposes it will be convenient to introduce new bases of tensor structures for them. There are two sets of orderings which are important. The first set reads as
\begin{equation}
\label{eq:neutral_set_1}
\<\cO^{(\ell,\ell)}_{\De,0}(\point_1)\bar\psi(\point_2)\psi(\point_3)\>,\quad
\<\cO^{(\ell+2,\ell)}_{\De,0}(\point_1)\bar\psi(\point_2)\psi(\point_3)\>,\quad
\<\cO^{(\ell,\ell+2)}_{\De,0}(\point_1)\bar\psi(\point_2)\psi(\point_3)\>.
\end{equation}
These orderings can be obtained by applying hermitian conjugation and $\pi_{13}$ permutation to~\eqref{eq:neutral_correlator_3}. We define the basis of the tensor structures for~\eqref{eq:neutral_set_1} by applying this procedure to tensor structures for~\eqref{eq:neutral_correlator_3}. Then the associated OPE coefficients are related to the ones in~\eqref{eq:neutral_correlator_3} in the following simple way
\begin{equation}
\label{eq:neutral_set1_ope_relations}
\lambda^{*a}_{\<\bar\psi\psi\cO^{(\ell,\ell)}_{\De,0}\>}=
\lambda^{a}_{\<\cO^{(\ell,\ell)}_{\De,0}\bar\psi\psi\>},\quad
\lambda^*_{\<\bar\psi\psi\cO^{(\ell,\ell+2)}_{\De,0}\>}=
\lambda_{\<{{\bar\cO}}^{(\ell+2,\ell)}_{\De,0}\bar\psi\psi\>},\quad
\lambda^*_{\<\bar\psi\psi\cO^{(\ell+2,\ell)}_{\De,0}\>}=
\lambda_{\<\bar\cO^{(\ell,\ell+2)}_{\De,0}\bar\psi\psi\>}.
\end{equation}
However, the two orderings can be related also by simply permuting the operators. Using permutations to relate three-point functions we find
\begin{equation}
\label{eq:neutral_permutation_relation_TS}
\lambda^{a}_{\<\bar\psi\psi\cO^{(\ell,\ell)}_{\De,0}\>}=(-1)^{\ell+1}
\lambda^{a}_{\<\cO^{(\ell,\ell)}_{\De,0}\bar\psi\psi\>}
\end{equation}
for TS operators and
\begin{equation}
\lambda_{\<\bar\psi\psi\cO^{(\ell+2,\ell)}_{\De,0}\>}=(-1)^\ell
\lambda_{\<\cO^{(\ell+2,\ell)}_{\De,0}\bar\psi\psi\>},\quad
\lambda_{\<\bar\psi\psi\cO^{(\ell,\ell+2)}_{\De,0}\>}=(-1)^\ell
\lambda_{\<\cO^{(\ell,\ell+2)}_{\De,0}\bar\psi\psi\>}
\end{equation}
for NTS operators. This is consistent with~\eqref{eq:neutral_ope_properties_TS}.

The second set of orderings reads as
\begin{equation}
\label{eq:neutral_set_2}
\<\cO^{(\ell,\ell)}_{\De,0}(\point_1)\psi(\point_2)\bar\psi(\point_3)\>,\quad
\<\cO^{(\ell+2,\ell)}_{\De,0}(\point_1)\psi(\point_2)\bar\psi(\point_3)\>,\quad
\<\cO^{(\ell,\ell+2)}_{\De,0}(\point_1)\psi(\point_2)\bar\psi(\point_3)\>.
\end{equation}
Theses are related to~\eqref{eq:neutral_set_1} by applying $\pi_{12}$ permutation and adding an overall minus sign coming from the anti-commutation of fermions. We use this procedure to obtain the basis of tensor structures for~\eqref{eq:neutral_set_2} from the basis for~\eqref{eq:neutral_set_1}. This leads to the following relations between the OPE coefficients
\begin{equation}
\label{eq:neutral_set2_ope_relations}
\lambda^{a}_{\<\cO^{(\ell,\ell)}_{\De,0}\psi\bar\psi\>}=
\lambda^{a}_{\<\cO^{(\ell,\ell)}_{\De,0}\bar\psi\psi\>},\quad
\lambda_{\<\cO^{(\ell+2,\ell)}_{\De,0}\psi\bar\psi\>}=
\lambda_{\<\cO^{(\ell+2,\ell)}_{\De,0}\bar\psi\psi\>},\quad
\lambda_{\<\cO^{(\ell,\ell+2)}_{\De,0}\psi\bar\psi\>}=
\lambda_{\<\cO^{(\ell,\ell+2)}_{\De,0}\bar\psi\psi\>}.
\end{equation}

\paragraph{Charged channel} The second class of three-point functions in~\eqref{eq:three_point_functions} reads as
\begin{align}
\label{eq:charged_correlator_1}
\<\psi(\point_1)\psi(\point_2)\cO^{(\ell,\ell)}_{\De,-}(\point_3)\>
&= \lambda_{\<\psi\psi\cO^{(\ell,\ell)}_{\De,-}\>}\times
K_3\left(\hat\II^{31}\hat\KK^{23}_1+(-1)^\ell\, \hat\II^{32}\hat\KK^{13}_2\right)
\left(\hat\JJ_{12}^3\right)^{\ell-1},\\
\label{eq:charged_correlator_2}
\<\psi(\point_1)\psi(\point_2)\cO^{(\ell+2,\ell)}_{\De,-}(\point_3)\>
&= \lambda_{\<\psi\psi\cO^{(\ell+2,\ell)}_{\De,-}\>}\times
K_3\;\hat{\KK}^{13}_2\hat{\KK}^{23}_1\left(\hat\JJ_{12}^3\right)^\ell
,\quad\ell\in\text{odd},\\
\label{eq:charged_correlator_3}
\<\psi(\point_1)\psi(\point_2)\cO^{(\ell,\ell+2)}_{\De,-}(\point_3)\>
&= \lambda_{\<\psi\psi\cO^{(\ell,\ell+2)}_{\De,-}\>}\times
K_3\;\hat{\II}^{31}\hat{\II}^{32}\left(\hat\JJ_{12}^3\right)^\ell,\quad\;\;\,\ell\in\text{odd},
\end{align}
The expression~\eqref{eq:charged_correlator_1} holds for $\ell\geq 1$ and contains in general two tensor structures. However an extra constraint must be imposted due to presence of identical fermions
\begin{equation}
\<\psi(\point_1)\psi(\point_2)\cO(\point_3)\>=
-\<\psi(\point_2)\psi(\point_1)\cO(\point_3)\>,
\end{equation}
which relates two structures. The same constraint is also responsible for removing even $\ell$ operators from the correlation functions~\eqref{eq:charged_correlator_2} and~\eqref{eq:charged_correlator_3}. In the special $\ell=0$ case there is a unique tensor structure in~\eqref{eq:charged_correlator_1} given by
\begin{equation}
\<\psi(\point_1)\psi(\point_2)\cO^{(0,0)}_{\De,-}(\point_3)\>^{(\uniq)}
\equiv K_3\hat\KK^{12}_3.
\end{equation}
When $\cO$ is a conserved TS operator, its OPE coefficients in~\eqref{eq:charged_correlator_1} must vanish to satisfy the conservation constraint.\footnote{This will lead to 
a fake primary effect when discussing upper bounds on scaling dimensions of TS charged operators, as we will see in sections~\ref{sec:topology_blocks} and \ref{sec:numerical_results}.} Thus, no conserved TS operators are allowed to appear in this channel. On the contrary, conserved NTS operators automatically satisfy the conservation condition.

Finally, the following set of three-point functions
\begin{equation}
\label{eq:charged_right}
\<\cO^{(\ell,\ell)}_{\De,+}(\point_1)\bar\psi(\point_2)\bar\psi(\point_3)\>,\quad
\<\cO^{(\ell+2,\ell)}_{\De,+}(\point_1)\bar\psi(\point_2)\bar\psi(\point_3)\>,\quad
\<\cO^{(\ell,\ell+2)}_{\De,+}(\point_1)\bar\psi(\point_2)\bar\psi(\point_3)\>
\end{equation}
is related to~\eqref{eq:charged_correlator_1}-\eqref{eq:charged_correlator_3} by complex conjugation and $\pi_{13}$ permutation. The tensor structures for~\eqref{eq:charged_right} are obtained by this procedure. This implies that the OPE coefficients of two sets of correlators are related as follows
\begin{equation}
\label{eq:charged_ope_relations}
\lambda^{*a}_{\<\psi\psi\cO^{(\ell,\ell)}_{\De,-}\>}=
\lambda^{a}_{\<{\bar\cO}^{(\ell,\ell)}_{\De,+}\bar\psi\bar\psi\>},\quad
\lambda^*_{\<\psi\psi\cO^{(\ell,\ell+2)}_{\De,-}\>}=
\lambda_{\<{\bar\cO}^{(\ell+2,\ell)}_{\De,+}\bar\psi\bar\psi\>},\quad
\lambda^*_{\<\psi\psi\cO^{(\ell+2,\ell)}_{\De,-}\>}=
\lambda_{\<\bar\cO^{(\ell,\ell+2)}_{\De,+}\bar\psi\bar\psi\>}.
\end{equation}

\subsection{Four-point tensor structures}
\label{sec:4ptstructures}

We now analyze in detail the four-point function~\eqref{eq:masterfourpt} and the crossing equations~\eqref{eq:s-t-crossing_equations} and~\eqref{eq:u-t-crossing_equations} it must satisfy. We begin by using the anticommutation properties of space-like separated fermions to rewrite the crossing equations in the following form 
\begin{align}
\label{eq:s-t-crossing_equations_final}
\contraction{\<}{\bar\psi}{(\point_1)}{\psi}
\contraction{\<\bar\psi(\point_1)\psi(\point_2)}{\bar\psi}{(\point_3)}{\psi}
\<\bar\psi(\point_1)\psi(\point_2)\bar\psi(\point_3)\psi(\point_4)\>
&=
-\pi_{13}\contraction{\<}{\bar\psi}{(\point_1)}{\psi}
\contraction{\<\bar\psi(\point_1)\psi(\point_2)}{\bar\psi}{(\point_3)}{\psi}
\<\bar\psi(\point_1)\psi(\point_2)\bar\psi(\point_3)\psi(\point_4)\>,\\
\label{eq:u-t-crossing_equations_final}
\contraction{\<}{\psi}{(\point_1)}{\psi}
\contraction{\<\bar\psi(\point_1)\psi(\point_2)}{\bar\psi}{(\point_3)}{\psi}
\<\psi(\point_1)\psi(\point_2)\bar\psi(\point_3)\bar\psi(\point_4)\>
&=
-\pi_{13}\contraction{\<}{\bar\psi}{(\point_1)}{\psi}
\contraction{\<\bar\psi(\point_1)\psi(\point_2)}{\bar\psi}{(\point_3)}{\psi}
\<\bar\psi(\point_1)\psi(\point_2)\psi(\point_3)\bar\psi(\point_4)\>,
\end{align}
where we have used the short-hand notation~\eqref{eq:definition_permutation} for permutation of points and the fact that $\pi_{41}\pi_{13}\pi_{34}=\pi_{13}$.
We have expressed these equations using the $\pi_{13}$ permutation because it acts on the standard cross-ratios $(z,\bar z)$ defined as
\be
\label{eq:definition_(z,zb)}
z\bar z \equiv u \equiv \frac{x_{12}^2x_{34}^2}{x_{13}^2x_{24}^2},\qquad
(1-z)(1-\bar z) \equiv v \equiv \frac{x_{14}^2x_{23}^2}{x_{13}^2x_{24}^2},
\ee
in a very simple way:
\be
\pi_{13}:\quad (z,\bar z)\mapsto (1-z,1-\bar z).
\ee
This fact allows us to study the crossing equations~\eqref{eq:s-t-crossing_equations_final} and~\eqref{eq:u-t-crossing_equations_final} in power series around the $\pi_{13}$ crossing symmetric point $z=\bar z=1/2$, as pioneered in~\cite{Rattazzi:2008pe}.

Instead of working with the $s$-, $t$- and $u$-channel conformal block expansions, the crossing equations~\eqref{eq:s-t-crossing_equations} and~\eqref{eq:u-t-crossing_equations} rewritten in the new form~\eqref{eq:s-t-crossing_equations_final} and~\eqref{eq:u-t-crossing_equations_final} require only the $s$-channel conformal block expansion. There are however three different four-point functions entering these equations
\be\label{eq:basicorderings}
\<\bar\psi(\point_1)\psi(\point_2)\bar\psi(\point_3)\psi(\point_4)\>,\quad
\<\bar\psi(\point_1)\psi(\point_2)\psi(\point_3)\bar\psi(\point_4)\>,\quad
\<\psi(\point_1)\psi(\point_2)\bar\psi(\point_3)\bar\psi(\point_4)\>.
\ee
Since all theses four-point functions are related by permutation we do not refer to them as three different four-point functions, but rather as three different orderings of operators in the four-point function~\eqref{eq:masterfourpt}.

In what follows we define a basis of tensor structures for three orderings~\eqref{eq:basicorderings} and study their properties. While these orderings are related by simple permutations, it is technically useful to introduce three different bases of tensor structures for them. Contrary to section~\ref{sec:three_point_functions} we will be working in the conformal frame~\cite{Osborn:1993cr,Kravchuk:2016qvl}.\footnote{We do not use the embedding formalism~\cite{SimmonsDuffin:2012uy,Elkhidir:2014woa} for four-point tensor structures since it suffers from redundancies which become worse when combined with permutation symmetries and crossing transformations. Conformal frame structures, on the other hand, are manifestly free of redundancies and transform in a simple way under permutations and crossing. For three-point functions in 4d the embedding formalism is, however, more convenient since it is manifestly covariant and has very little redundancies which can be tamed.} In this formalism we put the four operators in standard positions, parametrized by the cross-ratios $z$ and $\bar z$, 
\be\label{eq:cf}
	x_1=(0,0,0,0),\quad x_2=(\tfrac{\bar z-z}{2},0,0,\tfrac{\bar z+z}{2}),\quad x_3=(0,0,0,1),\quad x_4=(0,0,0,\oo),
\ee
and study the tensor structures as invariants of the  $\SO(2)$ little group. We refer the reader to~\cite{Cuomo:2017wme} for details of this formalism applied to 4d CFTs.

\paragraph{Ordering $\<\bar\psi\psi\bar\psi\psi\>$} 

The four-point function~\eqref{eq:masterfourpt} can be decomposed in a basis of four-point tensors structures. Before imposing any constraints other than conformal invariance, we find 6 structures,\footnote{Note that in the conformal frame the functions $g_i^0$ include the contribution $(z\bar z)^{-\Delta_\psi-1/2}$ coming from the covariant kinematic factor $(x_{12}^2x_{34}^2)^{-\Delta_\psi-1/2}$. As a matter of fact, this is the only term in $g_i^0$ that explicitly depends on $\Delta_\psi$.\label{foot:K4}}
\be\label{eq:g0}
	\<\bar\psi(\point_1)\psi(\point_2)\bar\psi(\point_3)\psi(\point_4)\>=\sum_{i=1}^6 \mathbb{T}^{0}_i g_i^0(z,\bar z).
\ee
We put the superscript ``0'' because we will shortly define a second basis. In the notation of~\cite{Cuomo:2017wme} (see in particular section 4.1.2 of~\cite{Cuomo:2017wme}) we define
\be
\T^0_1\equiv&\struct
	{0}{+\thalf}{0}{-\thalf}
	{+\thalf}{0}{-\thalf}{0},\qquad
\T^0_2\equiv\struct
	{0}{-\thalf}{0}{+\thalf}
	{-\thalf}{0}{+\thalf}{0},\nn\\
\T^0_3\equiv&\struct
	{0}{-\thalf}{0}{+\thalf}
	{+\thalf}{0}{-\thalf}{0},\qquad
\T^0_4\equiv\struct
	{0}{+\thalf}{0}{-\thalf}
	{-\thalf}{0}{+\thalf}{0},\label{eq:gchan4structures0}\\
\T^0_5\equiv&\struct
	{0}{+\thalf}{0}{+\thalf}
	{+\thalf}{0}{+\thalf}{0},\qquad
\T^0_6\equiv\struct
	{0}{-\thalf}{0}{-\thalf}
	{-\thalf}{0}{-\thalf}{0}.\nn
\ee
For analyzing further constraints it is convenient to introduce the following change of basis,
\be\label{eq:g4ptstructsconvenient}
\begin{matrix*}[l]
\T_{1,+}\equiv\frac{1}{z}\T^0_1+\frac{1}{\bar z}\T^0_2,&\qquad
\T_{1,-}\equiv\frac{i}{z}\T^0_1-\frac{i}{\bar z}\T^0_2,\\
\T_{2,+}\equiv\frac{1}{1-\bar z}\T^0_3+\frac{1}{1-z}\T^0_4,&\qquad
\T_{2,-}\equiv\frac{1}{1-\bar z}\T^0_3-\frac{1}{1-z}\T^0_4,\\
\T_{3,+}\equiv\T^0_5+\T^0_6,&\qquad
\T_{3,-}\equiv\T^0_5-\T^0_6,
\end{matrix*}
\ee
and the appropriate decomposition
\be
    \label{eq:ordering_1_tensor_structure_decomposition}
	\<\bar\psi(\point_1)\psi(\point_2)\bar\psi(\point_3)\psi(\point_4)\>=\sum_{i=1,\pm}^3 \mathbb{T}_{i,\pm} g_{i,\pm}(z,\bar z).
\ee

The functions $g_{i,\pm}(z,\bar z)$ entering~\eqref{eq:ordering_1_tensor_structure_decomposition} are not constrained by conformal symmetry, but should obey further non-trivial constraints coming from permutation symmetry, reality, parity invariance and smoothness
of the four-point function.

Let us first understand the permutation symmetry. Using the anti-commutation properties of space-like separated fermions we can write
\begin{equation}
\label{eq:1324invariance}
\<\bar\psi(\point_1)\psi(\point_2)\bar\psi(\point_3)\psi(\point_4)\>
=+\pi_{13}\pi_{24}
\<\bar\psi(\point_1)\psi(\point_2)\bar\psi(\point_3)\psi(\point_4)\>.
\end{equation}
Contrary to $\pi_{13}$, the permutation $\pi_{13}\pi_{24}$ does not change the cross-ratios $(z,\bar z)$. This property makes it a constraint on the functions $g_{i,\pm}(z,\bar z)$ at a single point in the $(z,\bar z)$-plane, rather than a relation between different points. We refer to permutations which do not change the cross-ratios as kinematic permutations.
Using the results of~\cite{Cuomo:2017wme} we can immediately infer the constraints which~\eqref{eq:1324invariance} implies for the functions $g_{i,\pm}(z,\bar z)$. We find
\be
\label{eq:permutation_constraints}
	g_{1,-}(z,\bar z)=g_{2,-}(z,\bar z)=0.
\ee
Note that permutations do not leave the conformal frame~\eqref{eq:cf} invariant, which leads to functions of $z$ and $\bar z$ appearing in the action of the permutations on conformal frame structures~\cite{Kravchuk:2016qvl,Cuomo:2017wme}. This explains the $z$- and $\bar z$-dependent prefactors in~\eqref{eq:g4ptstructsconvenient}.

Let us now address the reality constraints. The action of complex conjugation~\eqref{eq:CCdef} to the four-point function~\eqref{eq:ordering_1_tensor_structure_decomposition}
implies that the functions $g_{i,\pm}$ are real,
\be
	g_{i,\pm}^*(z,\bar z)=g_{i,\pm}(z,\bar z),\quad i=1,2,3. 
\ee
Imposing parity symmetry does not lead to further constraints on the four-point function. It gives however some constraints at the level of three-point functions, see appendix~\ref{app:parity}.

The last constraint to consider comes from the smoothness properties~\cite{Kravchuk:2016qvl,Dymarsky:2017yzx} of the four-point function~\eqref{eq:masterfourpt} which lead to the following
\begin{align}
\label{eq:gzzbparity}
&g_{i,\pm}(z,\bar z)=\pm g_{i,\pm}(\bar z,z),\\
\label{eq:gzzbregularity}
&\frac{1}{z}g_{1,+}(z,z)+\frac{1}{1-z}g_{2,+}(z,z)+g_{3,+}(z,z)=0.
\end{align}
The first constraint~\eqref{eq:gzzbparity} comes from the fact that $z$ and $\bar z$ can be exchanged by a rotation, while the second constraint~\eqref{eq:gzzbregularity} comes from analyzing how the basis of tensor structures~\eqref{eq:g4ptstructsconvenient} degenerates as $(z-\bar z)\to 0$. See appendix~\ref{app:analyticity} for details.

\paragraph{Orderings $\<\bar\psi\psi\psi\bar\psi\>$ and $\<\psi\psi\bar\psi\bar\psi\>$}
The decomposition of these orderings into tensor structures is as follows\footnote{Again the dependence on the external dimension $\Delta_\psi$ is hidden inside the functions $g'$ and $h$.}
\be
\label{eq:ordering_2_tensor_structure_decomposition}
\<\bar\psi(\point_1)\psi(\point_2)\psi(\point_3)\bar\psi(\point_4)\>&=\sum_{i,\pm} \T'_{i,\pm} g'_{i,\pm}(z,\bar z),\\
\label{eq:ordering_3_tensor_structure_decomposition}
\<\psi(\point_1)\psi(\point_2)\bar\psi(\point_3)\bar\psi(\point_4)\>&=\sum_{i,\pm}\Q_{i,\pm}h_{i,\pm}(z,\bar z).
\ee
The basis of structures entering~\eqref{eq:ordering_2_tensor_structure_decomposition} is defined as
\be
\T^{\prime 0}_1\equiv&\struct
{0}{+\thalf}{-\thalf}{0}
{+\thalf}{0}{0}{-\thalf},\qquad
\T^{\prime 0}_2\equiv\struct
{0}{-\thalf}{+\thalf}{0}
{-\thalf}{0}{0}{+\thalf},\\
\T^{\prime 0}_3\equiv&\struct
{0}{-\thalf}{+\thalf}{0}
{+\thalf}{0}{0}{-\thalf},\qquad 
\T^{\prime 0}_4\equiv\struct
{0}{+\thalf}{-\thalf}{0}
{-\thalf}{0}{0}{+\thalf},\\
\T^{\prime 0}_5\equiv&\struct
{0}{+\thalf}{+\thalf}{0}
{+\thalf}{0}{0}{+\thalf},\qquad
\T^{\prime 0}_6\equiv\struct
{0}{-\thalf}{-\thalf}{0}
{-\thalf}{0}{0}{-\thalf},
\ee
\be
\label{eq:gprime4ptstructsconvenient}
\begin{matrix*}[l]
	\T'_{1,+}\equiv\frac{1}{z}\T^{\prime 0}_1+\frac{1}{\bar z}\T^{\prime 0}_2,
	&\quad\T'_{1,-}\equiv\frac{1}{z}\T^{\prime 0}_1-\frac{1}{\bar z}\T^{\prime 0}_2,\\
	\T'_{2,+}\equiv\T^{\prime 0}_3+\T^{\prime 0}_4,
	&\quad\T'_{2,-}\equiv i\T^{\prime 0}_3-i\T^{\prime 0}_4,\\
	\T'_{3,+}\equiv\T^{\prime 0}_5+\T^{\prime 0}_6,
	&\quad\T'_{3,-}\equiv\T^{\prime 0}_5-\T^{\prime 0}_6.
\end{matrix*}
\ee
The basis of structures entering~\eqref{eq:ordering_3_tensor_structure_decomposition} is defined as
\be
\Q^{ 0}_1\equiv&\struct
{-\thalf}{+\thalf}{0}{0}
{0}{0}{+\thalf}{-\thalf},\qquad
\Q^{ 0}_2\equiv\struct
{+\thalf}{-\thalf}{0}{0}
{0}{0}{-\thalf}{+\thalf},\\
\Q^{ 0}_3\equiv&\struct
{+\thalf}{-\thalf}{0}{0}
{0}{0}{+\thalf}{-\thalf},\qquad
\Q^{ 0}_4\equiv\struct
{-\thalf}{+\thalf}{0}{0}
{0}{0}{-\thalf}{+\thalf},\\
\Q^{ 0}_5\equiv&\struct
{+\thalf}{+\thalf}{0}{0}
{0}{0}{+\thalf}{+\thalf},\qquad
\Q^{ 0}_6\equiv\struct
{-\thalf}{-\thalf}{0}{0}
{0}{0}{-\thalf}{-\thalf},
\ee
\be
\begin{matrix*}[l]
	\label{eq:h4ptstructsconvenient}
	\Q_{1,+}\equiv\frac{1}{1-z}\Q^{ 0}_1+\frac{1}{1-\bar z}\Q^{ 0}_2,
	&\quad\Q_{1,-}\equiv\frac{1}{1-z}\Q^{ 0}_1-\frac{1}{1-\bar z}\Q^{ 0}_2,\\
	\Q_{2,+}\equiv\Q^{ 0}_3+\Q^{ 0}_4,
	&\quad\Q_{2,-}\equiv i\Q^{ 0}_3-i\Q^{ 0}_4,\\
	\Q_{3,+}\equiv\Q^{ 0}_5+\Q^{ 0}_6,
	&\quad\Q_{3,-}\equiv\Q^{ 0}_5-\Q^{ 0}_6.
\end{matrix*}
\ee
Of course, the functions $g'$ and $h$ are not independent of the functions $g$ since they encode the same four-point function.

We omit the identical reasoning leading to the properties of $g'$- and $h$-functions analogous to those derived for the $g$-functions and provide the final summary only.

The same way we have derived the properties of $g$-functions in the previous paragraph, we can derive the analogous properties of $g'$- and $h$-functions. We omit the identical reasoning and provide the final summary only. Permutation symmetry requires
\be
g'_{1,-}(z,\bar z)=g'_{2,-}(z,\bar z)=h_{1,-}(z,\bar z)=h_{2,-}(z,\bar z)=0.
\label{eq:permutation_constraints_gph}
\ee
Complex conjugation implies
\be
g'^*_{i,\pm}(z,\bar z)=g'_{i,\pm}(z,\bar z),\quad
h^*_{i,\pm}(z,\bar z)=h_{i,\pm}(z,\bar z).
\ee
Finally the smoothness
of $g'$- and $h$-functions implies
\be\label{eq:gpzzbparity_and_hzzbparity}
g'_{i,\pm}(\bar z,z)=\pm g'_{i,\pm}(z,\bar z),\quad
h_{i,\pm}(\bar z,z)=\pm h_{i,\pm}(z,\bar z)
\ee
together with
\be\label{eq:gpzzbregularity}
\frac{1}{z}g'_{1,+}(z,z)+g'_{2,+}(z,z)+g'_{3,+}(z,z)&=0,\\
\frac{1}{1-z}h_{1,+}(z,z)+h_{2,+}(z,z)+h_{3,+}(z,z)&=0.\label{eq:hzzbregularity}
\ee

\subsection{Crossing equations}
\label{subsec:crossing_equations}

We can now plug the tensor structure decompositions~\eqref{eq:ordering_1_tensor_structure_decomposition}, \eqref{eq:ordering_2_tensor_structure_decomposition} and~\eqref{eq:ordering_3_tensor_structure_decomposition} into the crossing equations~\eqref{eq:s-t-crossing_equations_final} and~\eqref{eq:u-t-crossing_equations_final}. Applying the $\pi_{13}$ permutation it is easy to show that they translate into the following set of crossing equations:
\be	
\label{eq:crossing_equations_set_1}
	g_{1,+}(1-z,1-\bar z)&=g_{2,+}(z,\bar z),\nn\\
	g_{3,+}(1-z,1-\bar z)&=g_{3,+}(z,\bar z),\nn\\
	g_{3,-}(1-z,1-\bar z)&=g_{3,-}(z,\bar z),	
\ee
and 
\be
\label{eq:crossing_equations_set_2}
	g'_{1,+}(1-z,1-\bar z)&=h_{1,+}(z,\bar z),\nn\\
	g'_{2,+}(1-z,1-\bar z)&=h_{2,+}(z,\bar z),\nn\\
	g'_{3,\pm}(1-z,1-\bar z)&=h_{3,\pm}(z,\bar z).
\ee
Following~\cite{Rattazzi:2008pe} we study these crossing equations by expanding them in a power series around $z=\bar z=1/2$. For this purpose instead of $(z,\bar z)$ it is convenient to define new coordinates $x$, $y$ and $t$ as
\be
\label{eq:xyt_variables}
z\equiv x+y+\half,\qquad \bar z \equiv x-y+\half,\qquad t \equiv y^2.
\ee
Using these variables we define new functions
\be\label{eq:final4ptfunctions}
	\tl f_\ii(x,t)\equiv
\begin{cases}
f_{\ii,+}(z,\bar z),& \ii=1,2,3\\
\frac{1}{y}\,f_{\ii-3,-}(z,\bar z),& \ii=4,5,6
\end{cases},
\ee
where $f$ represents $g$, $g'$, or $h$. The new functions are smooth functions of $t$ due to the constraints~\eqref{eq:gzzbparity} and~\eqref{eq:gpzzbparity_and_hzzbparity}. This allows to rewrite the crossing equations \eqref{eq:crossing_equations_set_1} and \eqref{eq:crossing_equations_set_2} in the following way:
\begin{alignat}{3}
	&\ptl_x^m \ptl_t^n \tl g_1(0,0) &&= (-1)^m \ptl_x^m \ptl_t^n \tl g_2(0,0),  &&\qquad n,m\geq 0, \label{eq:crossingGfirst}
	\\
	&\ptl_x^m \ptl_t^n \tl g_3(0,0) &&=0, &&\qquad m\geq 0,\quad m\text{ odd},\;\; n\geq 1\label{eq:gredundant}\\
	&\ptl_x^m \ptl_t^n \tl g_6(0,0) &&=0, &&\qquad m,n\geq 0,\quad m\text{ even},\\
	& && && \nn \\
&\ptl_x^m \ptl_t^n \tl g'_1(0,0) &&= (-1)^m \ptl_x^m \ptl_t^n \tl h_1(0,0),  &&\qquad n,m\geq 0,
\\
&\ptl_x^m \ptl_t^n \tl g'_2(0,0) &&= (-1)^m \ptl_x^m \ptl_t^n \tl h_2(0,0),  &&\qquad n,m\geq 0,
\\
&\ptl_x^m \ptl_t^n \tl g'_3(0,0) &&= (-1)^m \ptl_x^m \ptl_t^n \tl h_3(0,0),  &&\qquad m\geq 0,\; n\geq 1\label{eq:gphredundant}
\\&\ptl_x^m \ptl_t^n \tl g'_6(0,0) &&= (-1)^{m+1} \ptl_x^m \ptl_t^n \tl h_6(0,0),  &&\qquad n,m\geq 0. \label{eq:crossingGlast}
\end{alignat}
Note that the constraints~\eqref{eq:gzzbregularity}, \eqref{eq:gpzzbregularity}, and~\eqref{eq:hzzbregularity} imply a linear relation between $\tl f_\ii$ for $\ii=1,2,3$ at $t=0$, which allows us to express $\tl f_3(x,0)$ in terms of  $\tl f_1(x,0)$ and $\tl f_2(x,0)$. The crossing equations involving $\tl g_3$, $\tl g_3'$ and $\tl h_3$ with no $\ptl_t$ derivatives are thus redundant. This explains why $n\geq 1$ in~\eqref{eq:gredundant} and~\eqref{eq:gphredundant}.

\subsection{Decomposition into conformal blocks}
\label{sec:fermion_blocks}
By using the OPE~\eqref{eq:neutral_ope} and~\eqref{eq:charged_ope} we can express the three orderings~\eqref{eq:basicorderings} as sums over contributions of individual primary operators. This allows to express the functions $g$, $g'$ and $h$ (or equivalently $\tl g$, $\tl g'$ and $\tl h$) in terms of the CFT data. In this section we discuss these decompositions.

\paragraph{Orderings $\<\bar\psi\psi\bar\psi\psi\>$ and $\<\bar\psi\psi\psi\bar\psi\>$.}
We start by studying the $s$-channel OPE decomposition of the first ordering in~\eqref{eq:basicorderings}. We apply the OPE~\eqref{eq:neutral_ope} twice to a pair of operators at positions 1, 2 and 3, 4. Using the properties of two-point functions one arrives at
\begin{align}
\contraction{\<}{\bar\psi}{(\point_1)}{\psi}
\contraction{\<\bar\psi(\point_1)\psi(\point_2)}{\bar\psi}{(\point_3)}{\psi}
\<\bar\psi(\point_1)\psi(\point_2)\bar\psi(\point_3)\psi(\point_4)\>
&=\sum_\cO\sum_{a,b} \lambda_{\<\bar\psi\psi\cO\>}^a \lambda_{\<\bar\cO\bar\psi\psi\>}^{b}
G_{\cO}^{ab}(\point_i),\label{eq:decomposition_1}
\end{align}
where the sum runs over the primary operators $\cO$ exactly as in~\eqref{eq:neutral_ope}, the OPE coefficients $\lambda$ are defined in section~\ref{sec:three_point_functions} and
the functions
\begin{equation}
\label{eq:conformal_block}
G_{\cO}^{ab}(\point_i)\equiv
C_{\<\bar\psi\psi\cO\>}^a
C_{\<\bar\cO\bar\psi\psi\>}^b
\<\bar\cO^{(\bar\ell,\ell)}_{\De,0}(x_1,s,\bar s)
\cO^{(\ell,\bar\ell)}_{\De,0}(x_3,t,\bar t)\>
\end{equation}
are called the conformal blocks. We omit here for brevity the arguments of the functions $C$. Since the $C$'s are completely fixed by the conformal symmetry, the conformal blocks are also completely fixed and represent the contribution of the primary operator $\cO$ and all its descendants.  
The conformal blocks depend on the scaling dimension $\De$ and the Lorentz representation $(\ell, \bar\ell)$ of $\cO$:\footnote{In general they also depend non-trivially on the scaling dimension of the ``external operators", but only through their differences.  
When all the external fermions have the same scaling dimension $\Delta_\psi$, as in our case, we are only left with the trivial dependence proportional to $(z\bar z)^{-\Delta_\psi-1/2}$ coming from the kinematic factor. See footnote~\ref{foot:K4}.}
we will then often use the following more explicit labelling
\begin{equation}
G^{ab}_{\De,(\ell,\bar\ell)}(\point_i)\equiv G_{\cO}^{ab}(\point_i). 
\end{equation}
Equivalently to~\eqref{eq:conformal_block} one can write the blocks as a certain gluing of three-point tensor structures~\cite{SimmonsDuffin:2012uy,Karateev:2017jgd}, which we denote by
\begin{equation}
\label{eq:conformal_block_1}
G_{\cO}^{ab}(\point_i)=
\<\bar\psi (\point_1)\psi (\point_2) \cO \>^{(a)} \bowtie \<\bar\cO\bar\psi (\point_3)\psi (\point_4)\>^{(b)},
\end{equation}
where the operation $\bowtie$ roughly corresponds to an integral over the coordinates and a sum over the polarizations of $\cO$ and $\bar\cO$. Its precise definition is unnecessary for the purposes of this paper. The calculation of the conformal blocks~\eqref{eq:conformal_block} or equivalently~\eqref{eq:conformal_block_1} presents the main technical challenge in this paper which we postpone to section~\ref{sec:conformal_blocks_approximations}.

All the conformal blocks we compute are normalized in such a way that the two-point function entering~\eqref{eq:conformal_block} is given by~\eqref{eq:two-point functions}. In case of the conserved currents $J$ and the stress tensor $T$ the correct normalization of the two-point functions is instead given by~\eqref{eq:definitions:central_charges}. 
Recalling that the $C$'s in \eqref{eq:conformal_block} also depend on the two point function normalization, we get that the associated conformal blocks should be rescaled as
\begin{equation}
\label{eq:conserved_block_rescaling}
G_J^{ab} \longrightarrow \frac{1}{C_J}\times G_J^{ab},\quad
G_T^{ab} \longrightarrow \frac{1}{C_T}\times G_T^{ab}.
\end{equation}

Let us now consider the products of OPE coefficients entering~\eqref{eq:decomposition_1} in more detail. So far the expansion is organized by individual local operators $\cO$. However, since the conformal blocks depend only on scaling dimension and spin of $\cO$, we cannot distinguish the contributions of operators which share these quantum numbers.\footnote{The degeneracies are to be expected if we consider the four-point function~\eqref{eq:masterfourpt} in a theory with a symmetry group sufficiently larger than $\mathrm{U}(1)$. If, on the other hand, $\mathrm{U}(1)$ is the only global symmetry, then one might argue that generically there should be no degeneracies. However, we would like to be agnostic about the complete global symmetry group, and we also do not want to rely on such expectations. Furthermore, from the point of view of our numerical approach, it is in any case impossible to impose a non-degeneracy condition on all operators at once. (It is only possible to do so for a finite number of operators, at the cost of a scan over the ratios of the OPE coefficients.)} This motivates defining the following quantity,
\be
	\p{P_{\De,(\ell,\bar\ell),0}}^{ba}\equiv\sum_{\cO^{(\ell,\bar\ell)}_{\De,0}}\lambda_{\<\bar\psi\psi\cO\>}^a \lambda_{\<\bar\cO\bar\psi\psi\>}^{b},
\ee
where we sum over all operators with given scaling dimension $\De$, spin $(\ell,\bar\ell)$ and the $\mathrm{U}(1)$ charge $Q=0$. Due to our choice of three-point tensor structures we have the property~\eqref{eq:neutral_set1_ope_relations}, which allows us to rewrite this as
\be
\label{eq:neutral_ope_product_with_degeneracy}
	\p{P_{\De,(\ell,\bar\ell),0}}^{ba}=\sum_{\cO^{(\ell,\bar\ell)}_{\De,0}}\lambda_{\<\bar\psi\psi\cO\>}^a \lambda_{\<\bar\psi\psi\cO\>}^{*b},
\ee
which in turn implies that 
\be
	P_{\De,(\ell,\bar\ell),0}\succeq 0
\ee
are positive-semidefinite hermitian matrices. This is the key to applying semidefinite programming to our setup.

As we discuss in section~\ref{sec:three_point_functions}, there are three families of operators contributing to $\bar\psi\psi$ OPE. These are operators of the type
\be
    \label{eq:operators_list_neutral}
	\cO^{(\ell,\ell)}_{\De,0},\quad  \cO^{(\ell+2,\ell)}_{\De,0}, \quad \cO^{(\ell,\ell+2)}_{\De,0}.
\ee
Correspondingly, we have three families of matrices $P_{\De,(\ell,\bar\ell),0}$. 
The matrices $P_{\De,(\ell,\ell),0}$ are $2\times2$ for $l\geq 1$, owing to the existence of two three-point tensor structures in the first correlator in~\eqref{eq:neutral_correlator_3}. 
The matrices $P_{\De,(\ell,\ell+2),0}$, $P_{\De,(\ell+2,\ell),0}$ and $P_{\De,(\ell=0,\ell=0),0}$ are $1\times 1$, since there is only one three-point tensor structure in the second and third correlators in~\eqref{eq:neutral_correlator_3} and in~\eqref{eq:neutral_correlator_3l0}.
Furthermore, from~\eqref{eq:neutral_ope_properties_TS} it follows that
\be\label{eq:PneutralPrimalDualRelation}
	P_{\De,(\ell,\ell+2),0}=P_{\De,(\ell+2,\ell),0}.
\ee
There is a corresponding relation between the NTS blocks
\begin{equation}
\label{eq:relation_between_NTS-blocks}
G_{\De,(\ell+2,\ell)}(\point_i) = \pi_{13}\pi_{24}
G_{\De,(\ell,\ell+2)}(\point_i),
\end{equation}
which follows from the ``left-right'' symmetry of the gluing operation
and the identities\footnote{Remember that the objects here are not the full correlators but rather their tensor structures.}
\begin{align}
\<\bar\psi (\point_1)\psi (\point_2) \cO (\point_0)\>^{(\uniq)} &=
(-1)^{\ell}\times\<\cO (\point_0)\bar\psi (\point_1)\psi (\point_2)\>^{(\uniq)},\\
\<\bar\cO (\point_0)\bar\psi (\point_3)\psi (\point_4)\>^{(\uniq)} &=
(-1)^{\ell}\times\<\bar\psi (\point_3)\psi (\point_4)\bar\cO (\point_0)\>^{(\uniq)},
\end{align}
which in turn follow from the definitions of tensor structures
given in section~\ref{sec:three_point_functions}.
One can then define a $\pi_{13}\pi_{24}$ symmetric block as
\begin{equation}
\label{eq:1324-symmetric_block}
G^{\pi_{13}\pi_{24}}_{\De,(\ell,\ell+2)}(\point_i)\equiv
G_{\De,(\ell,\ell+2)}(\point_i) + \pi_{13}\pi_{24}
G_{\De,(\ell+2,\ell)}(\point_i).
\end{equation}
Using the relations~\eqref{eq:1324-symmetric_block} and~\eqref{eq:relation_between_NTS-blocks} one can write the following improved form of~\eqref{eq:decomposition_1}:
\be
\contraction{\<}{\bar\psi}{(\point_1)}{\psi}
\contraction{\<\bar\psi(\point_1)\psi(\point_2)}{\bar\psi}{(\point_3)}{\psi}
\<\bar\psi(\point_1)\psi(\point_2)\bar\psi(\point_3)\psi(\point_4)\>
&=\sum_{\De,\ell}\tr\p{P_{\De,(\ell,\ell),0}G_{\De,(\ell,\ell)}(\point_i)}\nn\\
&+\sum_{\De,\ell}P_{\De,(\ell,\ell+2),0}
G^{\pi_{13}\pi_{24}}_{\De,(\ell,\ell+2)}(\point_i).
\label{eq:decomposition_1_final}
\ee
In this form the four-point function is manifestly symmetric under the kinematic permutation $\pi_{13}\pi_{24}$. 

The discussion above holds identically for the second ordering $\<\bar\psi\psi\psi\bar\psi\>$ in~\eqref{eq:basicorderings}. Due to the relation~\eqref{eq:neutral_set2_ope_relations} between the OPE coefficients we have
\be
\contraction{\<}{\bar\psi}{(\point_1)}{\psi}
\contraction{\<\bar\psi(\point_1)\psi(\point_2)}{\psi}{(\point_3)}{\bar\psi}
\<\bar\psi(\point_1)\psi(\point_2)\psi(\point_3)\bar\psi(\point_4)\>
&=\sum_{\De,\ell}\tr\p{P_{\De,(\ell,\ell),0}G^{\prime}_{\De,(\ell,\ell)}(\point_i)}\nn\\
&+\sum_{\De,\ell}P_{\De,(\ell,\ell+2),0}
G^{\prime\pi_{13}\pi_{24}}_{\De,(\ell,\ell+2)}(\point_i),
\label{eq:decomposition_2_final}
\ee
where the only difference with~\eqref{eq:decomposition_1_final} are the conformal blocks defined, contrary to~\eqref{eq:conformal_block_1}, as
\begin{equation}
\label{eq:conformal_block_2}
G_{\cO}^{\prime ab}(\point_i)\equiv
\<\bar\psi (\point_1)\psi (\point_2) \cO (\point_0)\>^{(a)} \bowtie \<\bar\cO (\point_0)\psi (\point_3)\bar\psi (\point_4)\>^{(b)}.
\end{equation}

Each conformal block $G$ in~\eqref{eq:decomposition_1_final} can be further expanded into the basis of four-point tensor structures defined in~\eqref{eq:ordering_1_tensor_structure_decomposition} as follows:
\begin{equation}
\label{eq:components_conformal_blocks_G}
G^{ab}_{\De,(\ell,\bar\ell)}(\point_i) =
\sum_{i=1,\pm}^3 \mathbb{T}_{i,\pm}\;
G^{ab}_{i,\pm,\De,(\ell,\bar\ell)}(z,\bar z).
\end{equation}
We refer to the objects $G^{ab}_{i,\pm}$ multiplying the tensor structures as the components of the conformal blocks $G^{ab}$.\footnote{In many works including~\cite{Cuomo:2017wme} the conformal blocks are referred to as conformal partial waves (CPWs). Instead the components of conformal blocks are referred to as conformal blocks.} One should not confuse the labeling $\pm$ with the charges of operators. The blocks do not contain information about charges and instead $\pm$ refers to the labeling of tensor structures, see the definition~\eqref{eq:ordering_1_tensor_structure_decomposition}. Similarly, the $G'$ blocks can be expanded in the basis of four-point structures $\mathbb{T}'_{i,\pm}$ defined in~\eqref{eq:ordering_2_tensor_structure_decomposition}.

Using the decompositions~\eqref{eq:decomposition_1_final} and~\eqref{eq:decomposition_2_final} and their expansions into four-point structures one can finally express the functions $g$ and $g'$ of section~\ref{sec:4ptstructures} as follows:
\be
	g_{i,\pm}(z,\bar z)
	&=\sum_{\De,\ell}\tr\p{P_{\De,(\ell,\ell),0}\,G_{i,\pm,\De,(\ell,\ell)}(z,\bar z)}\;+
	\sum_{\De,\ell}P_{\De,(\ell,\ell+2),0}\,G^{\pi_{13}\pi_{24}}_{i,\pm,\De,(\ell,\ell+2)}(z,\bar z),\nn\\
	g'_{i,\pm}(z,\bar z)
	&=\sum_{\De,\ell}\tr\p{P_{\De,(\ell,\ell),0}\, G'_{i,\pm,\De,(\ell,\ell)}(z,\bar z)}+
	\sum_{\De,\ell}P_{\De,(\ell,\ell+2),0}\,G^{\prime\pi_{13}\pi_{24}}_{i,\pm,\De,(\ell,\ell+2)}(z,\bar z).\label{eq:ggpexpansion}
\ee
The expression for the functions $\tl g_i(x,t)$ and $\tl g'_i(x,t)$ defined in~\eqref{eq:final4ptfunctions} follow straightforwardly from~\eqref{eq:ggpexpansion}. We note that since~\eqref{eq:decomposition_1_final} and \eqref{eq:decomposition_2_final} are $\pi_{13}\pi_{24}$ symmetric, the expressions~\eqref{eq:ggpexpansion} are also $\pi_{13}\pi_{24}$ symmetric and thus automatically satisfy the constraints~\eqref{eq:permutation_constraints} and~\eqref{eq:permutation_constraints_gph}.

\paragraph{Ordering $\<\psi\psi\bar\psi\bar\psi\>$.}

The same logic as above applies for the third ordering in~\eqref{eq:basicorderings}:
\be
\contraction{\<}{\psi}{(\point_1)}{\psi}
\contraction{\<\bar\psi(\point_1)\psi(\point_2)}{\bar\psi}{(\point_3)}{\psi}
\<\psi(\point_1)\psi(\point_2)\bar\psi(\point_3)\bar\psi(\point_4)\>=
\sum_{\cO}
\lambda_{\<\psi\psi\cO\>} \lambda_{\<\bar\cO\bar\psi\bar\psi\>}
H_{\cO}(\point_i),
\label{eq:Hblocks}
\ee
where now we sum over operators $\cO$ with $\mathrm{U}(1)$ charge $Q=-2$ and $H$ are the corresponding conformal blocks defined as\footnote{We use notation $H$ instead of $G$ to distinguish conformal blocks coming from $\psi\psi$ OPE from the blocks coming from $\bar\psi\psi$ OPE.}
\begin{equation}
\label{eq:conformal_block_3}
H_{\cO}(\point_i)\equiv
\<\psi (\point_1)\psi (\point_2) \cO (\point_0)\>^{(\uniq)} \bowtie
\<\bar\cO (\point_0)\bar\psi (\point_3)\bar\psi (\point_4)\>^{(\uniq)}.
\end{equation}
There are no indices in the conformal block and associated OPE coefficients since all the three-point functions always have a single tensor structure. As before we define
\be\label{eq:charged_ope_coeff}
P_{\De,(\ell,\bar\ell),-}\equiv\sum_{\cO^{(\ell,\bar\ell)}_{\De,-}}\lambda_{\<\psi\psi\cO\>} \lambda_{\<\bar\cO\bar\psi\bar\psi\>}=\sum_{\cO^{(\ell,\bar\ell)}_{\De,-}}|\lambda_{\<\psi\psi\cO\>}|^2\geq 0.
\ee
Here we sum over all operators with a given scaling dimension $\De$, spin $(\ell,\bar\ell)$ and $\mathrm{U}(1)$ charge $Q=-2$. In the last equality we have exploited the property~\eqref{eq:charged_ope_relations}. 

The operators $\cO$ come in three families,
\be
    \label{eq:operators_list_charged}
	\cO^{(\ell,\ell)}_{\De,-},\quad 	\cO^{(\ell,\ell+2)}_{\De,-},\quad 	\cO^{(\ell+2,\ell)}_{\De,-},
\ee
and we correspondingly have three families of $P_{\De,(\ell,\bar\ell),-}$ coefficients. The second and third families are restricted to have only odd spin due to~\eqref{eq:charged_correlator_2} and~\eqref{eq:charged_correlator_3}. Unlike the first two orderings, there is no relation between their contributions and the final form of the conformal block expansion reads as
\begin{multline}
\contraction{\<}{\psi}{(\point_1)}{\psi}
\contraction{\<\bar\psi(\point_1)\psi(\point_2)}{\bar\psi}{(\point_3)}{\psi}
\<\psi(\point_1)\psi(\point_2)\bar\psi(\point_3)\bar\psi(\point_4)\>=
\sum_{\De,\ell}
P_{\De,(\ell,\ell),-}H_{\De,(\ell,\ell)}(\point_i)\\
+\sum_{\De,\;\ell\in\text{odd}}
P_{\De,(\ell,\ell+2),-}H_{\De,(\ell,\ell+2)}(\point_i)
+\sum_{\De,\;\ell\in\text{odd}}
P_{\De,(\ell+2,\ell),-}H_{\De,(\ell+2,\ell)}(\point_i).
\label{eq:decomposition_H_final}
\end{multline}
We further expand the conformal blocks in the basis of four-point tensor structures~\eqref{eq:h4ptstructsconvenient} as
\begin{equation}
\label{eq:components_conformal_blocks_H}
H_{\De,(\ell,\bar\ell)}(\point_i) =
\sum_{i=1,\pm}^3 \Q_{i,\pm}\;
H_{i,\pm,\De,(\ell,\bar\ell)}(z,\bar z).
\end{equation}
Using~\eqref{eq:components_conformal_blocks_H} we can finally write the expansion of the $h$ functions from section~\ref{sec:4ptstructures} {in terms of the conformal block components} as
\begin{multline}
\label{eq:hexpansion}
h_{i,\pm}(z,\bar z)=
\sum_{\De,\ell}
P_{\De,(\ell,\ell),-} H_{i,\pm,\De,(\ell,\ell)}(z,\bar z)\\
+\sum_{\De,\;\ell\in\text{odd}}
P_{\De,(\ell,\ell+2),-} H_{i,\pm,\De,(\ell,\ell+2)}(z,\bar z)+\sum_{\De,\;\ell\in\text{odd}}
P_{\De,(\ell+2,\ell),-} H_{i,\pm,\De,(\ell+2,\ell)}(z,\bar z).
\end{multline}
The expression for the functions $\tl h_{\mathbf i}(x,t)$ defined in~\eqref{eq:final4ptfunctions} follow straightforwardly from~\eqref{eq:hexpansion}.

\subsection{Semidefinite problems}
\label{sec:sdp}
Given the crossing equations written in the final form~\eqref{eq:crossingGfirst}-\eqref{eq:crossingGlast} and the conformal block decompositions~\eqref{eq:ggpexpansion} and~\eqref{eq:hexpansion} one can obtain various bounds on scaling dimensions $\De$ of the operators~\eqref{eq:operators_list_neutral} and~\eqref{eq:operators_list_charged} and products of their OPE coefficients
\begin{equation}
\label{eq:list_product_ope}
P_{\De,(\ell,\ell),0}^{ab},\quad
P_{\De,(\ell,\ell+2),0},\quad
P_{\De,(\ell,\ell),-},\quad
P_{\De,(\ell,\ell+2),-},\quad
P_{\De,(\ell+2,\ell),-}
\end{equation}
in terms of the scaling dimension of the external  Weyl fermion $\De_\psi$. This is done in the standard fashion by setting up semidefinite problems~\cite{Kos:2014bka}. For previous studies of a single spinning correlator in 3d using this method see~\cite{Iliesiu:2015qra,Iliesiu:2017nrv,Dymarsky:2017xzb,Dymarsky:2017yzx}. 

To begin, we truncate the crossing equations~\eqref{eq:crossingGfirst}-\eqref{eq:crossingGlast} to a finite set by imposing 
\be
\label{eq:Lambda}
	m+2n\leq \Lambda,
\ee
where $\Lambda$ will be a parameter in our bounds. As usual, the bounds obtained at any finite $\Lambda$ are rigorous but not optimal. We expect the optimal bound to be recovered in the limit $\Lambda\to \oo$. Bringing all the terms in the truncated crossing equations~\eqref{eq:crossingGfirst}-\eqref{eq:crossingGlast} to the left-hand side, we can re-interpret them as a finite-dimensional vector  
equation
\be
    \label{eq:bootsrap_equations_vector_form}
	\vec F_\Lambda=0\,, \qquad {\text{dim}(\vec F_\Lambda) = \frac32\Lambda^2 + \frac92\Lambda + \left\{\begin{array}{l} 5 \quad \text{for $\Lambda$ even}\\3 \quad \text{for $\Lambda$ odd}\end{array}\right. \,.} 
\ee
Clearly, the components of $\vec F_\Lambda$ are certain linear combinations of functions $\tl g$, $\tl g'$ and $\tl h$ and their derivatives precisely specified by~\eqref{eq:crossingGfirst}-\eqref{eq:crossingGlast}. Using the conformal block expansions~\eqref{eq:ggpexpansion} and~\eqref{eq:hexpansion} one arrives at
\be\label{eq:Fexpansion}
	0=\vec F_\Lambda=&\sum_{\De,\ell}\tr\p{P_{\De,(\ell,\ell),0}\,{\vec G}_{\Lambda,\De,(\ell,\ell)}}+
	\sum_{\De,\ell}P_{\De,(\ell,\ell+2),0}\,{\vec G}_{\Lambda,\De,(\ell,\ell+2)}+
	\sum_{\De,\ell}
	P_{\De,(\ell,\ell),-}\vec H_{\Lambda,\De,(\ell,\ell)}\nn\\
	&+\sum_{\De,\;\ell\in\text{odd}}
	P_{\De,(\ell,\ell+2),-}\vec H_{\Lambda,\De,(\ell,\ell+2)}
	+\sum_{\De,\;\ell\in\text{odd}}
	P_{\De,(\ell+2,\ell),-}\vec H_{\Lambda,\De,(\ell+2,\ell)},
\ee
where $\vec G$ and $\vec H$ are vectors constructed from the appropriate linear combinations of conformal block components
\begin{equation}
 \label{eq:list_blocks_tildes}
 \tl G_{\ii,\De,(\ell,\bar\ell)},\quad
 \tl G'_{\ii,\De,(\ell,\bar\ell)},\quad
 \tl H_{\ii,\De,(\ell,\bar\ell)},\quad
 \ii=1,2,3,6,
\end{equation}
and their derivatives. The later objects are in turn obtained from the conformal block components
\begin{equation}
 \label{eq:list_blocks}
G_{i,\pm,\De,(\ell,\bar\ell)},\quad
G'_{i,\pm,\De,(\ell,\bar\ell)},\quad
H_{i,\pm,\De,(\ell,\bar\ell)},\quad
i=1,2,3
\end{equation}
defined in equations~\eqref{eq:components_conformal_blocks_G} and~\eqref{eq:components_conformal_blocks_H} by performing the change of variables~\eqref{eq:xyt_variables} and the redefinition~\eqref{eq:final4ptfunctions}. Notice that $\vec F_\Lambda$ implicitly depends on the scaling dimension $\De_\psi$ due to the implicit dependence of the conformal block components~\eqref{eq:list_blocks}.

Let us now zoom on the very first entry in~\eqref{eq:Fexpansion}. It contains several important terms which we should single out and discuss carefully. First, we have the identity operator with $\De=0$ and $\ell=0$ for which\footnote{Recall that $P_{\De,(\ell,\ell),0}$ is a $1\times 1$ matrix for $\ell=0$. Setting $\De=0$ and $\ell=0$ in the first entry of~\eqref{eq:neutral_correlator_3} one recovers the two-point function~\eqref{eq:two-point functions} if $\lambda^1_{\<\bar\psi\psi\cO\>}=i$.}
\begin{equation}
P_{0,(0,0),0} = 1.
\end{equation}
Second, we have the conserved current $J$ with $\De=3$ and $\ell=1$.\footnote{When bounding the $C_J$ central charge we assume that $J$ is the unique $U(1)$ conserved current. However in all the other bounds we are completely agnostic to the number of spin-1 currents.} Third, we have the stress tensor $T$ for $\De=4$ and $\ell=2$.\footnote{As usual, we assume that there is a unique conserved spin-2 operator $T$, although this will not be visible in our setup except for bounds on $C_T$.} In the last two cases one has to rescale the blocks according to~\eqref{eq:conserved_block_rescaling}. We now absorb the central charges coming from these rescaling in the definitions of $P$ and use the Ward identities to obtain the final form for the coefficients $P$. Utilizing~\eqref{eq:theta_parametrization} for the conserved current $J$ we get
\begin{equation}
\label{eq:PJ}
P^{ab}_{3,(1,1),0} = \frac{1}{8\pi^4\,C_J}\times \frac{1}{\left(1+\tan\theta\right)^2}
\begin{pmatrix}
1 & 2\tan\theta \\
2\tan\theta & 4\tan^2\theta
\end{pmatrix}.
\end{equation}
Utilizing~\eqref{eq:TmnOPEcoeff} for the stress tensor $T$ we get instead
\begin{equation}
\label{eq:PT}
P^{ab}_{4,(2,2),0} = \frac{1}{36\pi^4\,C_T}\times
\begin{pmatrix}
(2\De_\psi-3)^2  & 6\,(2\De_\psi-3) \\
6\,(2\De_\psi-3) & 36
\end{pmatrix}.
\end{equation}
In what follows will always treat the identity operator separately from other contributions, while $J$ and $T$ will be treated separately only in some bounds.

To proceed we consider vectors $\vec\alpha$ with real components and write the crossing equation~\eqref{eq:bootsrap_equations_vector_form} in the form
\be
\vec \alpha \cdot \vec F_\Lambda=0,
\ee
which reads in the expanded form as
\be\label{eq:FexpansionFunctional}
0=&\sum_{\De,\ell}\tr\p{P_{\De,(\ell,\ell),0}\,\vec\a\.{\vec G}_{\Lambda,\De,(\ell,\ell)}}+
\sum_{\De,\ell}P_{\De,(\ell,\ell+2),0}\,\vec\a\.{\vec G}_{\Lambda,\De,(\ell,\ell+2)}+
\sum_{\De,\ell}
P_{\De,(\ell,\ell),-}\vec\a\.\vec H_{\Lambda,\De,(\ell,\ell)}\nn\\
&+\sum_{\De,\;\ell\in\text{odd}}
P_{\De,(\ell,\ell+2),-}\vec\a\.\vec H_{\Lambda,\De,(\ell,\ell+2)}
+\sum_{\De,\;\ell\in\text{odd}}
P_{\De,(\ell+2,\ell),-}\vec\a\.\vec H_{\Lambda,\De,(\ell+2,\ell)}.
\ee
The objects entering the above equation
\begin{equation}
\label{eq:functionals}
\vec\alpha\cdot\vec G^{ab}_{\Lambda,\De,(\ell,\bar\ell)},\quad
\vec\alpha\cdot\vec H_{\Lambda,\De,(\ell,\bar\ell)}
\end{equation}
are functionals which map a set of conformal blocks to $1\times1$ or $2\times2$ matrices of real numbers. In what follows we look for functionals~\eqref{eq:functionals}, or equivalently for the vector $\vec\alpha$, satisfying certain conditions. If such functionals can be found we say that the problem is feasible. For performing this task in practice we use \texttt{SDPB}~\cite{Simmons-Duffin:2015qma}.

\paragraph{Bounds on the spectrum}
We first explain how to construct bounds on the spectrum of scaling dimensions. This is standard material in the numerical bootstrap literature, but we believe it can be useful to review the procedure  adopted to our case in what follows. We first single out the contribution of the identity operator in~\eqref{eq:FexpansionFunctional} and normalize $\vec\alpha$ in such a way that
\be
\label{eq:normalization_bound_spectrum}
\vec\a\.\vec G_{\Lambda,0,(0,0)}=1.
\ee
We then look for a vector $\alpha$ obeying the following properties
\be
\vec\a\.\vec G^{ab}_{\Lambda,\De,(\ell,\ell)}&\succeq 0,\quad \forall \ell\geq0, &&\forall \De\geq \De_\text{unitary}(\ell,\ell),\nn\\
\vec\a\.\vec G_{\Lambda,\De,(\ell,\ell+2)}&\geq 0,\quad \forall \ell\geq 0, 
&&\forall \De\geq \De_\text{unitary}(\ell,\ell+2),\nn\\
\vec\a\.\vec H_{\Lambda,\De,(\ell,\ell)}&\geq 0,\quad \forall \ell\geq0, 
&&\forall \De\geq \De_\text{unitary}(\ell,\ell),\nn\\
\vec\a\.\vec H_{\Lambda,\De,(\ell+2,\ell)}&\geq 0,\quad \forall \text{ odd }\ell>0, &&\forall \De\geq \De_\text{unitary}(\ell+2,\ell),\nn\\
\vec\a\.\vec H_{\Lambda,\De,(\ell,\ell+2)}&\geq 0,\quad \forall \text{ odd }\ell>0, &&\forall \De\geq \De_\text{unitary}(\ell,\ell+2),\label{eq:apositivity}
\ee
where we demand only the unitarity bounds~\eqref{eq:unitarity_bound} on the spectrum. If such $\vec\a$ is found, the crossing equation~\eqref{eq:FexpansionFunctional} cannot be satisfied, since
in the right-hand side we get one plus a non-negative contribution which cannot sum up to zero. 
Clearly, in the correct setup one will never be able to find $\vec\a$ satisfying~\eqref{eq:normalization_bound_spectrum} and~\eqref{eq:apositivity}, since otherwise we would prove that there exist no unitary CFTs with fermionic operators, which is clearly false. However things change if we introduce further assumptions on the scaling dimension of operators in the spectrum.

As an example let us assume that the CFTs we are looking for satisfy the constraint
\begin{equation}
\label{eq:one_assumption}
\De\geq \De{(\ell_*,\bar\ell_*)}= \De_\text{unitary}(\ell_*,\bar\ell_*)+ x,\quad x>0, 
\end{equation}
on the scaling dimension of operators (neutral or charged) in the spectrum with a given spin $(\ell_*,\bar\ell_*)$ and use it in~\eqref{eq:apositivity} instead of the unitarity bound. The parameter $x$ is often called the gap. It is very well possible that a CFT with a big enough value of $x$ is inconsistent or in other words does not satisfy the crossing equation~\eqref{eq:FexpansionFunctional}, in which case we will be able to find the vector $\vec \alpha$. If $\vec\a$ is not found one cannot draw any further conclusion, CFTs satisfying these assumptions might or might not exist.
The single gap assumption~\eqref{eq:one_assumption} can be trivially generalized to more complicated assumptions about the spectrum. {As we explain in detail in section~\ref{sec:topology_blocks}, however, one must be careful with the interpretation of these gaps.}

To construct a bound we assume~\eqref{eq:one_assumption} and perform a weighted binary search to find the smallest value of the parameter $x$ for which $\vec\a$ can be found at some fixed $\De_\psi$. 
As a result we find a value $x_\text{min}$ such that assumption~\eqref{eq:one_assumption} is inconsistent with crossing symmetry if $x>x_\text{min}$. This implies that any consistent CFT should have an operator with the dimension less or equal to $\De_\text{min}(\ell_*,\bar\ell_*)$ in the channel that we are studying. 

In case of the conserved current $J$ and the stress tensor $T$ one can instead put assumptions on the second $\ell=1$ and $\ell=2$ operators respectively. In order to do that one must single out and treat separately contributions of these operators in~\eqref{eq:FexpansionFunctional} similarly to the identity operator. We can also single out some non-conserved operators but this brings an extra parameter to the search, namely the scaling dimension of this operator.

\paragraph{Bounds on OPE coefficients}
Another type of bounds we consider are bounds on OPE coefficients or more precisely on the $P$ coefficients~\eqref{eq:list_product_ope} of an operator with a given charge, spin $(\ell_*,\bar\ell_*)$, and scaling dimension $\Delta_*$. This is also standard material,  but it can be useful to review here how such kind of bounds are imposed in our setup. We focus only on cases where $P$ is a $1\times 1$ matrix. This restriction does not exclude bounds on $C_J$ and $C_T$, since in those cases $P$ has the special form~\eqref{eq:PJ} or~\eqref{eq:PT}. In other words, the effective $P$ matrix is $1\times 1$ and given by $1/C_J$ or $1/C_T$ (assuming $\theta$ is fixed in the $J$ case). Thus, the upper/lower bound on OPE coefficients described below in these cases translates into lower/upper bound on the central charges $C_J$ and $C_T$. Note that in the case of the current we have to scan over different values of $\theta$.\footnote{It is possible to bound $C_J$ over all values of $\theta$ at once since there is a relation $C_J^{-1}=\mathrm{tr}(P_{3,(1,1),0} W)$ for a suitable $\theta$-independent choice of matrix $W$. This relation is simply the matrix analog of the Ward identity~\eqref{eq:J_ope_relation}. However, we will not use this approach here.}

In what follows let us focus for concreteness on the OPE coefficient of a neutral scalar with scaling dimension $\De_*$. The discussion below trivially applies to other cases.
We start by deriving an upper bound. Consider vectors $\vec \alpha$ satisfying the following normalization
\be
\label{eq:normalization_bound_OPE}
\vec\a\.\vec G_{\Lambda,\De_*,(0,0)}=1.
\ee
Using~\eqref{eq:normalization_bound_OPE} one can rewrite~\eqref{eq:FexpansionFunctional} as
\begin{align}
P_{\De_*,(\ell,\ell),0} &= - \vec\a\cdot\vec G_{\Lambda,0,(0,0)}
-\sum_{\De\neq \De_*,\;\ell}\tr\p{P_{\De,(\ell,\ell),0}\,\vec\a\.{\vec G}_{\Lambda,\De,(\ell,\ell)}}\nn\\
&-\sum_{\De,\ell}P_{\De,(\ell,\ell+2),0}\,\vec\a\.{\vec G}_{\Lambda,\De,(\ell,\ell+2)}
-\sum_{\De,\ell}
P_{\De,(\ell,\ell),-}\vec\a\.\vec H_{\Lambda,\De,(\ell,\ell)}\nn\\
&-\sum_{\De,\;\ell\in\text{odd}}
P_{\De,(\ell,\ell+2),-}\vec\a\.\vec H_{\Lambda,\De,(\ell,\ell+2)}
-\sum_{\De,\;\ell\in\text{odd}}
P_{\De,(\ell+2,\ell),-}\vec\a\.\vec H_{\Lambda,\De,(\ell+2,\ell)}.
\label{eq:ope_bound_equation}
\end{align}
We will now look for the vector $\vec\a$ satisfying~\eqref{eq:normalization_bound_OPE} and~\eqref{eq:apositivity} with neutral scalars of dimension $\De_*$ excluded.\footnote{Notice that one can also use~\eqref{eq:apositivity} with stronger assumptions for more advanced bounds.} Let us assume that such an $\a$ is found. Since except for the very first term in the right-hand side of~\eqref{eq:ope_bound_equation} all the terms are non-positive, we find that
\begin{equation}
\label{eq:upper_bound_ope}
P_{\De_*,(\ell,\ell),0} \leq - \vec\a\cdot\vec G_{\Lambda,0,(0,0)}.
\end{equation}
To obtain the strongest possible bound~\eqref{eq:upper_bound_ope} we also require that the vector $\vec\a$ minimizes the right-hand side of~\eqref{eq:upper_bound_ope}. We thus finally obtain
\be
P_{\De_*,(0,0),0}\leq \min_{\vec\a}\p{-\vec\a\.\vec G_{\Lambda,0,(0,0)}}.
\ee
To summarize, one can construct an upper bound on $P_{\De_*,(0,0),0}$ by solving the following problem:
\textit{minimize 
\begin{equation}
-\vec\a\.\vec G_{\Lambda,0,(0,0)}
\end{equation}
with $\vec\a$ satisfying the normalization condition~\eqref{eq:normalization_bound_OPE} and subject to the positivity conditions~\eqref{eq:apositivity} where the neutral scalar operator with $\De_*$ is excluded.
}

In order to obtain a lower bound, instead of the normalization condition~\eqref{eq:normalization_bound_OPE} we have to use
\be\label{eq:OPEminnormalization}
	\vec\a\.G_{\Lambda,\De_*,(0,0)}=-1.
\ee
Repeating the above arguments one arrives at
\be
P_{\De_*,(0,0),0}\geq \max_{\vec\a}\p{\vec\a\.\vec G_{\Lambda,0,(0,0)}}.
\ee
To summarize, one can construct a lower bound on $P_{\De_*,(0,0),0}$ by solving the following problem:
\textit{maximize
	\begin{equation}
	\vec\a\.\vec G_{\Lambda,0,(0,0)}
	\end{equation}
	with $\vec\a$ satisfying the normalization condition~\eqref{eq:OPEminnormalization} and	subject to positivity conditions~\eqref{eq:apositivity} where the neutral scalar with $\De_*$ is excluded. 
}

There is a subtle problem however with constructing a lower bound if the dimension $\De_*$ is not separated from all the other neutral scalars by a gap. The positivity condition~\eqref{eq:apositivity} and continuity of the blocks in $\De$ imply in such a case
\be
	\vec\a \cdot \vec G_{\Lambda,\De_*,(0,0)}\geq 0,
\ee
which is in a direct conflict with the normalization condition~\eqref{eq:OPEminnormalization}. No nontrivial lower bound can then be obtained. An intuitive explanation for this fact is that the contribution to the OPE from the exact scaling dimension $\De_*$ can always be reduced to $0$ at the cost of increasing infinitesimally close contributions.

\section{Computation of conformal blocks}
\label{sec:conformal_blocks_approximations}
The goal of this section is to compute the fermion blocks~\eqref{eq:conformal_block_1}, \eqref{eq:conformal_block_2} and~\eqref{eq:conformal_block_3}. We derive their analytic expressions in section~\ref{sec:fermion_block_computation} by relating them to the known 4d seed blocks described in section~\ref{sec:seed_blocks}. We then explain in section~\ref{sec:rational_approximation} our method for obtaining their rational approximations at the crossing-symmetric point $z=\bar z =1/2$ as required {for the numerical analysis}.

\subsection{Seed blocks}
\label{sec:seed_blocks}
The seed conformal blocks are the simplest conformal blocks with an internal operator in the $(\ell,\ell+p)$ or $(\ell+p,\ell)$ spin representations~\cite{Echeverri:2015rwa}. Following~\cite{Echeverri:2016dun,Cuomo:2017wme} we define them as follows
\begin{align}
\label{eq:primal_seed_block}
G^{(p)\;\text{primal}}_{\De,\ell,\De_i}(\point_i) &\equiv
\<\cO_{\De_1}^{(0,0)}(\point_1) \cO_{\De_2}^{(p,0)}(\point_2)\cO_{\De}^{(\ell,\ell+p)}\>^{(\uniq)}
\bowtie
\<{\bar\cO}_{\De}^{(\ell+p,\ell)}\cO_{\De_3}^{(0,0)}(\point_3) \cO_{\De_4}^{(0,p)}(\point_4)\>^{(\uniq)},\\
\label{eq:dual_seed_block}
G^{(p)\;\text{dual}}_{\De,\ell,\De_i}(\point_i) &\equiv
\<\cO_{\De_1}^{(0,0)}(\point_1) \cO_{\De_2}^{(p,0)}(\point_2)
\cO_{\De}^{(\ell+p,\ell)}\>^{(\uniq)}
\bowtie
\<\bar\cO_{\De}^{(\ell,\ell+p)}\cO_{\De_3}^{(0,0)}(\point_3) \cO_{\De_4}^{(0,p)}(\point_4)\>^{(\uniq)}.
\end{align}
We refer to the blocks~\eqref{eq:primal_seed_block} and~\eqref{eq:dual_seed_block} as the primal seed and the dual seed blocks. In the $p=0$ case the primal and dual blocks coincide by definition.
Note that in the above definitions we have removed all the charge labels since the seeds block are purely kinematic objects and do not depend on representations of global symmetries. In what follows we will adopt this convention whenever we refer to seeds blocks.

The left and right three-point structures appearing in the definitions~\eqref{eq:primal_seed_block} and~\eqref{eq:dual_seed_block} are called the seed three-point structures. The convention for them was chosen in~\cite{Echeverri:2016dun}, we summarize it here for convenience. The left seed structures are\footnote{In the formulas that follow $\point_3$ denotes the position and polarization of the 
exchanged operator $\cO$ and should not be confused with the position and polarization of the 
external $\cO_{\Delta_3}^{(0,0)}$ operator in~\eqref{eq:primal_seed_block} and~\eqref{eq:dual_seed_block}.}
\begin{align}
\label{seed_structure_1}
\<\cO_{\De_1}^{(0,0)}(\point_1) \cO_{\De_2}^{(p,0)}(\point_2)\cO_{\De}^{(\ell,\ell+p)}(\point_3)\>^{(\uniq)}
&\equiv K_{\text{left}}^{\text{seed}}
\left(\hat\II^{32}\right)^p\left(\hat\JJ^{3}_{12}\right)^\ell,\\
\label{seed_structure_2}
\<\cO_{\De_1}^{(0,0)}(\point_1) \cO_{\De_2}^{(p,0)}(\point_2)
\cO_{\De}^{(\ell+p,\ell)}(\point_3)\>^{(\uniq)}
&\equiv K_{\text{left}}^{\text{seed}}
\left(\hat\KK^{23}_1\right)^p\left(\hat\JJ^{3}_{12}\right)^\ell.
\end{align}
The right seed structures are
\begin{align}
\label{seed_structure_3}
\<{\bar\cO}_{\De_1}^{(\ell+p,\ell)}(\point_1)\cO_{\De_2}^{(0,0)}(\point_2) \cO_{\De_3}^{(0,p)}(\point_3)\>^{(\uniq)}
&\equiv K_{\text{right}}^{\text{seed}}
\left(\hat\II^{31}\right)^p\left(\hat\JJ^{1}_{23}\right)^\ell, \\
\label{seed_structure_4}
\< \bar \cO_{\De_1}^{(\ell,\ell+p)}(\point_1)\cO_{\De_2}^{(0,0)}(\point_2) \cO_{\De_3}^{(0,p)}(\point_3)\>^{(\uniq)}
&\equiv K_{\text{right}}^{\text{seed}}
\left(\hat{\bar\KK}^{13}_2\right)^p\left(\hat\JJ^{1}_{23}\right)^\ell.
\end{align}
The dependence on scaling dimensions hides in the kinematic factors $K$ which read as
\begin{align}
K_{\text{left}}^{\text{seed}}  &\equiv
x_{12}^{-\De_1-\De_2+\De_3+\ell}\,
x_{13}^{-\De_1+\De_2-\De_3-\ell}\,
x_{23}^{\De_1-\De_2-\De_3-\ell-p},\\
K_{\text{right}}^{\text{seed}} &\equiv
x_{12}^{-\De_1-\De_2+\De_3-\ell}\,
x_{13}^{-\De_1+\De_2-\De_3-\ell-p}\,
x_{23}^{\De_1-\De_2-\De_3+\ell}.
\end{align}

\paragraph{Analytic expressions} The analytic expressions for the seed blocks were found in~\cite{Echeverri:2016dun}
and are implemented in the ``CFTs4D'' package~\cite{Cuomo:2017wme} for $p\leq4$. The seed blocks can be further expanded into the basis of four-point structures. {Following \cite{Echeverri:2016dun,Cuomo:2017wme} we write
\begin{equation}
\label{eq:expansion_components_seed_block}
G^{(p)\;\text{seed}}_{\De,\ell,\De_i}(\point_i)  = K_\text{seed} \sum_{e=0}^p (-2)^{p-e} G_{e,\De,\ell}^{(p)\;\text{seed}}(z,\bar z) (\hat\II^{42})^e(\hat\II^{42}_{31})^{p-e}
\end{equation}
where in conformal frame $ K_\text{seed} \rightarrow (z\bar z)^{-(\Delta_1+\Delta_2)/2-p/4}$.}
The components of the seed blocks in this expansion are labeled by the index $e=0,\ldots,p$ and have the following form (here the left-hand side can either stand for the primal or the dual blocks)
\begin{equation}
\label{eq:components_seed_block}
G_{e,\De,\ell}^{(p)\;\text{seed}}(z,\bar z) = \left(\frac{z \bar z}{z-\bar z}\right)^{2p+1}\sum_{m,n}
c_{m,n}^e \mathcal{F}^{(a_e,b_e;c_e)}_{\rho_1+m,\rho_2+n}(z,\bar z),
\end{equation}
where $c_{m,n}^e$ are some rational functions of the parameters and
\begin{equation}
\mathcal{F}^{(a,b;c)}_{\rho_1,\rho_2}(z,\bar z)\equiv
k_{\rho_1}^{(a,b;c)}(z)k_{\rho_2}^{(a,b;c)}(\bar z)
-(z\leftrightarrow \bar z),
\end{equation}
while $a_e,b_e,c_e,\rho_1$, and $\rho_2$ have simple expressions in terms of the parameters of the conformal blocks. 
{We stress that the seed block components depend on the scaling dimensions of the external operators only through the quantities}
\be
	a=-\frac{\De_1-\De_2-p/2}{2},\quad 	b=\frac{\De_3-\De_4-p/2}{2}.
\ee
{while the full seed conformal block also depend on $\Delta_1+\Delta_2$, as shown in  \eref{eq:expansion_components_seed_block}.}
The $k$-functions appearing above are given in turn by the hypergeometric function as
\begin{equation}
\label{eq:k-functions}
k_\rho^{(a,b;c)}(z)\equiv z^{\rho}{}_2F_1(a+\rho,b+\rho,c+2\rho;z).
\end{equation}

For $p=0$ the expression~\eqref{eq:components_seed_block} reduces to the Dolan and Osborn result for scalar blocks~\cite{Dolan:2000ut,Dolan:2003hv,Dolan:2011dv}
\begin{equation}
\label{eq:Dolan_Osborn_block}
G_{{e=0},\De,\ell}^{(0)\;\text{seed}}(z,\bar z) =
(-1)^{\ell}\times
\frac{z \bar z}{z-\bar z}\;
\mathcal{F}^{(-\frac{\De_1-\De_2}{2},\frac{\De_3-\De_4}{2};0)}_{\frac{\De+\ell}{2},\frac{\De-\ell-2}{2}}(z,\bar z).
\end{equation}
Unfortunately, the coefficients $c_{m,n}^e$ for $p>0$ are rather complicated and it is challenging to construct rational approximations of seed blocks based on the analytic solutions. Instead, we obtain the $p>0$ seed blocks starting from the $p=0$ case.

\paragraph{Recursion relations}
In~\cite{Karateev:2017jgd} it was shown that any two conformal blocks in a given number of spacetime dimensions can be related to each other by means of differential operators. 
In particular, differential operators were found which relate the seed blocks for $p$ and $p-1$.  When decomposed into components, they take the schematic form
\be\label{eq:seed_recursion_relation}
	G_{e,\De,\ell}^{(p)\;\text{seed}}(z,\bar z)=D_0 G_{e,\De,\ell}^{(p-1)\;\text{seed}}(z,\bar z)+D_1 G_{e-1,\De,\ell}^{(p-1)\;\text{seed}}(z,\bar z)+D_2 G_{e-2,\De,\ell}^{(p-1)\;\text{seed}}(z,\bar z),
\ee
where the parameters $a$ and $b$ in the blocks appearing in the right-hand side of~\eqref{eq:seed_recursion_relation} coincide with those of $G_{e,\De,\ell}^{(p)\;\text{seed}}$ in the left-hand side, and the operators $D_0,D_1$, and $D_2$ are some explicit differential operators in $(z,\bar z)$ with coefficients which are rational functions of $\De,a,b,p,\ell,e$. The explicit expressions for these operators are different for primal and dual blocks and are given in~\cite{Karateev:2017jgd}. They are also implemented in~\texttt{CFTs4D} package.

\paragraph{Permuted seed blocks}
It will be convenient for us to have the seed blocks with a different 
orderings 
of spinning operators. In addition to the seed blocks~\eqref{eq:primal_seed_block} and \eqref{eq:dual_seed_block} we then define permuted seed blocks
\begin{align}
\label{eq:primal_seed_block_alt}
G^{(p)\;\text{primal}}_{\De,\ell,\De_i;\;\pi_{34}}(\point_i) &\equiv
\<\cO_{\De_1}^{(0,0)}(\point_1) \cO_{\De_2}^{(p,0)}(\point_2)\cO_{\De}^{(\ell,\ell+p)}\>^{(\uniq)}
\small{\bowtie}
\<{\bar\cO}_{\De}^{(\ell+p,\ell)}
\cO_{\De_3}^{(0,p)}(\point_3)
\cO_{\De_4}^{(0,0)}(\point_4)
\>^{(\uniq)},\\
\label{eq:dual_seed_block_alt}
G^{(p)\;\text{dual}}_{\De,\ell,\De_i;\;\pi_{34}}(\point_i) &\equiv
\<\cO_{\De_1}^{(0,0)}(\point_1) \cO_{\De_2}^{(p,0)}(\point_2)
\cO_{\De}^{(\ell+p,\ell)}\>^{(\uniq)}
\small{\bowtie}
\<
{\bar\cO}_{\De}^{(\ell,\ell+p)}
\cO_{\De_3}^{(0,p)}(\point_3)
\cO_{\De_4}^{(0,0)}(\point_4)\>^{(\uniq)}.
\end{align}
In the above formulas the only change compared to~\eqref{eq:primal_seed_block} and \eqref{eq:dual_seed_block} was made in the position of the last two operators in the right-hand three-point structures. We use a convention for them such that the seed blocks~\eqref{eq:primal_seed_block_alt} and \eqref{eq:dual_seed_block_alt} are related to the original seed blocks~\eqref{eq:primal_seed_block} and \eqref{eq:dual_seed_block} in the following way
\begin{align}
G^{(p)\;\text{primal}}_{\De,\ell,\De_i;\;\pi_{34}}(\point_i) &= 
(-1)^\ell\times\pi_{34}G^{(p)\;\text{primal}}_{\De,\ell,\De_i}(\point_i)
\big|_{\De_3\leftrightarrow\De_4},\\
G^{(p)\;\text{dual}}_{\De,\ell,\De_i;\;\pi_{34}}(\point_i) &=
(-1)^\ell\times\pi_{34}G^{(p)\;\text{dual}}_{\De,\ell,\De_i}(\point_i)
\big|_{\De_3\leftrightarrow\De_4}.
\end{align}
The $\pi_{34}$ permutation changes tensor structures in a straightforward way and transforms the conformal cross-ratios as follows
\begin{equation}
\label{eq:permutation34_zzb}
(z,\,\bar z) \longrightarrow \left(\frac{z}{z-1},\,\frac{\bar z}{\bar z-1}\right).
\end{equation}
This implies a simple transformation rule for the components of the tensor conformal blocks \eqref{eq:components_seed_block}. Conjugating the differential operators $D_i$ in~\eqref{eq:seed_recursion_relation} by this transformation, we immediately obtain recursion relations for the permuted seed blocks.

\subsection{Fermion blocks}
\label{sec:fermion_block_computation}
One can construct the fermion three-point tensor structures defined in section~\ref{sec:three_point_functions} from the seed three-point structures~\eqref{seed_structure_1}-\eqref{seed_structure_4} using differential operators~\cite{Costa:2011dw,Echeverri:2015rwa,Karateev:2017jgd}. The latter do not interfere with the $\bowtie$ operation in~\eqref{eq:primal_seed_block} and~\eqref{eq:dual_seed_block} and allow us to express the fermion blocks~\eqref{eq:conformal_block_1}, \eqref{eq:conformal_block_2} and~\eqref{eq:conformal_block_3} in  terms of the seed blocks.

The differential operators can be constructed as products of basic differential operators defined in~\cite{Echeverri:2015rwa}. In the case of Weyl fermions it is simpler and more transparent however to build them directly from the fundamental and anti-fundamental weight-shifting operators~\cite{Karateev:2017jgd}.\footnote{The operators constructed in~\cite{Echeverri:2015rwa} can in turn be written as products of weight-shifting operators corresponding to higher-dimensional representations of the conformal group.}
These differential operators change both the spin and the scaling dimensions of external operators. For this reason it is convenient to define the following notation for shifted scaling dimensions: 
\begin{equation}
\label{eq:shifted_scaling_dimensions}
\De_\psi^{+ (a)}\equiv \De_\psi+(3/2-a),\quad
\De_\psi^{- (a)}\equiv \De_\psi-(3/2-a),\quad
\De^{+}_\psi\equiv\De_\psi^{+ (1)},\quad
\De^{-}_\psi\equiv\De_\psi^{- (1)}.
\end{equation}

In what follows we provide the details of this procedure. We will split the discussion of fermion blocks~\eqref{eq:conformal_block_1}, \eqref{eq:conformal_block_2} and~\eqref{eq:conformal_block_3} into neutral channel and charged channel subsections respectively.

\subsubsection{Neutral channel}
We compute here the $s$-channel conformal blocks for the $\<\bar\psi\psi\bar\psi\psi\>$ and $\<\bar\psi\psi\psi\bar\psi\>$ orderings. By looking at the differential operators available, the spin structure of these four-point functions and the (permuted) seed blocks available, it is clear that the first ordering should be expressed in terms of the permuted seed blocks~\eqref{eq:primal_seed_block_alt}, \eqref{eq:dual_seed_block_alt}, whereas the second ordering should be expressed instead in terms of the standard seeds~\eqref{eq:primal_seed_block}, \eqref{eq:dual_seed_block}.

\paragraph{Ordering $\<\bar\psi\psi\bar\psi\psi\>$.}
We start by considering the conformal blocks~\eqref{eq:conformal_block_1}. For TS exchanged operators the left three-point structures can be written as
\begin{align}
\label{eq:left_neutral_TS}
\<\bar\psi(\point_1)\psi(\point_2)\cO_{\De,0}^{(\ell,\ell)}(\point_3)\>^{(a)}
&=\sum_{b=1}^2M^{ab}\, D^{(b)}_{12,\,p=0}\,
\<\cO_{\De_\psi^{+(b)}}^{(0,0)}(\point_1) \cO_{\De_\psi^{+(b)}}^{(0,0)}(\point_2)\cO_{\De}^{(\ell,\ell)}(\point_3)\>^{(\uniq)},\\
\label{eq:right_neutral_TS}
\<\cO^{(\ell,\ell)}_{\De,0}(\point_1)\bar\psi(\point_2)\psi(\point_3)\>^{(a)}
&=\sum_{b=1}^2M^{\prime ab}\, D^{(b)}_{23,\,p=0}\,
\<\cO^{(\ell,\ell)}_{\De}(\point_1)
\cO_{\De_\psi^{+(b)}}^{(0,0)}(\point_2) \cO_{\De_\psi^{+(b)}}^{(0,0)}(\point_3)\>^{(\uniq)}.
\end{align}
In the above expression, the shifted scaling dimensions in the right-hand side are defined in~\eqref{eq:shifted_scaling_dimensions} and the differential operators are given in terms of the weight-shifting operators as
\begin{equation}
\label{eq:neutral_TS_diff_operators}
 D^{(b=1)}_{ij,\,p=0}\equiv(\cD_{-0+}^i\cdot\bar\cD^{-+0}_j),\quad
 D^{(b=2)}_{ij,\,p=0}\equiv(\cD_{++0}^j\cdot\bar\cD^{+0+}_i).
\end{equation}
The matrices $M$ and $M'$ entering~\eqref{eq:left_neutral_TS} and~\eqref{eq:right_neutral_TS} are given by
\begin{equation}
\label{eq:neutral_matrix_TS}
M^{ab}=\begin{pmatrix}
1 & 0\\
\frac{(2\De_\psi+\De-\ell-5)(2\De_\psi-\De+\ell-1)}{4\ell(\De-1)} &
\frac{1}{\ell(\De-1)(2\De_\psi-3)^2}
\end{pmatrix},\qquad
M^{\prime ab}=(-1)^{\ell+1}M^{ab}.
\end{equation}
For $\ell=0$ we have a single tensor structure represented by $a=1$ and thus only the first differential operator in~\eqref{eq:neutral_TS_diff_operators} is needed. The matrices $M$ and $M'$ in~\eqref{eq:neutral_matrix_TS} collapse to their first entries which are $+1$ and $-1$ respectively. Using~\eqref{eq:left_neutral_TS} and \eqref{eq:right_neutral_TS} the fermion conformal blocks in the neutral channel for traceless symmetric exchanged operators  $(p=0)$ can be written as\footnote{For $p=0$ there is no distinction between primal, dual or permuted seed conformal blocks.}
\begin{equation}
\label{eq:p=0_block_in_g_channel}
G^{ab}_{\De,(\ell,\ell)}(\point_i)=
\sum_{c,d=1}^2
M^{ac}M^{\prime bd}
D^{(c)}_{12,p=0}D^{(d)}_{34,p=0}\;
G^{(0)\;\text{primal}}_{\De,\ell,\De_i^{(cd)};\;\pi_{34}}(\point_i),
\end{equation}
where the shifted external scaling dimensions are given by
\begin{equation}
\label{eq:shifted_dimensions_p0}
\De_i^{(cd)} \equiv\{
\De_\psi^{+ (c)},\De_\psi^{+ (c)},
\De_\psi^{+ (d)},\De_\psi^{+ (d)}\}.
\end{equation}

For NTS operators there is a single differential operator for the left and right three-point structures:
\begin{equation}
D_{ij,\,p=2}^{\text{left}} \equiv -\frac{1}{2}(\cD_{-0+}^i\cdot\bar\cD^{--0}_j),\quad
D_{ij,\,p=2}^{\text{right}}\equiv -\frac{1}{2}(-1)^\ell(\cD_{-0-}^i\cdot\bar\cD^{-+0}_j).
\end{equation}
Using these we relate the fermion structures to the permuted seed structures as
\begin{align}
\label{eq:left_neutral_NTS}
\<\bar\psi(\point_1)\psi(\point_2)\cO_{\De,0}^{(\ell,\bar\ell)}(\point_3)\>^{(\uniq)}
&= \frac{\ell-\bar\ell}{2}\times D_{12,\,p=2}^{\text{left}}
\<\cO_{\De^+_\psi}^{(0,0)}(\point_1) \cO_{\De^+_\psi}^{(2,0)}(\point_2)\cO_{\De,0}^{(\ell,\bar\ell)}(\point_3)\>^{(\uniq)},\\
\label{eq:right_neutral_NTS}
\<{\bar\cO^{(\bar\ell,\ell)}_{\De,0}}(\point_1)\bar\psi(\point_2)\psi(\point_3)\>^{(\uniq)}
&= {\frac{\bar\ell-\ell}{2}}\times D_{23,\,p=2}^{\text{right}}
\<{\cO^{(\bar\ell,\ell)}_{\De}}(\point_1)
\cO_{\De^+_\psi}^{(2,0)}(\point_2) \cO_{\De^+_\psi}^{(0,0)}(\point_3)\>^{(\uniq)},
\end{align}
where $|\ell-\bar\ell|=2$ and the shifted scaling dimensions are defined in~\eqref{eq:shifted_scaling_dimensions}. 
Here by $(\ell,\bar\ell)$ we mean either $(\ell+2,\ell)$ or $(\ell,\ell+2)$.
Using~\eqref{eq:left_neutral_NTS} and \eqref{eq:right_neutral_NTS} we get the fermion blocks for the NTS ($p=2$) operators
\begin{equation}
\label{eq:p=2_block_in_g_channel}
G_{\De,(\ell,\bar\ell)}(\point_i)=
-D_{12,\,p=2}^{\text{left}}D_{34,\,p=2}^{\text{right}}
G^{(2)\;\text{seed}}_{\De,\ell,\De_i;\;\pi_{34}},\quad
\De_i =\{\De_\psi^+,\De_\psi^+,\De_\psi^+,\De_\psi^+\}.
\end{equation}
In this expression we use the dual permuted seed blocks for  $(\ell+2,\ell)$ operators and primal permuted seed blocks for $(\ell,\ell+2)$ operators.

\paragraph{Ordering $\<\bar\psi\psi\psi\bar\psi\>$.}
We now move to the conformal blocks~\eqref{eq:conformal_block_2}. The only difference with respect to the previous case is in the form of the right three-point structure. For TS exchanged operators ($p=0$)  instead of~\eqref{eq:right_neutral_TS} we have
\begin{align}
\<\cO^{(\ell,\ell)}_{\De,0}(\point_1)\psi(\point_2)\bar\psi(\point_3)\>^{(a)}
&=\sum_{b=1}^2M^{ab}\, D^{(b)}_{32,\,p=0}\,
\<\cO^{(\ell,\ell)}_{\De,0}(\point_1)
\cO_{\De_\psi^{+(b)}}^{(0,0)}(\point_2) \cO_{\De_\psi^{+(b)}}^{(0,0)}(\point_3)\>^{(\uniq)},
\end{align}
where the matrix $M$ and the differential operator are exactly the ones given in the previous paragraph. The $p=0$ conformal block is thus given by
\begin{equation}
\label{eq:block_in_g'_channel}
G^{\prime ab}_{\De,(\ell,\ell)}(\point_i)=
\sum_{c,d=1}^2
M^{ac}M^{bd}
D^{(c)}_{12,p=0}D^{(d)}_{43,p=0}\;
G^{(0)\;\text{primal}}_{\De,\ell,\De_i^{(cd)}}(\point_i),
\end{equation}
with the shifted scaling dimensions defined in~\eqref{eq:shifted_dimensions_p0}.
For NTS ($p=2$) operators instead of~\eqref{eq:right_neutral_NTS} we have
\begin{align}
\nn
\<{\bar\cO^{(\bar\ell,\ell)}_{\De,0}}(\point_1)\psi(\point_2)\bar\psi(\point_3)\>^{(\uniq)}
&= (-1)^{{\bar\ell}+1}\,{\frac{\bar\ell-\ell}{2}}\times D_{32,\,p=2}^{\text{right}}
\<{\cO^{(\bar\ell,\ell)}_{\De}}(\point_1)
\cO_{\De^+_\psi}^{(0,0)}(\point_2) \cO_{\De^+_\psi}^{(0,2)}(\point_3)\>^{(\uniq)}.
\end{align}
This leads to the following conformal block
\begin{equation}
G^\prime_{\De,(\ell,\bar\ell)}(\point_i)=
(-1)^{\ell}\,D_{12,\,p=2}^{\text{left}}D_{43,\,p=2}^{\text{right}}
G^{(2)\;\text{seed}}_{\De,\ell,\De_i}(\point_i),\quad
\De_i =\{\De_\psi^+,\De_\psi^+,\De_\psi^+,\De_\psi^+\}.
\end{equation}
We use the dual seed blocks for $(\ell+2,\ell)$ exchanged operators and primal seed blocks for $(\ell,\ell+2)$ exchanged operators.

\subsubsection{Charged channel}
Finally we compute the conformal blocks~\eqref{eq:conformal_block_3} 
for the third ordering $\<\psi\psi\bar\psi\bar\psi\>$.
In case of TS exchanged operators there are in general two independent differential operators one can use to generate tensor structures. For the left three-point functions they read as
\begin{equation}
\label{eq:left_charged_TS}
L^{(a=1)}_{ij,p=0}\equiv(\cD_{++0}^i\cdot\bar\cD^{-+0}_j),\quad
L^{(a=2)}_{ij,p=0}\equiv(\cD_{++0}^j\cdot\bar\cD^{-+0}_i).
\end{equation}
For the right three-point functions they read as
\begin{equation}
\label{eq:right_charged_TS}
R^{(a=1)}_{ij,p=0}\equiv(\cD_{-0+}^i\cdot\bar\cD^{+0+}_j),\quad
R^{(a=2)}_{ij,p=0}\equiv(\cD_{-0+}^j\cdot\bar\cD^{+0+}_i).
\end{equation}
For $\ell=0$ it is enough to use only the first differential operators in both~\eqref{eq:left_charged_TS} and~\eqref{eq:right_charged_TS}. One can write
\begin{align}
\<\psi(\point_1)\psi(\point_2)\cO_{\De,-}^{(0,0)}(\point_3)\>^{(\uniq)}
&=+\mathcal{C}\times L^{(a=1)}_{12,p=0}
\<\cO_{\De^{-}_\psi}^{(0,0)}(\point_1) \cO_{\De^{+}_\psi}^{(0,0)}(\point_2)\cO_{\De}^{(0,0)}(\point_3)\>^{(\uniq)},\\
\<{\bar\cO}_{\De,+}^{(0,0)}(\point_1)\bar\psi(\point_2)\bar\psi(\point_3)\>^{(\uniq)}
&=-\mathcal{C}\times R^{(a=1)}_{23,p=0}
\<{\cO}_{\De}^{(0,0)}(\point_1)\cO_{\De^{-}_\psi}^{(0,0)}(\point_2) \cO_{\De^{+}_\psi}^{(0,0)}(\point_3)\>^{(\uniq)},
\end{align}
where the coefficient $\mathcal{C}$ is defined as
\begin{equation}
\mathcal{C}\equiv\frac{2}{(2\De_\psi-3)(\De-1)}.
\end{equation} 
For $\ell\geq 1$ there is a single structure~\eqref{eq:charged_correlator_1} which consists however of two different pieces which can be generated by the above operators. We find
\begin{align}
\<\psi(\point_1)\psi(\point_2)\cO_{\De,-}^{(\ell,\ell)}(\point_3)\>^{(\uniq)} &=
\sum_{a=1}^2 N^a L^{(a)}_{12,p=0}
\<
\cO_{\De_\psi^{- (a)}}^{(0,0)}(\point_1)
\cO_{\De_\psi^{+ (a)}}^{(0,0)}(\point_2)
\cO_{\De}^{(\ell,\ell)}(\point_3)\>^{(\uniq)},\\
\<{\bar\cO}_{\De,+}^{(\ell,\ell)}(\point_1)\bar\psi(\point_2)\bar\psi(\point_3)\>^{(\uniq)} &=
\sum_{a=1}^2 N^{\prime a} R^{(a)}_{23,p=0}
\<
\cO_{\De}^{(\ell,\ell)}(\point_1)
\cO_{\De_\psi^{- (a)}}^{(0,0)}(\point_2)
\cO_{\De_\psi^{+ (a)}}^{(0,0)}(\point_3)
\>^{(\uniq)},
\end{align}
where the shifted external scaling dimension are defined in~\eqref{eq:shifted_scaling_dimensions} and the matrices $N$ and $N'$ are given by the expressions
\begin{equation}
N^a\equiv \mathcal{E}\times \{+1,\;(-1)^{\ell+1}\},\quad
N^{\prime a}\equiv \mathcal{E}\times \{(-1)^{\ell+1},\;+1\},\quad
N^{\prime a} = (-1)^{\ell+1}\times N^a,
\end{equation}
with the coefficient $\mathcal{E}$ defined as
\begin{equation}
\mathcal{E}\equiv
-\frac{(\De+\ell-1)+(-1)^{\ell+1}\times(\De-\ell-1)}{2\ell(\De-1)(2\De_\psi-3)}.
\end{equation}
As a result the conformal block in the charged channel for TS ($p=0$) exchanged operators are given by
\begin{equation}
\label{eq:block_in_h_channel}
H_{\De,(\ell,\ell)}(\point_i)=
\sum_{a,b=1}^2
N^{a}N^{\prime b}L^{(a)}_{12,p=0}R^{(b)}_{34,p=0}\;
G^{(0)\;\text{primal}}_{\De,\ell,\De_i^{+-(ab)}}(\point_i),
\end{equation}
where the shifted external scaling dimensions are defined as 
\begin{equation}
\De_i^{+-(ab)} \equiv 
\{\De_\psi^{- (a)}, \De_\psi^{+ (a)},\De_\psi^{- (b)},\De_\psi^{+ (b)}\}.
\end{equation}

For NTS operators the left and right three-point functions can be generated as follows
\begin{align}
\<\psi(\point_1)\psi(\point_2)\cO^{(\ell,\bar\ell)}_{\De,-}(\point_3)\>^{(\uniq)}
&=\mathcal{F}^L_{\ell,\bar\ell}\times L_{12,p=2}
\<\cO_{\De^{-}_\psi}^{(0,0)}(\point_1) \cO_{\De^{+}_\psi}^{(2,0)}(\point_2)\cO_{\De}^{(\ell,\bar\ell)}(\point_3)\>^{(\uniq)},\\
\<\bar\cO^{{(\bar\ell,\ell)}}_{\De,+}(\point_1)\bar\psi(\point_2)\bar\psi(\point_3)\>^{(\uniq)}
&=\mathcal{F}^R_{{\bar\ell,\ell}}\times R_{12,p=2}
\<\cO_{\De}^{{(\bar\ell,\ell)}}(\point_1)\cO_{\De^{-}_\psi}^{(0,0)}(\point_2) \cO_{\De^{+}_\psi}^{(0,2)}(\point_3)\>^{(\uniq)},
\end{align}
where as in the neutral case by $(\ell,\bar\ell)$ we mean either $(\ell+2,\ell)$ or $(\ell,\ell+2)$. The differential operators $L$ and $R$ are defined as
\begin{equation}
L_{ij,p=2}\equiv(\cD_{++0}^i\cdot\bar\cD^{--0}_j),\quad
R_{ij,p=2}\equiv (-1)^\ell \times (\cD_{-0+}^j\cdot\bar\cD^{+0-}_i)
\end{equation}
and the coefficients $\mathcal{F}^L$ and $\mathcal{F}^R$ are given by
\begin{equation}
\mathcal{F}^L_{\ell,\bar\ell} \equiv \frac{\ell-\bar\ell}{2\,\kappa\,(2\De_\psi-3)},\quad
\mathcal{F}^R_{{\bar\ell,\ell}} \equiv \frac{\ell-\bar\ell}{2\,\kappa\,(2\De_\psi-7)},\quad
\kappa\equiv \De-3-\frac{1}{4}\,\ell\,(\ell+2)+\frac{1}{4}\,\bar\ell\,(\bar\ell+2).
\end{equation}
As a result the fermion conformal blocks for the NTS ($p=2$) operators read as
\begin{equation}
\label{eq:block_in_h_channel_p=2}
H_{\De,(\ell,\bar\ell)}(\point_i)=
\mathcal{F}\times
L_{12,p=2}R_{34,p=2}\;
G^{(p=2)\;\text{seed}}_{\De,\,\De_i^{+-}}(\point_i),\quad
\mathcal{F} \equiv
\mathcal{F}^L_{\ell,\bar\ell} \mathcal{F}^R_{\bar\ell,\ell},
\end{equation}
where we use the dual seed blocks for $(\ell+2,\ell)$ exchanged operators and primal seed blocks for $(\ell,\ell+2)$ exchanged operators.
The shifted external scaling dimensions are defined as
\begin{equation}
\De_i^{+-}\equiv \{\De^{-}_\psi,\,\De^{+}_\psi,\,\De^{-}_\psi,\,\De^{+}_\psi\}.
\end{equation}

\subsection{Rational approximation}
\label{sec:rational_approximation}

Our basic strategy for computing rational approximations of fermion blocks will be to start with rational approximations for scalars conformal blocks, then successively obtain from them the approximations for seed blocks using the recursion relations~\eqref{eq:seed_recursion_relation}, and finally get the rational approximations of fermion blocks by using the construction of section~\ref{sec:fermion_block_computation}.

While in principle this procedure is conceptually straightforward, each step involves a number of subtle points, which we clarify in this section. For concreteness, we will focus on obtaining the fermion blocks for a $p=2$ NTS exchange for the neutral channel ordering $\<\bar\psi\psi\bar\psi\psi\>$. All other cases can be treated in a completely analogous way.

The order of the computation can be summarized as follows
\be\label{eq:rationalstrategy}
	G^{(0)\,\text{seed}}_{e=0,\De,\ell}\to G^{(1)\,\text{seed}}_{e,\De,\ell}\to G^{(2)\,\text{seed}}_{e,\De,\ell} \to  G_{i,\De,(\ell,\ell+2)},
\ee
where $ G_{i,\De,(\ell,\ell+2)}$ are the blocks which enter the expansion of functions $ g_i$ in~\eqref{eq:ggpexpansion}.
We start by expressing the scalar blocks $G^{(0)\,\text{seed}}_{e=0,\De,\ell}$ in terms of the parameters $x$ and $t$ defined in~\eqref{eq:xyt_variables}
Since scalar blocks are invariant under $z\leftrightarrow \bar z$, they are holomorphic functions of $x,t$ near $x=t=0$.\footnote{Recall that $x=t=0$ is the crossing-symmetric point.} We then compute rational approximations for their derivatives in the form
\be\label{eq:scalarrational}
	\ptl_x^m\ptl_t^n G^{(0)\,\text{seed}}_{e=0,\De,\ell}(0,0)\approx \frac{(4r_0)^\De}{\prod_i (\De-\De_i)^{\kappa_i}}P^{m,n}_\ell(\De),
\ee
where $\De_i\leq \De_\text{unitary}(\ell,\ell)$, $\kappa_i\in\{1,2\}$, $r_0=3-2\sqrt 2$, and $P^{m,n}_\ell$ are some polynomials.
The set of poles $\De_i$ depends on $\ell$, and their number depends on the desired precision of the approximation. The positions of the poles $\De_i$ and their orders $\kappa_i$ are dictated by representation theory~\cite{Penedones:2015aga}
. Approximations such as~\eqref{eq:scalarrational} are usually constructed using Zamolodchikov-like recursion relations~\cite{Kos:2013tga,Kos:2014bka,Penedones:2015aga}, but those are tricky to implement in even dimensions. While it is possible to adapt these recursion relations to $d=4$, we choose a more simple-minded approach to obtain the approximation~\eqref{eq:scalarrational} directly from the Dolan-Osborn formulas~\eqref{eq:Dolan_Osborn_block}. This is described in appendix~\ref{app:rationalscalar}.

Our goal is now to start from the approximations~\eqref{eq:scalarrational} and make our way through~\eqref{eq:rationalstrategy}. In~\eqref{eq:rationalstrategy} every step is performed by applying differential operators in $(z,\bar z)$ as described in sections~\ref{sec:seed_blocks} and~\ref{sec:fermion_block_computation}. We always rewrite these differential operators in $(x,t)$ coordinates. This is important because the differential operators in $(z,\bar z)$ contain inverse powers of $z-\bar z$, which make the result apparently singular at the crossing-symmetric point $z=\bar z=\half$. For example, when we write out the first step of~\eqref{eq:rationalstrategy} using~\eqref{eq:seed_recursion_relation}
\be\label{eq:p=1example}
	G_{0,\De,\ell}^{(1)\;\text{seed}}(z,\bar z)&=D_0 G_{e=0,\De,\ell}^{(0)\;\text{seed}}(z,\bar z),\nn\\
	G_{1,\De,\ell}^{(1)\;\text{seed}}(z,\bar z)&=D_1 G_{e=0,\De,\ell}^{(0)\;\text{seed}}(z,\bar z),
\ee
we find that the differential operator $D_1$ acting on the scalar conformal block contains a term proportional to
\be\label{eq:troublesometerm}
	\propto\frac{z\bar z}{z-\bar z}\p{(1-z)\ptl_z-(1-\bar z)\ptl_{\bar z}}G_{e=0,\De,\ell}^{(0)\;\text{seed}}(z,\bar z)
\ee
which naively appears singular at $z=\bar z$. However, the singularity goes away if we remember that the scalar block $G_{\De,\ell}^{(0)\;\text{seed}}(z,\bar z)$ is symmetric under $z\leftrightarrow \bar z$. When we express everything in terms of $(x,t)$ coordinates, this symmetry is automatically taken into account, and such apparent singularities go away. For instance, the term~\eqref{eq:troublesometerm} becomes
\be\label{eq:xtniceterm}
	\propto\p{\half(t+(x+\thalf)^2)\ptl_x+(x-\thalf)(t-(x+\thalf)^2)\ptl_t}G_{e=0,\De,\ell}^{(0)\;\text{seed}}(x,t).
\ee
In fact the differential operators often become polynomial in $(x,t)$. In these variables it is straightforward to find relations of the form
\be
	\ptl_x^m\ptl_t^n G_{e,\De,\ell}^{(1)\;\text{seed}}(0,0)=\sum_{m',n'}\cM^{m,n}_{e,\ell;m',n'}(\De)\ptl_x^{m'}\ptl_t^{n'}G^{(0)\,\text{seed}}_{e=0,\De,\ell}(0,0)
\ee
by simply differentiating expressions such as~\eqref{eq:xtniceterm} with respect to $(x,t)$ and setting $x=t=0$. Since the differential operators have coefficients which are rational functions in $\De$, the same is true for the matrices $\cM^{m,n}_{e,\ell;m',n'}(\De)$, and we can use~\eqref{eq:scalarrational} to obtain the rational approximation for $G_{e,\De,\ell}^{(1)\;\text{seed}}$. In this way we find the approximations
\be\label{eq:approx_to_be_fixed}
	\ptl_x^m\ptl_t^n G_{e,\De,\ell}^{(1)\;\text{seed}}(0,0)\approx \frac{(4r_0)^\De}{\prod_{\hat i} (\De-\De_{\hat i})^{\kappa_{\hat i}}}\hat P^{m,n}_{(1),e,\ell}(\De),
\ee
where the set of poles $\De_{\hat i}$ now includes both the poles from the scalar blocks~\eqref{eq:rationalstrategy} and the poles from the matrices $\cM_{e,\ell}(\De)$. We temporarily put a hat on $\hat P^{m,n}_{(1),e,\ell}(\De)$ and the index $i$ because there is a problem with the approximation~\eqref{eq:approx_to_be_fixed} which we now discuss and fix.

The problem is that the matrices $\cM^{m,n}_{e,\ell;m',n'}(\De)$ occasionally have poles which are not allowed to appear in $G_{e,\De,\ell}^{(1)\;\text{seed}}$ by representation theory. In principle this is not so problematic, but sometimes these poles are above the unitarity bound and this ruins the numerics~\cite{Simmons-Duffin:2015qma}. But even if they are below the unitarity bound, it is desirable to get rid of them since they are only making the approximation more complicated, without improving the accuracy. Let $\De_0$ be such a pole. Being forbidden by representation theory, consistency requires that the polynomials $\hat P^{m,n}_{(1),e,\ell}(\De)$ have a zero at $\De=\De_0$. This would mean that $\hat P^{m,n}_{(1),e,\ell}(\De)=(\De-\De_0) P^{m,n}_{(1),e,\ell}(\De)$, but since the scalar blocks are only approximate at this point, $\hat P^{m,n}_{(1),e,\ell}(\De_0)$ is not exactly $0$.
The solution is to divide these polynomials by $(\De-\De_0)$ and discard the remainder,
\be
	\hat P^{m,n}_{(1),e,\ell}(\De)=(\De-\De_0) P^{m,n}_{(1),e,\ell}(\De) + \text{remainder}.
\ee
We then arrive at the approximation
\be\label{eq:finalrationalapprox}
\ptl_x^m\ptl_t^n G_{e,\De,\ell}^{(1)\;\text{seed}}(0,0)\approx \frac{(4r_0)^\De}{\prod_i (\De-\De_i)^{\kappa_i}} P^{m,n}_{(1),e,\ell}(\De),
\ee
where now the poles $\De_i$ do not include the spurious pole $\De_0$. If there is more then one spurious pole, we perform this procedure for every one of them.

Another problem that arises occasionally is as follows. For $m=n=0$ we just get the conformal block, and it is known that it has the large $\De$ asymptotic $r_0^\De\times O(1)$. This implies that the degree of $P$ is bounded by
\be
	\mathrm{deg}\, P^{0,0}_{(1),e,\ell}\leq\sum_{i}\kappa_i.
\ee
For non-zero $m,n$ each derivative brings down at most a power of $\De$ and we find
\be\label{eq:degreebound}
	\mathrm{deg}\, P^{m,n}_{(1),e,\ell}\leq m+n+\sum_{i}\kappa_i.
\ee
This condition is indeed obeyed by the polynomials $P^{m,n}_{(1),e,\ell}$ found numerically, with a small caveat. In the intermediate steps the degree can be larger, but the extraneous leading powers of $\De$ cancel in the end. However, as in any numerical calculation with floating-point numbers, this cancellation is not exact, and~\eqref{eq:degreebound} ends up being violated by powers of $\De$ with extremely small coefficients. These terms should be removed by hand, since even a small coefficient can potentially alter the analysis of positivity at large $\De$.\footnote{In practice these coefficients are so small that~\texttt{Mathematica} treats them as $0$ when producing input files for~\texttt{SDPB}, but it is still useful to bear in mind that there is room for a numerical error here.}

The procedure we just described is completely generic and works for all steps in the sequence~\eqref{eq:rationalstrategy}, as well as for the other fermion blocks $G'$ and $H$.

\section{Generalized free theory}
\label{sec:GFT}
In this section we study the generalized free theory (GFT),  also known as mean field theory, of a Weyl fermion $\psi$. For $\De_\psi>3/2$ this is a 
unitary CFT which does not have a conserved stress-energy tensor, while for $\De_\psi = 3/2$ the fermion GFT reduces to the theory of a free Weyl fermion.\footnote{See~\cite{Elkhidir:2017iov} for an example of a scalar-fermion GFT.} We refer here to $\psi$ and its conjugate $\bar\psi$ as fundamental fields. The fermion GFT is defined by a set of $n$-point functions which are computed using Wick contractions of the fundamental fields $\psi$ and $\bar \psi$.

We focus here on two four-point functions, which due to Wick contractions split into products of two-point functions~\eqref{eq:two-point functions} as follows
\begin{align}
\label{eq:4pf_gft_neutral}
\<\bar\psi(\point_1)\psi(\point_2)\bar\psi(\point_3)\psi(\point_4)\>=
\<\bar\psi(\point_1)\psi(\point_2)\>\<\bar\psi(\point_3)\psi(\point_4)\>-
\<\bar\psi(\point_1)\psi(\point_4)\>\<\bar\psi(\point_3)\psi(\point_2)\>,\\
\label{eq:4pf_gft_charged}
\<\psi(\point_1)\psi(\point_2)\bar\psi(\point_3)\bar\psi(\point_4)\>=
\<\bar\psi(\point_3)\psi(\point_2)\>\<\bar\psi(\point_4)\psi(\point_1)\>-
\<\bar\psi(\point_3)\psi(\point_1)\>\<\bar\psi(\point_4)\psi(\point_2)\>.
\end{align}
Besides the identity operator, the only operators which give a non-zero contribution to the conformal block expansion of these correlators are the double-twist operators of the following schematic form
\begin{align}
\label{eq:gft_operators_neutral}
\cO_{\De,0}  &\propto 
:\!\bar\psi^{\dot\a} \ptl^{\mu_1}\ldots \ptl^{\mu_\ell} \partial^{2n} \psi^\a\!:+\text{descendants},\\
\label{eq:gft_operators_charged}
\cO_{\De,-} &\propto 
:\!\bar\psi^{\dot\a} \ptl^{\mu_1}\ldots \ptl^{\mu_\ell} \partial^{2n} \bar\psi^{\dot\beta}\!:+\text{descendants}.
\end{align}
By descendants in the above equations we mean terms that are total derivatives and are needed to make the operators in the left-hand side to be primaries. Here $\ell$ and $n$ are non-negative integers. Notice that the operators~\eqref{eq:gft_operators_neutral} and~\eqref{eq:gft_operators_charged} are generically in a reducible spin representation. Their scaling dimensions are given by
\begin{equation}
\De=2\De_\psi+2n+\ell.
\end{equation}

Using the $s$-channel conformal block decomposition of section~\ref{sec:fermion_blocks} we can interpret~\eqref{eq:4pf_gft_neutral} and~\eqref{eq:4pf_gft_charged} as equations for the OPE data of the exchanged double-twist operator and compute all the products of OPE coefficients~\eqref{eq:list_product_ope} between two fundamental Weyl fermions and the double-twist operators~\eqref{eq:gft_operators_neutral} and~\eqref{eq:gft_operators_charged} order by order in $\sqrt{z\bar z}$. Concretely, this is done by using the explicit expressions for the fermion conformal blocks found in section~\ref{sec:fermion_block_computation}, decomposing~\eqref{eq:4pf_gft_neutral} and~\eqref{eq:4pf_gft_charged} into six independent equations spanned by six independent four-point tensor structures, making the replacement
\begin{equation}
z \rightarrow \epsilon z,\quad
\bar z \rightarrow \epsilon \bar z 
\end{equation}
and expanding in $\epsilon$. At order $\epsilon^N$, only a finite number of operators~\eqref{eq:gft_operators_neutral} and~\eqref{eq:gft_operators_charged} contribute. 
Matching all the coefficients proportional to $z^{N-k} \bar z^k$, $k=0,\ldots, N$, gives rise to an over-determined system of equations for the products of OPE coefficients~\eqref{eq:list_product_ope}, from which we find the OPE data fo the low-lying operators.
Having obtained the coefficients for several values of $\ell$ and $n$ we can guess the general result.

The main reasons for studying the spectrum of the double-twist operators in the fermion GFT are the following. First, it provides a consistency check for our setup: finding a solution for an over-determined system is non-trivial, the solution must obey all the properties of OPE coefficients from section~\ref{sec:three_point_functions} and the correctly approximated blocks at the crossing-symmetric point $z=\bar z=1/2$ must still reproduce the GFT correlation functions.
Second, in the numerical analysis the fermion GFT provides a reference point on all the plots and should always lie in the allowed region.\footnote{Unless of course an assumption 	on the CFT spectrum is made that is not respected by the GFT.}
Finally, the result is important on its own, since the double-twist operators describe approximately part of the spectrum for generic CFTs which consists of large spin operators~\cite{Fitzpatrick:2012yx,Komargodski:2012ek}.

In what follows we will discuss the spectrum of the operators~\eqref{eq:gft_operators_neutral} and~\eqref{eq:gft_operators_charged} in more detail and provide the final expressions for the products of OPE coefficients~\eqref{eq:list_product_ope}. We also derive the free fermion CFT data as a $\De_\psi\rightarrow 3/2$ limit of our results.

One could in principle compute directly the CFT data associated to the operators~\eqref{eq:gft_operators_neutral} and~\eqref{eq:gft_operators_charged}. 
First, this would require to fix their precise form by demanding that these operators are primaries. Second, one would have to normalize and diagonalize their basis by computing their two-point functions using Wick contractions. Third, one would need to compute their three-point functions with $\psi$ and $\bar \psi$. This procedure gives more information compared to the one we use here, namely it provides the individual OPE coefficients rather then their products. However, it is rather tedious, and we will not pursue this direction.

There exists yet another method of computing the products of OPE coefficients~\cite{Karateev:2018oml}. It is based on the harmonic analysis of the conformal group~\cite{Dobrev:1977qv} which allows one to derive an Euclidean inversion formula.\footnote{The interest in harmonic analysis in CFTs was recently revived by the derivation of the Lorentzian inversion formula~\cite{Caron-Huot:2017vep}, see~\cite{Simmons-Duffin:2017nub,Kravchuk:2018htv} for its further developments.} It expresses the CFT data in terms of the four-point function, and is especially easy to apply to four-point functions of GFT fundamental operators.

\subsection{Neutral channel}
\label{sec:GFTneutral}
We address here the double-twist operators~\eqref{eq:gft_operators_neutral}. 
We start by decomposing them into irreducible spin representations. We have\footnote{We treat the $\mu_i$ indices as traceless, because traces are taken care of by the $\ptl^{2n}$ factor.}
\be
\label{eq:spin_decomposition_neutral}
(1,0)\otimes (\ell,\ell) \otimes (0,1) = (\ell-1,\ell-1)\oplus (\ell+1,\ell+1)\oplus (\ell+1,\ell-1)\oplus (\ell-1,\ell+1).
\ee
As we see, there are four types of double-twist operators which we refer to as ``towers'': two towers of TS operators and two towers of NTS operators related by hermitian conjugation. Notice that the case of scalar TS operators is special since they are contained only in the first entry of the right-hand side of~\eqref{eq:spin_decomposition_neutral}.
Using an obvious redefinition of the spin parameter $\ell$ we write the schematic form of all four towers in the right-hand side of~\eqref{eq:spin_decomposition_neutral} respectively as\footnote{The derivatives are defined as follows $\bar\partial^{\dot\alpha \beta}\equiv \bar\sigma^{\dot\alpha \beta}_\mu \partial^\mu$ and $\partial_{\alpha\dot\beta}\equiv \sigma_{\alpha\dot\beta}^\mu \partial_\mu$. We also use the standard convention for contracting the Lorentz indices.
}
\begin{align}
\label{eq:gft_neutral_TS_1}
\cO^{(\ell,\ell)}_{\De,0} &=
:\!\psi^\a(x)\partial_{\a\dot\beta} (\bar s \bar\partial s)^{\ell}\partial^{2n}\bar\psi^{\dot\beta}(x)\!:,     &&\De=2\De_\psi+2n+\ell+1, &&&\ell\geq 0,\\
\label{eq:gft_neutral_TS_2}
\cO^{\prime(\ell,\ell)}_{\De,0} &=
:\!\psi(x,s) (\bar s \bar\partial s)^{\ell-1} \partial^{2n'} \bar\psi(x,\bar s)\!:, &&\De=2\De_\psi+2n'+\ell-1,   &&&\ell\geq 1,\\
\label{eq:gft_neutral_NTS}
\cO^{(\ell+2,\ell)}_{\De,0} &=
 :\!\psi(x,s) (s\partial)_{\dot\beta} (\bar s \bar\partial s)^\ell \partial^{2n} \bar\psi^{\dot\beta}(x)\!:,   &&\De=2\De_\psi+2n+\ell+1, &&&\ell\geq 0,\\
\label{eq:gft_neutral_NTSb}
\cO^{(\ell,\ell+2)}_{\De,0} &=
:\!\psi_\a(x) (\bar s\bar\partial)^\a(\bar s \bar\partial s)^\ell \partial^{2n} \bar\psi(x,\bar s)\!:,   &&\De=2\De_\psi+2n+\ell+1, &&&\ell\geq 0.
\end{align} 
In order not to clutter the notation we have suppressed the dependence of the operators on the non-negative integers $n$ and $n'$, omitted the contributions of the descendants
 needed to make these operators primaries, and ignored their normalization.
For $\ell\geq 1$ the TS operators~\eqref{eq:gft_neutral_TS_1} and~\eqref{eq:gft_neutral_TS_2} have degenerate scaling dimensions when $n'=n+1$. This implies that from the four-point function~\eqref{eq:4pf_gft_neutral} we cannot extract the products of individual OPE coefficients and instead we can only compute their combined contribution~\eqref{eq:neutral_ope_product_with_degeneracy}, where we sum over the two degenerate operators.

The lowest dimensional $\ell\geq 1$ TS operators in the fermion GFT spectrum appear in the~\eqref{eq:gft_neutral_TS_2} tower with $n'=0$. They saturate the unitarity bound~\eqref{eq:unitarity_bound} only if $\De_\psi=3/2$. We thus see explicitly that the fermion GFT has neither the conserved current nor the stress tensor for $\De_\psi>3/2$.

\paragraph{Results} We summarize here the analytic expressions found for the products of the neutral OPE coefficients. Let us first focus on the two TS operators~\eqref{eq:gft_neutral_TS_1} and~\eqref{eq:gft_neutral_TS_2}. As we already mentioned, for $\ell=0$ only the first tower of operators contributes and their squared OPE coefficients are found to be 
\be
P_{\De,(0,0),0}=
\frac{4}{n!(n+2)!}
\frac{
	\left(\Delta_\psi-\frac{3}{2}\right)_{n+2}^2
	\left(\Delta_\psi-\frac{1}{2}\right)_{n}
	\left(\Delta_\psi-\frac{1}{2}\right)_{n+1}}
{(2\Delta_\psi-1)^2\left(2\Delta_\psi+n- 3\right)_{n+1}\left(2\Delta_\psi +n - 1\right)_{n+1}},
\ee
where $\De=2\De_\psi+2n+1$. The TS operators $\cO'$ in~\eqref{eq:gft_neutral_TS_2} with $n'=0$ are also non-degenerate and their products of OPE coefficients read as
\begin{equation}
\label{eq:gft_neutral_TS_n=0}
P^{22}_{\De,(\ell,\ell),0} =
\frac{1}{(\ell-1)!}
\frac{\left(\De_\psi+\frac{1}{2}\right)_{\ell-1}^2}
{\left(2\De_\psi+\ell-1\right)_{\ell-1}}, \quad P^{11}_{\De,(\ell,\ell),0} = P^{12}_{\De,(\ell,\ell),0} = P^{21}_{\De,(\ell,\ell),0} =0\,,
\end{equation}
where $\Delta = 2\Delta_\psi +\ell-1$. When $\ell\geq 1$ and $n'>0$,  the sums of product of OPE coefficients of $\cO$ and $\cO'$ read as
\begin{align}
\label{eq:neutral_gft_TS_11}
P^{11}_{\De,(\ell,\ell),0} & =
\frac{n}{\ell}\frac{n+(\ell+1)(2\De_\psi+2n+\ell-3)(2\De_\psi+n-4)}{2\De_\psi+2n+\ell-2} \times \mathcal{C},\\
\label{eq:neutral_gft_TS_12}
P^{12}_{\De,(\ell,\ell),0} & =P^{21}_{\De,(\ell,\ell),0} =
-n\,(2\De_\psi+n-5) \times \mathcal{C},\\
\label{eq:neutral_gft_TS_22}
P^{22}_{\De,(\ell,\ell),0} & =
\left((2+\ell)(2\De_\psi+\ell-4)+2n\,(2\De_\psi+\ell-3)+2n^2\right) \times \mathcal{C},
\end{align}
where the common factor $\mathcal{C}$ is defined as 
\begin{align}
\nn
\mathcal{C} &\equiv
\frac{2\De_\psi+2n-5}{4^n(\ell+2)(\ell-1)!n!\, \left(\ell+2\right)_n}\\
\label{eq:neutral_gft_TS_common_factor}
&\times
\frac{
	\left(\De_\psi-\frac{3}{2}\right)_{n-1}
	\left(\De_\psi+\frac{1}{2}\right)_{n+\ell-1}^2
	\left(2\De_\psi-3\right)_{n-1}
	\left(2\De_\psi+\ell+n-3\right)_{n-1}
}{
	\left(\De_\psi-1\right)_{n-1}
	\left(2\De_\psi+\ell+n-3\right)_{n}^2
	\left(2\De_\psi+\ell+2n-1\right)_{\ell-1}
}
\end{align}
and the scaling dimension is $\Delta = 2\Delta_\psi +2n+\ell+1$. It turns out that the coefficient~\eqref{eq:gft_neutral_TS_n=0} is identical to~\eqref{eq:neutral_gft_TS_22} for $n=0$. 
For NTS operators~\eqref{eq:gft_neutral_NTS} and \eqref{eq:gft_neutral_NTSb} we get
\begin{align}
\nn
P_{\De,(\ell+2,\ell),0} & =P_{\De,(\ell,\ell+2),0} =
\frac{2}{\ell! n! \left(2+\ell\right)_{n+1}}\\
&\times\frac{
	\left(\De_\psi-\frac{3}{2}\right)_{n+1}^2
	\left(\De_\psi+\frac{1}{2}\right)_{\ell+n}
	\left(\De_\psi+\frac{1}{2}\right)_{\ell+n+1}}
{\left(2\De_\psi+n-3\right)_{n+1}
	\left(2\De_\psi+\ell+n-1\right)_{n+1}
	\left(2\De_\psi+\ell+2n+1\right)_\ell},
\end{align}
where $\Delta = 2\Delta_\psi +2n+\ell$.

Equations ~\eqref{eq:neutral_gft_TS_11} - \eqref{eq:neutral_gft_TS_common_factor} fully agree with (3.141), (3.142) - (3.145) of ~\cite{Karateev:2018oml}, where these results  were obtained independently using harmonic analysis.

\paragraph{Free theory}
In the special $\Delta_\psi=3/2$ case the GFT reduces to the theory of a free Weyl fermion which satisfies the usual equations of motion
\begin{equation}
\label{eq:equations_of_motion_Weyl}
\partial^{\dot\alpha\beta}\psi_\beta = 0
\quad\Longrightarrow\quad\partial^2\psi_\beta = 0\,.
\end{equation}
This immediately implies that the tower of operators~\eqref{eq:gft_neutral_TS_1}, \eqref{eq:gft_neutral_NTS} and~\eqref{eq:gft_neutral_NTSb} must vanish. The only non-vanishing tower of operators is given by~\eqref{eq:gft_neutral_TS_2} with $n = 0$.  
The result for the product of OPE coefficients follows from~\eqref{eq:gft_neutral_TS_n=0} and reads as
\begin{eqnarray}
\nn
&&P^{11}_{\De,(\ell,\ell),0} = P^{12}_{\De,(\ell,\ell),0} = P^{21}_{\De,(\ell,\ell),0} = 0,\\
\label{P22FreeTheory}
&&P^{22}_{\De,(\ell,\ell),0} =\frac{\ell}{\left(2+\ell\right)_{\ell-1}}\times \Gamma(\ell+1),\quad\De=\ell+2,\quad\ell\geq 1.
\end{eqnarray}
In other words, there are no scalar or NTS operators which can appear in the neutral channel in the free fermion theory, there are only conserved spin $\ell\geq 1$ currents satisfying the unitarity bound~\eqref{eq:unitarity_bound}. 
The OPE coefficients~\eqref{P22FreeTheory} have been already derived (in an arbitrary number of dimensions) in \cite{Giombi:2017rhm} by a direct computation of three-point functions. 
Their result (2.28) perfectly matches \eqref{P22FreeTheory}.\footnote{In matching the results one has to pay attention to the different normalizations of the two point-functions:
	in detail we have $P^{22}_{\De,(s,s),0} = C_{s\psi \psi}  C_{\psi\psi}^2/C_{ss}$, where $C_{s\psi \psi}$, $C_{ss}$ and $C_{\psi\psi}$  are defined in (2.24), (2.26) and (2.28) of~\cite{Giombi:2017rhm} respectively.}

To conclude, let us also compute here $C_J$ and $C_T$ central charges. This is done by equating~\eqref{P22FreeTheory} with~\eqref{eq:PJ} and~\eqref{eq:PT} for $\ell=1$ and $\ell=2$ respectively. In the former case one gets
\begin{equation}
\label{eq:CJ_free_fermion}
\theta=\pi/2,\quad
\lambda^{1}_{\<\bar\psi\psi J\>}=0,\quad
\lambda^{2}_{\<\bar\psi\psi J\>}=\frac{1}{\sqrt{2}\pi^2},\quad
C_J=\frac{1}{2\pi^4}.
\end{equation}
In the later case one gets
\begin{equation}
\label{eq:CT_free_fermion}
\lambda^1_{\<\bar\psi\psi T\>}=0,\quad
\lambda^2_{\<\bar\psi\psi T\>}=-\frac{i}{\pi^2},\quad
C_T=\frac{1}{\pi^4}.
\end{equation}
These central charges agree with~\cite{Osborn:1993cr}.

\subsection{Charged channel}
\label{sec:GFTcharged}
We address now the double-twist operators~\eqref{eq:gft_operators_charged}. As before we start by decomposing them into irreducible spin representations which read as
\be
\label{eq:spin_decomposition_charged}
(0,1)\otimes (\ell,\ell) \otimes (0,1) = (\ell,\ell) \oplus (\ell,\ell)
\oplus(\ell,\ell+2)\oplus (\ell,\ell-2).
\ee
As before there are generically four towers of operators. Notice however that in the right-hand side of the decomposition~\eqref{eq:spin_decomposition_charged} the second entry can appear only for $\ell\geq 1$ and the last entry only for $\ell\geq 2$.
Again using a redefinition of the spin parameter $\ell$ we write the schematic form of all four towers in the right-hand side of~\eqref{eq:spin_decomposition_charged} respectively as
\begin{align}
\label{eq:gft_charged_TS_1}
\cO^{(\ell,\ell)}_{\De,-} &=
 :\!\bar\psi_{\dot\a}(x) (\bar s \bar\partial s)^{\ell} \partial^{2n} \bar\psi^{\dot\a}(x)\!:,     &&\De=2\De_\psi+2n+\ell, &&&
 {\ell\in \text{Even},}\\
\label{eq:gft_charged_TS_2}
\cO^{\prime(\ell,\ell)}_{\De,-} &=
:\!\bar\psi^{\dot\a}(x)(s\partial)_{(\dot\a} \bar s_{\dot\b)} (\bar s \bar\partial s)^{\ell-1} \partial^{2n} \bar\psi^{\dot\b}(x)\!:,   &&\De=2\De_\psi+2n+\ell, &&&
 {\ell \in\text{Odd},}\\
\label{eq:gft_charged_NTS}
\cO^{(\ell+2,\ell)}_{\De,-} &=
:\!\bar\psi^{\dot\a}(x) (s\partial)_{\dot\a}(s\partial)_{\dot\b} (\bar s \bar\partial s)^\ell \partial^{2n}\bar\psi^{\dot\b}(x)\!:,   &&\De=2\De_\psi+2n+\ell+2, &&&
 {\ell \in\text{Odd},}\\
\label{eq:gft_charged_NTSb}
\cO^{(\ell,\ell+2)}_{\De,-} &=
:\!\bar\psi(x,\bar s) (\bar s \bar\partial s)^{\ell} \partial^{2n'}\bar\psi(x,\bar s)\!:, &&\De=2\De_\psi+2n+\ell,   &&&
{\ell \in \text{Odd}.}
\end{align} 
The same comments apply to the notation here as below~\eqref{eq:gft_neutral_TS_1}-\eqref{eq:gft_neutral_NTSb}.
Contrary to the neutral case, the NTS operators in \eqref{eq:gft_charged_NTS} and \eqref{eq:gft_charged_NTSb} are not related by hermitian conjugation.
Additionally the restriction to even or odd $\ell$ in the above expressions is due to the identity
\be
	:\!\bar\psi_{\dot\a} \ptl_{\mu_1}\cdots\ptl_{\mu_\ell} \ptl^{2n}\bar\psi_{\dot\b}\!:=(-1)^{\ell+1}:\!\bar\psi_{\dot\b} \ptl_{\mu_1}\cdots\ptl_{\mu_\ell} \ptl^{2n}\bar\psi_{\dot\a}\!:+\text{descendants} , 
\ee
which effectively implies (anti-)symmetry in $\a\b$ modulo {terms that are descendants of other operators.  We thus find that} for each $\ell$ one of the two TS operators is a descendant, and for even $\ell$ both NTS operators are descendants. In particular the primary~\eqref{eq:gft_charged_TS_1} exists only for even $\ell$, while the primaries~\eqref{eq:gft_charged_TS_2}-\eqref{eq:gft_charged_NTSb} exist only for odd $\ell$.
Note that for $\ell\geq 1$ the traceless symmetric operators $\cO$ and $\cO'$ with the same $n$ would seem to have degenerate scaling dimensions, but since only one of them exists at any given $\ell$, we can actually extract squares of their individual OPE coefficients.

\paragraph{Results}
We start from TS operators~\eqref{eq:gft_charged_TS_1} and~\eqref{eq:gft_charged_TS_2}. 
For any $\ell\geq0$ the product of OPE coefficients is always a number and not a $2\times 2$ matrix, like in the neutral case, due to presence of the $\mathbb{Z}_2$ permutation symmetry exchanging two identical fermions which
relates the two otherwise independent tensor structures, recall \eqref{eq:charged_correlator_1}. Since depending on the parity of $\ell$ we are studying~\eqref{eq:gft_charged_TS_1} or~\eqref{eq:gft_charged_TS_2}, we expect two different expressions for the product of OPE coefficients
for even and odd spin operators. We get
\begin{align}
\nn
P_{\De,(\ell,\ell),-}
& = \frac{\ell(\ell+1)}{2^{\ell-1}n!\,(2+\ell)\Gamma(n+\ell+2)} \\
\label{eq:charged_TS_spinOdd}
& \times\frac{
	\left(\De_\psi-\frac{3}{2}\right)_{n+1}^2
	\left(\De_\psi+\frac{1}{2}\right)_{n+\frac{\ell-1}{2}} \left(\Delta _{\psi
	}+\frac{1}{2}\right)_{n+\ell}}
{\left(2\De_\psi+n-3\right)_{n+1}
	\left(\De_\psi+\frac{\ell-1}{2}+n\right)_{\frac{\ell+1}{2}}
	\left(2 \De_\psi+\ell+n-2\right)_n},\quad&&\ell\in\text{Odd},\\
\nn
P_{\De,(\ell,\ell),-}
& = \frac{\ell+1}{2^\ell n!\, \Gamma(\ell+n+2)} \frac{2\De_\psi+2n-3}{2\De_\psi+2n+\ell-3}\\
\label{eq:charged_TS_spinEven}
&\times\frac{
	\left(\De_\psi-\frac{3}{2}\right)_{n}^2
	\left(\De_\psi+\frac{1}{2}\right)_{n+\frac{\ell}{2}} \left(\Delta _{\psi
	}+\frac{1}{2}\right)_{n+\ell}}
{\left(2\De_\psi+n-3\right)_{n}
	\left(\De_\psi+n+\frac{\ell}{2}\right)_{\frac{\ell}{2}}
	\left(2 \De_\psi+\ell+n-2\right)_n},\quad&&\ell\in\text{Even}\,,
\end{align}
where $\De=2\De_\psi+2n+\ell$.

The fact that NTS operators only exist for odd $\ell$ is consistent with the  $\mathbb{Z}_2$ permutation symmetry of the external fermions, see \eqref{eq:charged_correlator_2} and  \eqref{eq:charged_correlator_3}. The products of their OPE coefficients read
\begin{align}
\nn
P_{\De,(\ell+2,\ell),-}
&= \frac{1}{2^\ell \ell! n!\left(\ell+2\right)_{n+2}} \frac{\Gamma(\De_\psi+n+\frac{\ell+3}{2})}{\Gamma(\De_\psi+n+\ell+1)} \nn \\
&\times\frac{
	\left(\De_\psi-\frac{3}{2}\right)_{n+1}^2
	\left(\De_\psi+\frac{1}{2}\right)_{n+\frac{\ell-1}{2}}
	\left(\Delta _{\psi
	}+\frac{1}{2}\right)_{n+\ell+1}}
{\left(2\De_\psi+n-2\right)_{n}
	\left(2 \De_\psi+\ell+n-1\right)_{n+1}},\quad\ell\in\text{Odd}\,,
\end{align}
where $\De=2\De_\psi+2n+\ell+2$, and
\begin{align}
\nn
P_{\De,(\ell,\ell+2),-}
&= \frac{1}{2^{\ell-1} \ell! n!\left(\ell+2\right)_{n}} \frac{\Gamma(\De_\psi+n+\frac{\ell+1}{2})}{\Gamma(\De_\psi+n+\ell)} \nn \\
&\times\frac{
	\left(\De_\psi-\frac{3}{2}\right)_{n+1}^2
	\left(\De_\psi+\frac{1}{2}\right)_{n+\frac{\ell-1}{2}}
	\left(\Delta _{\psi}+\frac{1}{2}\right)_{n+\ell}}
{\left(2\De_\psi+n-4\right)_{n}
	\left(2 \De_\psi+\ell+n-1\right)_{n}},\quad\ell\in\text{Odd},
\end{align}
where $\De=2\De_\psi+2n+\ell$.

\paragraph{Free theory}

Due to the free fermion equations of motions~\eqref{eq:equations_of_motion_Weyl}, the operators~\eqref{eq:gft_charged_TS_2} and~\eqref{eq:gft_charged_NTS} vanish and the operators~\eqref{eq:gft_charged_TS_1} and~\eqref{eq:gft_charged_NTSb} can be non-zero only for $n=0$.  Interestingly enough, the only TS operator which has a non-vanishing OPE coefficient is the scalar $\De=3$ operator $\psi^\alpha(x)\psi_\alpha(x)$ with\footnote{It appears that for free fermions the TS primaries~\eqref{eq:gft_charged_TS_1} with $\ell>0$ do not exist. For example, it is easy to see that~\eqref{eq:gft_charged_TS_1} with $\ell=2$ is absent by studying the first few terms in the character of the relevant tensor product of Verma modules.}
\begin{equation}
P_{\De,(0,0),-}=1\,.
\end{equation}
The NTS operators have the following products of OPE coefficients
\begin{equation}
\label{eq:ope_charged_nts_free}
P_{\De,(\ell,\ell+2),-} = \frac{
	\Gamma(\ell+2)\Gamma(\ell+3)}{\Gamma(2\ell+2)},\quad\De=3+\ell,\quad
\ell\in\text{Odd}.
\end{equation}
The operators $(\ell,\ell+2)$ with $\ell\geq 1$ saturating the unitarity bound as in~\eqref{eq:ope_charged_nts_free} are conserved NTS currents. These operators have been discussed in~\cite{Giombi:2017rhm}, but their OPE coefficients with $\psi$ and $\bar \psi$ have not been derived.

\section{Fake primary effect}
\label{sec:topology_blocks}

In this section we discuss the fake primary effect alluded to in the introduction.
It originates from the peculiar properties of conformal blocks in a given bootstrap setup. This is best formulated in a slightly formal but very convenient way in terms of the space of functionals entering the setup and its topology.

The crossing equations~\eqref{eq:Fexpansion} are expressed in terms of the functionals~\eqref{eq:functionals} which we list here again explicitly for the reader convenience
\be
\label{eq:full_list_functionals}
	\vec\a \. \vec G^{ab}_{\L,\De,(\ell, \ell)},\quad \vec\a \. \vec G_{\L,\De,(\ell, \ell+2)},\quad \vec\a \. \vec H_{\L,\De,(\ell, \ell)},\quad \vec\a \. \vec H_{\L,\De,(\ell+2, \ell)},\quad \vec\a \. \vec H_{\L,\De,(\ell, \ell+2)}.
\ee
Given a vector $\vec\alpha$ the entries in~\eqref{eq:full_list_functionals} can be seen as functions of the scaling dimension $\De$ and spin $(\ell,\bar \ell)$.
Let us denote the full set of functionals~\eqref{eq:full_list_functionals} by
\be
\label{eq:collective_functional}
\vec\a\.\vec\cG_{\L,\sigma},\quad
\sigma = (\De,(\ell,\bar\ell),Q),
\ee
where $\sigma$ is the collective label specifying the scaling dimension $\De$, the spin representation $(\ell,\bar\ell)$ and the block type $G$ or $H$. More precisely, the blocks $G$ and $H$ appear in the neutral and charged channels respectively and we use the labels $Q=0$ and $Q=-$ here to distinguish between them. 
For a given vector $\vec\alpha$ one can now treat the objects~\eqref {eq:collective_functional} as functions of $\s\in \Sigma$, where the space $\Sigma$ includes all values of $\sigma$ which enter in~\eqref{eq:apositivity} (when no assumptions on the spectrum are made).

When defining the semi-definite problems of section~\ref{sec:sdp} we require various positivity constraints. Implications of these constraints can be strongly affected by the continuity properties of $\vec\cG_{\L,\sigma}$ in $\sigma$ -- a continuous function which is positive at some point has to remain positive in an open neighborhood of this point.
A convenient way to describe these properties is to specify the topology of the space $\Sigma$.

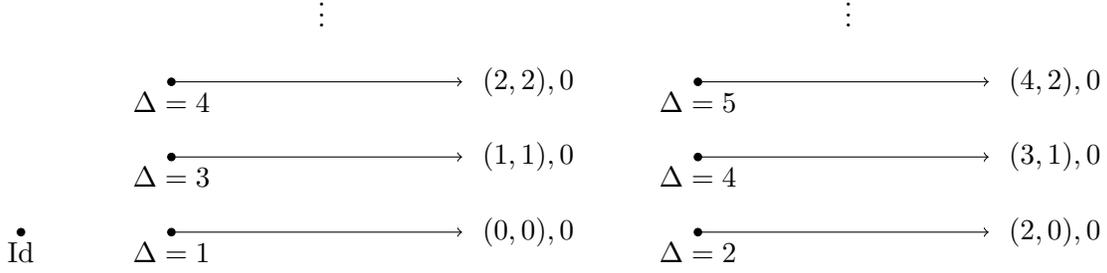
\begin{figure}[t]
	\begin{center}
		\begin{tikzpicture}
			\node[twopt] (identity) at (0,0) {};
			\node[twopt] (spin0U) at (2,0) {};
			\node (spin0Inf) at (6,0) {};
			\node[twopt] (spin1U) at (2,1) {};
			\node (spin1Inf) at (6,1) {};
			\node[twopt] (spin2U) at (2,2) {};
			\node (spin2Inf) at (6,2) {};
			\draw[->] (spin0U) -- (spin0Inf);
			\draw[->] (spin1U) -- (spin1Inf);
			\draw[->] (spin2U) -- (spin2Inf);
			\node[below] at (spin0U) {$\De=1$};
			\node[below] at (spin1U) {$\De=3$};
			\node[below] at (spin2U) {$\De=4$};
			\node[below] at (identity) {Id};
			\node[right] at (spin0Inf) {$(0,0),0$};
			\node[right] at (spin1Inf) {$(1,1),0$};
			\node[right] at (spin2Inf) {$(2,2),0$};
			\node[twopt] (spin0NSU) at (9,0) {};
			\node (spin0NSInf) at (13,0) {};
			\node[twopt] (spin1NSU) at (9,1) {};
			\node (spin1NSInf) at (13,1) {};
			\node[twopt] (spin2NSU) at (9,2) {};
			\node (spin2NSInf) at (13,2) {};
			\draw[->] (spin0NSU) -- (spin0NSInf);
			\draw[->] (spin1NSU) -- (spin1NSInf);
			\draw[->] (spin2NSU) -- (spin2NSInf);
			\node[below] at (spin0NSU) {$\De=2$};
			\node[below] at (spin1NSU) {$\De=4$};
			\node[below] at (spin2NSU) {$\De=5$};
			\node[right] at (spin0NSInf) {$(2,0),0$};
			\node[right] at (spin1NSInf) {$(3,1),0$};
			\node[right] at (spin2NSInf) {$(4,2),0$};
			\node at (4,3) {$\vdots$};s
			\node at (11,3) {$\vdots$};s
		\end{tikzpicture}
		\caption{Naive topology of $\Sigma$ in charge $q=0$ sector.}
		\label{fig:SigmaNaiveTopology}
	\end{center}
\end{figure}

The intuitive picture of this topology for charge $0$ is shown in figure~\ref{fig:SigmaNaiveTopology}. For each value of $(\ell,\bar\ell)$, we have a half-line $\R_+$ of operators of spin $(\ell,\bar\ell)$, parametrized by $\Delta$. For $\ell=\bar\ell=0$ we have an extra disconnected point for the identity operator. In other words, we would expect that 
$\Sigma$ is a disconnected sum of a point and an infinite countable tower of half-lines $\R_+$. We have indeed tacitly assumed this intuitive picture when we wrote~\eqref{eq:apositivity} 
in reviewing the way upper bounds on operator dimensions are obtained in numerical bootstrap studies. 
However, as we will soon see, this intuitive topology does not capture all the continuity properties of~$\vec\cG_{\L,\sigma}$. In other words, there is a coarser (``more connected'') topology on $\Sigma$ with respect to which~$\vec\cG_{\L,\sigma}$ is continuous. With this improved topology not only does $\Sigma$ have much fewer connected components, but these components are not even simply-connected.

\subsection{Unitary poles in conformal blocks}
\label{sec:poles-in-cb}

It is known that conformal blocks and thus $\vec\cG_{\L,\s}$ have poles in $\Delta$~\cite{Zamolodchikov:1987,Kos:2013tga,Penedones:2015aga}.
For concreteness let us take $\s=(\De,(\ell,\ell),-)$, in which case $\cG_{\L,\s}$ is related to the blocks $H_{\De,(\ell,\ell)}$. There exists an expansion of $H_{\De,(\ell,\ell)}$ which converges for all $\De\in \C$ and has the following schematic form
\be\label{eq:blockschematic}
	H_{\De,(\ell,\ell)}=\sum_{n}\frac{\<\psi\psi|n\>\<n|\bar\psi\bar\psi\>}{\<n|n\>},
\ee
where the sum is over an orthogonal set of states $|n\rangle$ related by the operator-state correspondence to the descendants of $\cO_{\De,-}^{(\ell,\ell)}$. Instead of assuming that the descendants are unit-normalized we explicitly divide by their norms. These norms are polynomial functions of $\De$ and vanish at a discrete set of scaling dimensions, leading to poles in $H_{\De,(\ell,\ell)}$. One can furthermore check that this is the only way in which singularities can arise, provided that the conformal blocks are appropriately normalized.\footnote{Indeed we can trivially add poles by changing the normalization as $H_{\De,(\ell,\ell)}\to (\De-\De_*)^{-1}H_{\De,(\ell,\ell)}$. The statement that the only poles come from null descendants is true if the three-point structures which are used to define the blocks are entire functions of $\De$, and the two-point functions do not have zeros in $\De$, as is the case in our conventions.}

Therefore, poles in $\cG_{\L,\s}$ are associated with some descendants becoming null. All null descendants have been classified~\cite{Penedones:2015aga}. They occur for $\De<\De_\text{unitary}(\ell,\bar\ell)$ and generally give rise to simple poles in $\Delta$ for CFTs defined in $d$ dimensions. Poles can and do occur also at $\De=\De_\text{unitary}(\ell,\bar \ell)$, because the unitarity bound itself is determined by some descendant developing negative norm~\cite{Mack:1975je}. When $d$ approaches an even integer value, some simple poles 
can collide and give rise to double poles, but this effect can only occur for values of $\Delta$ strictly below the unitarity bound $\De<\De_\text{unitary}(\ell,\bar\ell)$. We then conclude
that in all dimensions, including $d=4$, the poles at the unitarity bound are simple. These are the poles we will focus on in what follows.

As can be seen from~\eqref{eq:blockschematic}, poles will not appear if either 
\be\label{eq:vanishing_conditions}
	\<\psi\psi|n\>=0\text{  or  }\<n|\bar\psi\bar\psi\>=0,
\ee
where $|n\>$ is the descendant which becomes null.\footnote{For $\s$ in the neutral channel we should check $\<\psi\bar\psi|n\>$ instead.} Importantly, if the pole does appear, the residue is known to be proportional to the conformal block for exchange of a primary with the same quantum numbers as $|n\>$~\cite{Zamolodchikov:1987,Penedones:2015aga}. For $\ell\bar\ell\neq 0$ the null descendant is the ``conservation'' operator
\be\label{eq:conservednull}
	(\ptl_s\ptl_x\ptl_{\bar s}) {\cO_{\De,q}^{(\ell,\bar\ell)}}(x,s,\bar s),
\ee
which has dimension $\De_\text{unitary}(\ell,\bar\ell)+1$ and spin $(\ell-1,\bar\ell-1)$. For $\bar\ell=0$, $\ell\neq 0$, it is given by
\be\label{eq:diracnull}
		(\ptl_s\ptl_x\bar s) {\cO_{\De,q}^{(\ell,0)}}(x,s),
\ee
with dimension $\De_\text{unitary}(\ell,0)+1$ and spin $(\ell-1,1)$. The null descendant for $\ell=0$, $\bar\ell\neq 0$ is constructed analogously. For scalars $\ell=\bar\ell=0$ the null descendant is the Laplace operator
\be
	\ptl^2\cO_{\De,q}^{(0,0)},
\ee
and has dimension $\De_\text{unitary}(0,0)+2$ and spin $(0,0)$.

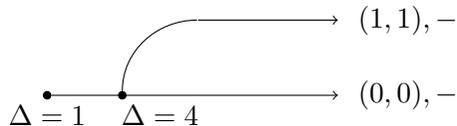
\begin{figure}[t]
	\begin{center}
		\begin{tikzpicture}
		\node[twopt] (spin0U) at (2,0) {};
		\node (spin0Inf) at (6,0) {};
		\node[twopt] (spin0Conn) at (3,0) {};
		\node (spin1Inf) at (6,1) {};
		\node[inner sep=0pt,minimum size=0pt] (spin1Bend) at (4,1) {};
		\draw[->] (spin0U) -- (spin0Inf);
		\draw (spin0Conn) to[out=90,in=180] (spin1Bend);
		\draw[->] (spin1Bend) -- (spin1Inf);
		\node[below] at (spin0U) {$\De=1$};
		\node[below] at (3.5,0) {$\De=4$};
		\node[right] at (spin0Inf) {$(0,0),-$};
		\node[right] at (spin1Inf) {$(1,1),-$};
		\end{tikzpicture}
		\caption{Merging of scalar and vector lines due to a pole in vector blocks. Scaling dimensions indicate dimensions of the scalar blocks.}
		\label{fig:Spin1Spin0ConnectionTopology}
	\end{center}
\end{figure}

There is nothing interesting to say when there is no pole at the unitarity bound. We thus focus on the case when there is a pole and study for concreteness the case of charged vector operators.
Applying the differential operator in~\eqref{eq:conservednull} to the tensor structure $\<\psi\psi\cO^{(1,1)}_\De\>$ in~\eqref{eq:charged_correlator_1} we find
\be\label{eq:consspin1}
(\ptl_{s_3}\ptl_{x_3}\ptl_{\bar s_3})K_3\left(\hat\II^{31}\hat\KK^{23}_1+(-1)^1 \, \hat\II^{32}\hat\KK^{13}_2\right)
\propto K_3  \, \hat\KK^{12}_3
\ee
with a non-zero proportionality coefficient.
We see that the three-point functions in~\eqref{eq:vanishing_conditions} are non-vanishing.
The conformal block $H_{\De,(1,1)}$ then behaves near the unitarity bound $\De_\text{unitary}(1,1)=3$ as
\be
	H_{\De,(1,1)}\sim \frac{c}{\De-3}H_{4,(0,0)},
\ee
where in the right hand side we have a block exchanging a scalar operator of dimension $4$, which are the quantum numbers of the null state~\eqref{eq:conservednull}.  Using unitarity one can show that the coefficient $c$ must be positive.
If we now define a rescaled conformal block
\be
	\hat H_{\De,(1,1)}=c^{-1}(\De-3)H_{\De,(1,1)},
\ee
we conclude 
\be
	\hat H_{3,(1,1)}=H_{4,(0,0)}.
\ee
Since replacing $H$ by $\hat H$ is equivalent to a positive rescaling of the OPE coefficients, which is inessential, we must conclude that the line of conformal blocks for charged spin-1 operators joins the line of scalar blocks at dimension $4$, see figure~\ref{fig:Spin1Spin0ConnectionTopology}. In other words, from the point of view of our numerical setup, vector contributions infinitesimally close to the unitarity bound are indistinguishable from scalar contributions at $\De=4$. This is the fake primary effect -- the limit of vector primaries at the unitarity bound produces a fake scalar primary with $\De=4$.

It is straightforward to see that this phenomenon persists to higher-spin TS conformal blocks in the charged sector. For generic $\ell$ we have
\be
	\hat H_{\De_\text{unitary}(\ell,\ell),(\ell,\ell)}=H_{\De_\text{unitary}(\ell,\ell)+1,(\ell-1,\ell-1)}.
\ee
As we discussed above, the nature of the pole for $\ell=0$ TS blocks is different, and is due to the Laplace operator. We can again check that the three-point tensor structures at $\ell=0$ do not satisfy Laplace equation and thus we have a pole at the unitarity bound $\De=1$. However, the residue is now again an $\ell=0$ block, and we have
\be
	\hat H_{1,(0,0)}=H_{3,(0,0)},
\ee
and hence the charged scalar line reconnects into itself. It turns out that charged NTS blocks with odd $\ell$ (recall that even $\ell$ is forbidden) do not have poles at unitarity bound because the three-point functions satisfy the appropriate equations, and thus these blocks remain isolated. 

In the neutral sector we find that TS blocks have no poles at the unitarity bound, except for $\ell=0$ which behaves exactly as in the charged sector. However, now the NTS blocks have poles and for $\ell>0$ we get
\be
	\hat G_{\De_\text{unitary}(\ell+2,\ell),(\ell+2,\ell)}= G_{\De_\text{unitary}(\ell+2,\ell)+1,(\ell+1,\ell-1)}.
\ee
For $\ell=0$ we again have a pole, but the type of the null descendant is different, see~\eqref{eq:diracnull}, and we find 
\be
	\hat G_{2,(2,0)}=G_{3,(1,1)}.
\ee
This case needs a special clarification. The equation above cannot be literally true because the block on the left hand side is $1\times 1$ while the one on the right is $2\times 2$. In other words, there is only one tensor structure
\be
	\<\bar\psi\psi\cO^{(2,0)}_2\>^{(\uniq)}
\ee
but two tensor structures
\be
	\<\bar\psi\psi\cO^{(1,1)}_3\>^{(a)}.
\ee
The precise statement is instead
\be
	\hat G_{2,(2,0)}=\l_a\l_b^* G^{ab}_{3,(1,1)},
\ee
where $\lambda^a$ is determined by\footnote{There is a simple characterization of $\l_a$. These coefficients are such that they do not contribute to the Ward identity~\eqref{eq:J_ope_relation}, i.e. $2\l_1+\l_2=0$, because the left-hand side of~\eqref{eq:topologytwist} is identically annihilated by $(\ptl_{s_3}\ptl_x\ptl_{\bar s_3})$, as opposed to giving some contact terms. 
The Ward identity essentially counts the coefficient of the contact term, and hence this structure does not contribute to it.}
\be\label{eq:topologytwist}
	(\ptl_{s_3}\ptl_{x_3}\bar s_3)\<\bar\psi\psi\cO^{(2,0)}_2\>^{(\uniq)}\propto \l_a \<\bar\psi\psi\cO^{(1,1)}_3\>^{(a)}.
\ee
This gives more refined information than simply the topology of $\Sigma$, it would be interesting to find the appropriate mathematical object which captures also this additional structure.

Collecting all these observations together, we find the topology of $\Sigma$ which is shown in figure~\ref{fig:SigmaTrueTopologyNeutral} for the neutral sector and in figure~\ref{fig:SigmaTrueTopologyCharged} for the charged sector. As promised, it is far from the naive expectation in figure~\ref{fig:SigmaNaiveTopology}.

\begin{figure}[t]
	\begin{center}
		\begin{tikzpicture}
		\node[twopt] (identity) at (0,0) {};
		\node[twopt] (spin0U) at (2,0) {};
		\node (spin0Inf) at (6,0) {};
		\node[twopt] (spin2U) at (2,1) {};
		\node (spin2Inf) at (6,1) {};
		\node[twopt] (spin3U) at (2,2) {};
		\node (spin3Inf) at (6,2) {};
		\draw[->] (spin0U) -- (spin0Inf);
		\draw[->] (spin2U) -- (spin2Inf);
		\draw[->] (spin3U) -- (spin3Inf);
		\draw (spin0U) to[out=180, in=90] (1,-0.5) to[out=-90,in=180] (1.5,-1) to[out=0,in=-90] (spin0U);
		\node[above] at (spin0U) {$\De=3$};
		\node[below] at (spin2U) {$\De=4$};
		\node[below] at (spin3U) {$\De=5$};
		\node[below] at (identity) {Id};
		\node[right] at (spin0Inf) {$(0,0),0$};
		\node[right] at (spin2Inf) {$(2,2),0$};
		\node[right] at (spin3Inf) {$(3,3),0$};
		\node[twopt] (spin1U) at (9,-1) {};
		\node (spin1Inf) at (13,-1) {};
		\node[twopt] (spin0NSBend) at (10,0) {};
		\node (spin0NSInf) at (13,0) {};
		\node[twopt] (spin1NSBend) at (11,1) {};
		\node (spin1NSInf) at (13,1) {};
		\node[twopt] (spin2NSBend) at (12,2) {};
		\node (spin2NSInf) at (13,2) {};
		\draw[->] (spin1U) -- (spin1Inf);
		\draw[->] (spin1U) to[out=90,in=180] (spin0NSBend) -- (spin0NSInf);
		\draw[->] (spin0NSBend) to[out=90, in=180] (spin1NSBend) -- (spin1NSInf);
		\draw[->] (spin1NSBend) to[out=90, in=180] (spin2NSBend) -- (spin2NSInf);
		\draw (spin2NSBend) -- (12,2.3);
		\draw[dotted] (12,2.3) -- (12,2.6);
		\node[below] at (spin1U) {$\De=3$};
		\node[below] at (spin0NSBend) {$\De=5$};
		\node[below] at (spin1NSBend) {$\De=6$};
		\node[below] at (spin2NSBend) {$\De=7$};
		\node[right] at (spin1Inf) {$(1,1),0$};
		\node[right] at (spin0NSInf) {$(2,0),0$};
		\node[right] at (spin1NSInf) {$(3,1),0$};
		\node[right] at (spin2NSInf) {$(4,2),0$};
		\node at (4,3) {$\vdots$};
		\node at (11,3) {$\vdots$};
		\end{tikzpicture}
		\caption{Topology of $\Sigma$ in neutral sector after taking into account poles. The dimensions shown near intersections correspond to the block which appears as the residue.}
		\label{fig:SigmaTrueTopologyNeutral}
	\end{center}
\end{figure}
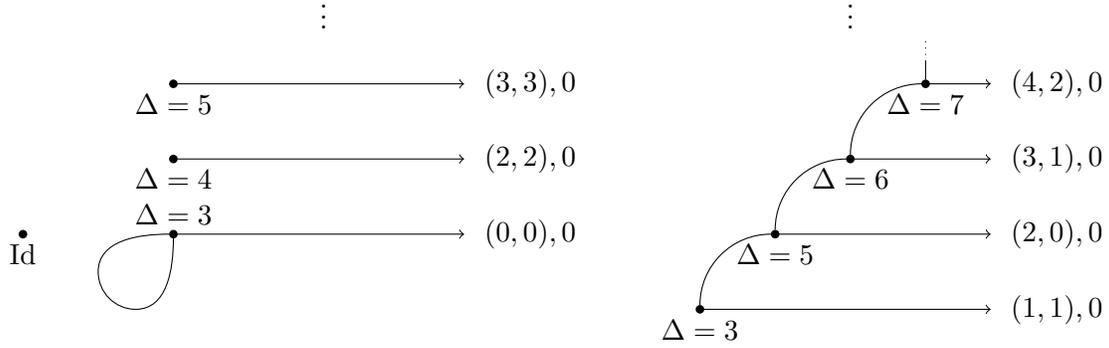

\begin{figure}[t]
	\begin{center}
		\begin{tikzpicture}
		\node[twopt] (spin1NSU) at (9,0-1) {};
		\node (spin1NSInf) at (13,0-1) {};
		\node[twopt] (spin3NSU) at (9,1-1+0.5) {};
		\node (spin3NSInf) at (13,1-1+0.5) {};
		\node[twopt] (spin5NSU) at (9,2-1+1) {};
		\node (spin5NSInf) at (13,2-1+1) {};
		\node[twopt] (spin1NSbU) at (9,0.4-1) {};
		\node (spin1NSbInf) at (13,0.4-1) {};
		\node[twopt] (spin3NSbU) at (9,1.4-1+0.5) {};
		\node (spin3NSbInf) at (13,1.4-1+0.5) {};
		\node[twopt] (spin5NSbU) at (9,2.4-1+1) {};
		\node (spin5NSbInf) at (13,2.4-1+1) {};
		\draw[->] (spin1NSU) -- (spin1NSInf);
		\draw[->] (spin3NSU) -- (spin3NSInf);
		\draw[->] (spin5NSU) -- (spin5NSInf);
		\draw[->] (spin1NSbU) -- (spin1NSbInf);
		\draw[->] (spin3NSbU) -- (spin3NSbInf);
		\draw[->] (spin5NSbU) -- (spin5NSbInf);
		\node[below] at (spin1NSU) {\small $\De=4$};
		\node[below] at (spin3NSU) {\small $\De=6$};
		\node[below] at (spin5NSU) {\small $\De=8$};
		\node[right] at (spin1NSInf) {\small $(3,1),-$};
		\node[right] at (spin3NSInf) {\small $(5,3),-$};
		\node[right] at (spin5NSInf) {\small $(7,5),-$};
		\node[right] at (spin1NSbInf) {\small $(1,3),-$};
		\node[right] at (spin3NSbInf) {\small $(3,5),-$};
		\node[right] at (spin5NSbInf) {\small $(5,7),-$};
		\node[twopt] (spin0U) at (2,-1) {};
		\node[twopt] (spin0UReal) at (1,-1) {};
		\node (spin0Inf) at (6,-1) {};
		\node[twopt] (spin1Bend) at (3,0) {};
		\node (spin1Inf) at (6,0) {};
		\node[twopt] (spin2Bend) at (4,1) {};
		\node (spin2Inf) at (6,1) {};
		\node[twopt] (spin3Bend) at (5,2) {};
		\node (spin3Inf) at (6,2) {};
		\draw[->] (spin0U) -- (spin0Inf);
		\draw[->] (spin0U) to[out=90,in=180] (spin1Bend) -- (spin1Inf);
		\draw[->] (spin1Bend) to[out=90, in=180] (spin2Bend) -- (spin2Inf);
		\draw[->] (spin2Bend) to[out=90, in=180] (spin3Bend) -- (spin3Inf);
		\draw (spin3Bend) -- (5,2.3);
		\draw[dotted] (5,2.3) -- (5,2.6);
		\draw (spin0UReal) to[out=180, in=-90] (0,0.5-1) to[out=90,in=180] (0.5,+1-1) to[out=0,in=90] (spin0UReal);
		\draw (spin0UReal) -- (spin0U);
		\node[below] at (2.3,-1) {$\De=4$};
		\node[below] at (0.7,-1) {$\De=3$};
		\node[below] at (spin1Bend) {$\De=5$};
		\node[below] at (spin2Bend) {$\De=6$};
		\node[below] at (spin3Bend) {$\De=7$};
		\node[right] at (spin0Inf) {$(0,0),-$};
		\node[right] at (spin1Inf) {$(1,1),-$};
		\node[right] at (spin2Inf) {$(2,2),-$};
		\node[right] at (spin3Inf) {$(3,3),-$};
		\node at (4,3) {$\vdots$};
		\node at (11,3) {$\vdots$};
		\end{tikzpicture}
		\caption{Topology of $\Sigma$ in charged sector after taking into account poles. The dimensions shown near intersections correspond to the block which appears as the residue.}
		\label{fig:SigmaTrueTopologyCharged}
	\end{center}
\end{figure}
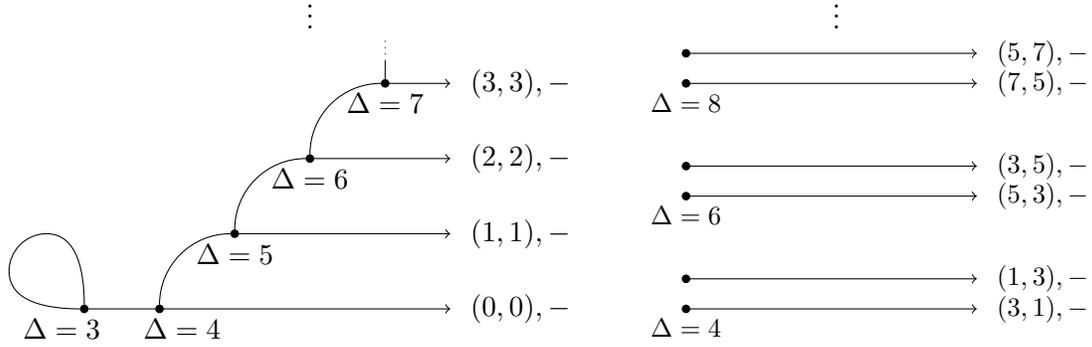

\subsection{Implications for numerics}
\label{sec:implications_numerics}

The fact that $\Sigma$ has a non-trivial topology has strong implications for traditional numerical bounds. With the benefit of hindsight, let us consider the bound on the dimension of the first charged scalar. As we discussed in section~\ref{sec:sdp}, in order to construct such a bound we remove all the charged scalars of dimension below $\De_*$ from $\Sigma$, and try to disprove the existence of solutions to crossing which only contain contributions from this reduced space. In practice of course we consider many different values of $\De_*$ to find the smallest value for which we can disprove the existence of solutions to crossing. Let us denote this minimum value by $\De_*^{\text{min}}$ and denote by $\Sigma_{\De_*^{\text{min}}}$ the associated reduced space.

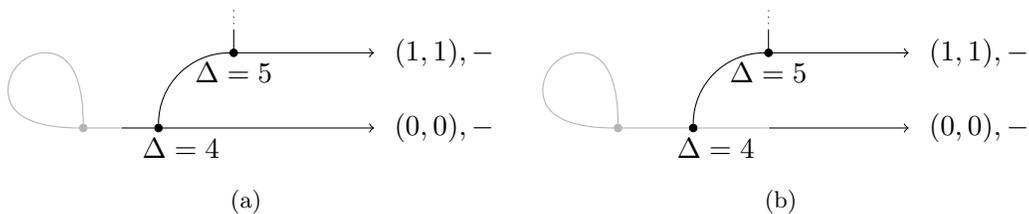
\begin{figure}[t]
	\centering
	\subfigure[]{
		\begin{tikzpicture}
		\node[twopt] (spin0U) at (2,-1) {};
		\node[twopt,draw=black!30, fill=black!30] (spin0UReal) at (1,-1) {};
		\node[inner sep=0pt,minimum size=0pt] (spin0Bound) at (1.5,-1) {};
		\node (spin0Inf) at (5,-1) {};
		\node[twopt] (spin1Bend) at (3,0) {};
		\node (spin1Inf) at (5,0) {};
		\draw[->] (spin0U) -- (spin0Inf);
		\draw[->] (spin0U) to[out=90,in=180] (spin1Bend) -- (spin1Inf);
		\draw (spin1Bend) -- (3,0.3);
		\draw[dotted] (3,0.3) -- (3,0.6);
		\draw[draw=black!30] (spin0UReal) to[out=180, in=-90] (0,0.5-1) to[out=90,in=180] (0.5,+1-1) to[out=0,in=90] (spin0UReal);
		\draw[draw=black!30] (spin0UReal) -- (spin0Bound);
		\draw (spin0Bound) -- (spin0U);
		\node[below] at (2.3,-1) {$\De=4$};
		\node[below] at (spin1Bend) {$\De=5$};
		\node[right] at (spin0Inf) {$(0,0),-$};
		\node[right] at (spin1Inf) {$(1,1),-$};
		\end{tikzpicture}}~~
	\subfigure[]{
		\begin{tikzpicture}
		\node[twopt] (spin0U) at (2,-1) {};
		\node[twopt,draw=black!30, fill=black!30] (spin0UReal) at (1,-1) {};
		\node[inner sep=0pt,minimum size=0pt] (spin0Bound) at (3,-1) {};
		\node (spin0Inf) at (5,-1) {};
		\node[twopt] (spin1Bend) at (3,0) {};
		\node (spin1Inf) at (5,0) {};
		\draw[->] (spin0Bound) -- (spin0Inf);
		\draw[->] (spin0U) to[out=90,in=180] (spin1Bend) -- (spin1Inf);
		\draw (spin1Bend) -- (3,0.3);
		\draw[dotted] (3,0.3) -- (3,0.6);
		\draw[draw=black!30] (spin0UReal) to[out=180, in=-90] (0,0.5-1) to[out=90,in=180] (0.5,+1-1) to[out=0,in=90] (spin0UReal);
		\draw[draw=black!30] (spin0UReal) -- (spin0U);
		\draw[draw=black!30] (spin0Bound) -- (spin0U);
		\node[below] at (2.3,-1) {$\De=4$};
		\node[below] at (spin1Bend) {$\De=5$};
		\node[right] at (spin0Inf) {$(0,0),-$};
		\node[right] at (spin1Inf) {$(1,1),-$};
		\end{tikzpicture}}
	\caption{Topology of $\Sigma$ near charged scalar sector with a gap. Left: the gap in scalar sector is less than $4$. Right: the gap in scalar sector is greater than $4$.}
	\label{fig:ChargedScalarBoundSurprise}
\end{figure}

The crucial observation is that $\Sigma_{{\De_*}^{\text{min}}}$ looks very differently depending on whether $\De_*^{\text{min}}$ is greater or less than $4$. 
The two situations are shown in figure~\ref{fig:ChargedScalarBoundSurprise}. We see immediately that for $\De_*^{\text{min}}>4$ there is in fact no way to exclude contributions of charged scalars at dimension $4$ by imposing a gap in this sector only, since the dimension $4$ scalars can be obtained as a limit of spin-1 contributions. This implies that for $\De_*^{\text{min}}>4$ we are not actually studying the problem of bounding the dimension of the first charged scalar, but rather the dimension of the second charged scalar, assuming that the first scalar has dimension $4$. However, for $\De_*^{\text{min}}< 4$ we are indeed bounding the dimension of the first charged scalar. Therefore, as $\De_*^{\text{min}}$ crosses dimension $4$, the problem we are studying changes. This change is discontinuous since we expect the bound on dimension of the second scalar to be much weaker than the bound on the dimension of the first scalar.

This leads to a striking prediction that the bound on the gap in the charged scalar sector should jump discontinuously as soon as it reaches $\De_*^{\text{min}}=4$, at any value of $\L$. Similarly, this can happen in all other sectors where a topology similar to figure~\ref{fig:ChargedScalarBoundSurprise} is observed. In particular, we expect such jumps in bounds on gaps in charged TS sectors, and neutral NTS sectors. The critical value of the scaling dimension at which the jump should occur in the charged TS and neutral NTS sectors respectively is
\begin{align}
\De_{-\;\text{jump}}^{(\ell,\ell)}   &=\De_{\text{unitary}}(\ell+1,\ell+1)+1=4+\ell,\\
\De_{0\;\text{jump}}^{(\ell+2,\ell)} &=\De_{\text{unitary}}(\ell+3,\ell+1)+1=5+\ell.
\end{align}
The jumps are of course only expected if the bound ever crosses this value. In section~\ref{sec:numerical_results} we will confirm these predictions and perform some further tests.

As a final comment, we should note that the fake primary effect could also work in the opposite direction. Without further assumptions, in principle we should interpret bounds on charged TS operators and neutral NTS operators as bounds on the dimension of the {\it second} allowed operator, with the first one being almost at unitarity.
For instance a scalar charged operator of dimension $\Delta\simeq4$ could mimic an almost conserved charged operator in the $(1,1)$ representation.
In practice however we observe that the solutions of crossing extremizing the gap in a given sector do not contain operators close to the unitarity bound.\footnote{The extremal functional is strictly positive at the unitarity bound, even if it was not required to.} The only exception is the bound on the dimension of the first neutral operator $(2,0)$. As shown in figure \ref{fig:SigmaTrueTopologyNeutral}, this branch of $\Sigma$ connects with the neutral NTS operator in the (3,1) and the neutral TS operators in the $(1,1)$ representation at the unitarity bound, i.e.\ a $\mathrm{U}(1)$ conserved current.
Since the latter generically is present in a solution of crossing, the bound on $(2,0)$ is actually a bound on the next operator after 2. If this bounds happens to be above 5, then in reality it becomes a bound on the next operator after 5. We will see in \sref{sec:numerical_results} that indeed this bound does not display any jump and it starts approximatively at 10.

\subsection{Topology of $\Sigma$ in other setups}
\label{sec:topology_blocks:other}
In what follows we discuss other conformal bootstrap setups where the jump-like behavior was also observed. In section~\ref{sec:topology_blocks:ising} we consider the scalar mixed correlator bootstrap in 3d and discuss the implications of the fake primary effect for the 3d Ising model. In section~\ref{sec:3dfermion} we address the 3d Majorana fermion bootstrap.

\subsubsection{Scalar mixed-correlator bootstrap in 3d}
\label{sec:topology_blocks:ising}

As mentioned in \sref{sec:poles-in-cb}, $\Sigma$ has a non-trivial topology only if some of the conformal blocks have a pole at the unitarity bound. This is not the case for the correlation function of identical scalars.\footnote{Except for the pole in the $\ell=0$ block.} Indeed the three-point function of two scalars and a TS operator with $\Delta=\ell+d-2$ is automatically conserved whenever the scalars have equal dimension. The minimal example then requires correlation functions of scalars with different dimensions. The prototypical example is then the Ising model in 3d, where one considers the mixed system of the fields $\sigma$ and $\epsilon$.  We will not describe the technical setup here, referring to~\cite{Kos:2014bka} for details.

Before continuing the discussion let us make a disclaimer: the mechanism presented in this work does not not affect in any way the precision measurements of the 3d Ising critical exponents of \cite{Kos:2016ysd,Simmons-Duffin:2016wlq}. At best it can help in shrinking the size of the allowed region.

By studying the correlation function of the $\sigma$ field only, one can obtain an upper bound $\Delta_\epsilon^{\text{max}}(\Delta_\sigma)$ on the dimension of the lowest $\Z_2$-even scalar appearing in the  OPE $\sigma\times\sigma\sim 1+ \epsilon+\ldots$. This bound has a nice kink coinciding with the expected values of $(\Delta_\sigma,\Delta_\epsilon)$ for the 3d Ising model, see for instance figure 3 of \cite{Kos:2014bka}.
\begin{figure}[t]
	\begin{center}
		\includegraphics[scale=0.6]{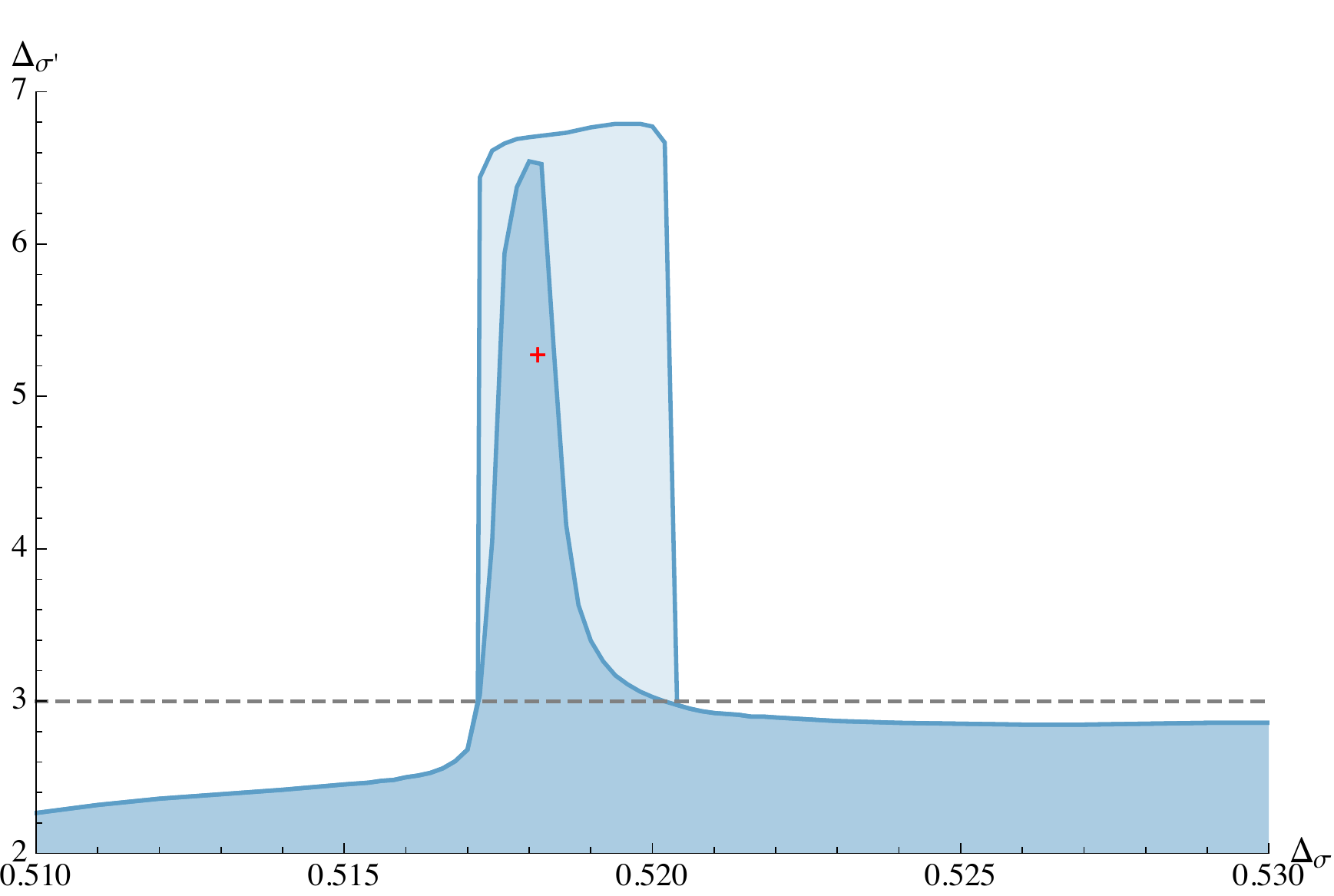}
		\caption{Upper bound on the dimension of the first parity odd scalar $\sigma'$ appearing in the OPE \eqref{eq:sigma-eps-OPE} assuming that $\Delta_\epsilon$ saturates the bound shown in figure 3 of  \cite{Kos:2014bka}. Light blue has not further assumptions. Dark blue assumes a gap in the $\Z_2$-odd spin-1 sector $\Delta^{\ell=1}_-\geq 2.3$. The red cross corresponds to the values $(\Delta_{\sigma},\Delta_{\sigma'})=(0.5181489, 5.2906)$ determined in \cite{Simmons-Duffin:2016wlq}.}
		\label{fig:isingsigmaprime}
	\end{center}
\end{figure}
Next, let us consider the OPE $\sigma\times\epsilon$; it contains $\Z_2$-odd operators of all spins, schematically:
\be\label{eq:sigma-eps-OPE}
\sigma \times \epsilon \sim \sigma +\sigma'+\ldots 
\ee
where the dots stand for higher dimensional scalars and higher spin operators. By considering the mixed system $\langle\sigma\sigma\sigma\sigma\rangle,\langle\sigma\epsilon\sigma\epsilon\rangle, \langle\epsilon\epsilon\epsilon\epsilon\rangle$ and assuming for instance $\Delta_\epsilon=\Delta_\epsilon^\text{max}(\Delta_\sigma)$, we can obtain an upper bound on the $\sigma'$ dimension. This was first done in \cite{Kos:2014bka}. However, since the conformal blocks of $\Z_2$-odd vectors are singular at the unitarity bound,  the residue mimics the contribution of a $\Z_2$-odd scalar of dimension 3. Without any further assumptions, the bound obtained on $\Delta_{\sigma'}$ is then a bound on the next $\Z_2$-odd scalar after 3. This effect can be straightforwardly eliminated by introducing a small gap in the spin-1 $\Z_2$-odd sector $\Delta^{\ell=1}_-\geq 2.3$.\footnote{We are grateful to Ning Su for making a preliminary plot and for checking that larger gaps $\Delta^{\ell=1}_-\geq 3, 4$ give similar bounds as $\Delta^{\ell=1}_-\geq 2.3$. In the 3d Ising model the first $\Z_2$-odd vector is expected to have dimension $\sim8$ \cite{Simmons-Duffin:2016wlq}.} The results with and without the gap are shown in figure \ref{fig:isingsigmaprime}. The two lines agree whenever the bounds are below 3 and differ substantially above. In particular, on the right side the jump disappears, while on the left side the bound still grows rapidly but it gets smoother. We expect that a similar phenomenon is responsible for the jumps present in figure 3 of \cite{Nakayama:2016jhq}.

\subsubsection{Majorana fermion bootstrap in 3d}
\label{sec:3dfermion}

Jumps similar to ours have been observed in 3d fermion bootstrap~\cite{Iliesiu:2015qra,Iliesiu:2017nrv}. For simplicity we will discuss~\cite{Iliesiu:2015qra}, although similar conclusions apply to~\cite{Iliesiu:2017nrv}. 

In their setup one studies a four-point function of a single Majorana fermion $\psi=\psi^\dagger$ operator. There is only one type of OPE $\psi\times\psi$, and the operators appearing in it are characterized by spin $\ell$ and $P$-parity. For even spin both $P$-even and $P$-odd operators can be exchanged, while for odd spin only $P$-odd operators are exchanged. This immediately implies that $P$-even $\ell>0$ three-point tensor structures are automatically conserved at the unitarity bounds: if they were not, then the action of the conservation operator would produce a valid $P$-even odd-spin tensor structure, which does not exist. However, the $P$-odd tensor structures with $\ell>0$ can potentially be not conserved at the unitarity bound. And indeed, an explicit calculation shows that the conservation equation is not satisfied, and we have a pole at the unitarity bound for all $P$-odd exchanges. Both $P$-even and $P$-odd scalar exchanges have the usual pole at the unitarity bound due to a violation of the Laplace equation. 

This means that the topology in the $P$-even sector is similar to the topology of neutral TS operators in our setup as shown in the left panel of figure~\ref{fig:SigmaTrueTopologyNeutral}, and the topology in the $P$-odd sector is similar to that of our TS operators in the charged sector, as shown in the left panel of figure~\ref{fig:SigmaTrueTopologyCharged}.

In~\cite{Iliesiu:2015qra,Iliesiu:2017nrv} jumps were observed in the upper bound on the dimension of the first $P$-odd scalar operator. We now recognize that these jumps are completely explained by fake $P$-odd scalar primaries at $\De=3$ coming from the unitarity bound pole of $P$-odd vector exchanges. We can furthermore predict the existence of such jumps in all $P$-odd bounds (assuming that these bounds are ever below the fake primary dimension). 
There is however one important difference between \cite{Iliesiu:2015qra} and \cite{Iliesiu:2017nrv}:  in the former a kink at the same value of $\De_\psi$ as the jump is observed in the upper bound on the leading $P$-even scalar. This kink does not have a straightforward explanation in terms of the topology of the blocks. Moreover the jump seems to happen before 3. Instead, in \cite{Iliesiu:2017nrv}  there are no kinks in the $P$-even sector and the jumps are exactly at 3. This suggests that with no global symmetry the situation is very much like the bound on $\sigma'$ in the Ising model: the bound on the $P$-odd scalar would rapidly grow above 3 for other reasons (real CFT?) and when it reaches 3 it jumps because of the fake primary effect.  It would be therefore interesting to redo the analysis of~\cite{Iliesiu:2015qra} with a small gap in $P$-odd vector sector.

\section{Numerical results}
\label{sec:numerical_results}

We now present various numerical bounds obtained by solving the optimization problems of section~\ref{sec:sdp}. We start by considering bounds on scaling dimensions of the first charged and neutral operators in sections~\ref{sec:resultscharged} and~\ref{sec:resultsneutral} respectively. 
We will use the following short-hand notation for their scaling dimensions
\begin{equation}
\Delta_-^{(\ell,\bar\ell)},\quad
\Delta_0^{(\ell,\bar\ell)}.
\end{equation}
In section~\ref{sec:CentralCharge} we show bounds on the central charges $C_J$ and $C_T$. Finally, in section~\ref{subsec:OPE_scalars} we address bounds on the product of OPE coefficients for the neutral and charged scalar operators. 

\begin{figure}[t]
	\begin{center}
		\includegraphics[scale=0.6]{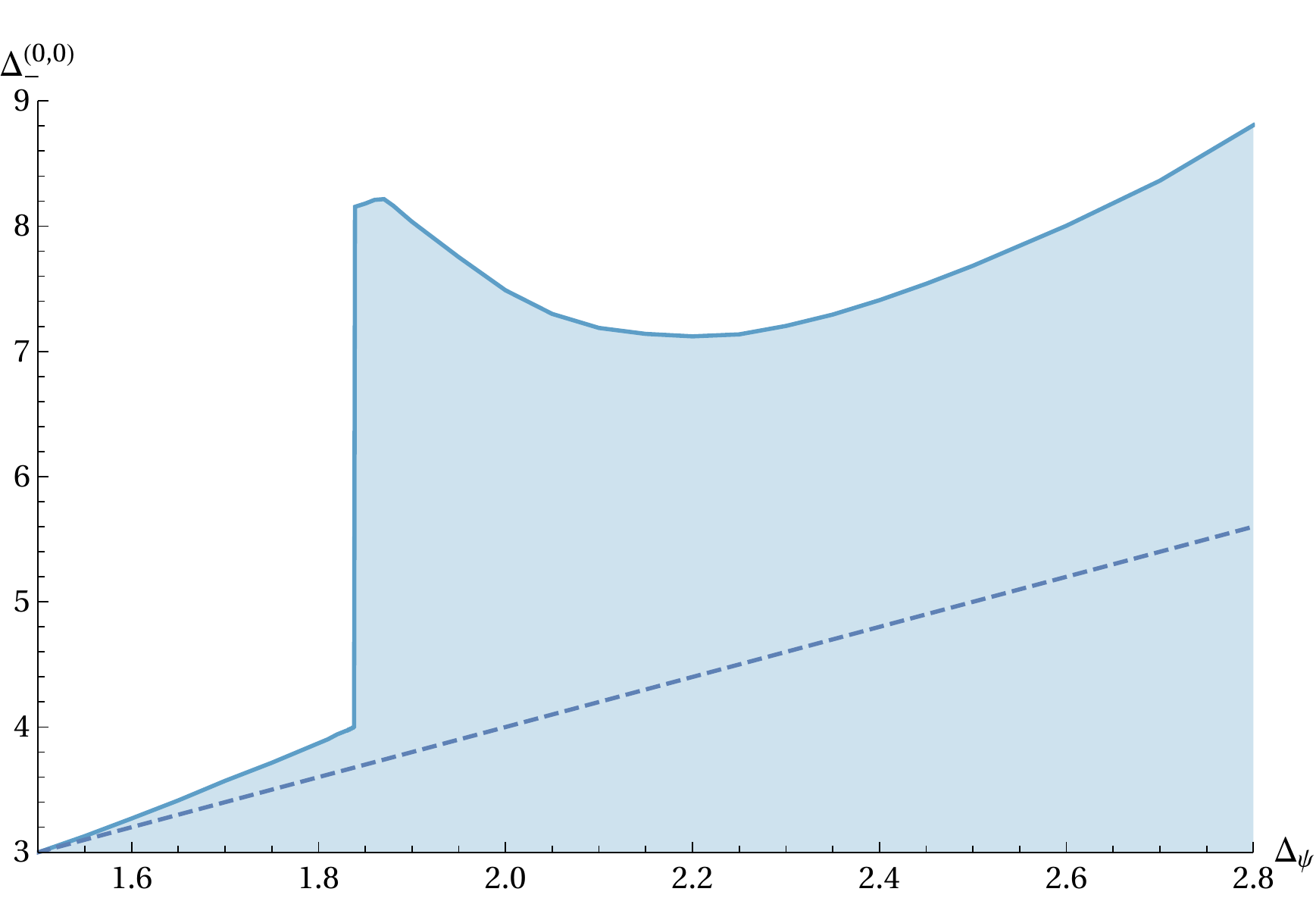}
		\caption{Upper bound on the dimension of the first charged scalar TS operator. The shaded region is the allowed one. The bound has been computed at $\Lambda=20$. The dashed line represents the GFT line. The bound has a discontinuity at $\De_\psi\approx1.84$.}
		\label{fig:charged00}
	\end{center}
\end{figure}

\subsection{Bounds on scaling dimensions: charged channel}
\label{sec:resultscharged}

In what follows we construct upper bounds on the scaling dimensions of the lowest dimensional charged operators (denoted ``lightest" for short in the following) as a function of $\De_\psi$. We will consider $(0,0)$, $(1,1)$, $(2,2)$ TS operators and $(1,3)$, $(3,1)$, $(3,5)$, $(5,3)$ NTS operators.

The fermion GFT defined in section~\ref{sec:GFT} gives an example of a consistent (non-local) CFT. Thus the operators in this theory should always lie in the allowed region of the bounds. According to section~\ref{sec:GFTcharged}, the lightest charged GFT operators have the following scaling dimensions: 
\begin{equation}
\label{eq:gft_first_charged_scalings}
\De^{(\ell,\ell)}_{-\;\text{GFT}} = 2\De_\psi+\ell,\quad
\De^{(\ell+2,\ell)}_{-\;\text{GFT}} = 2\De_\psi+\ell+2,\quad
\De^{(\ell,\ell+2)}_{-\;\text{GFT}} = 2\De_\psi+\ell.
\end{equation}
We depict their values by dashed lines on all the plots. One can try to remove the GFT in the attempt to make the bounds stronger and probe CFTs with operators lighter than the ones in~\eqref{eq:gft_first_charged_scalings}. This can be done for example by requiring the central charge $C_T$ to be finite when constructing the bounds. We found in practice that this requirement does not bring strong constraints unless $C_T$ is taken to be very small and starts violating bounds found later in section~\ref{sec:CentralCharge}. We will not therefore discuss such bounds in this work.

\begin{figure}[t]
	\begin{center}
		\includegraphics[scale=0.6]{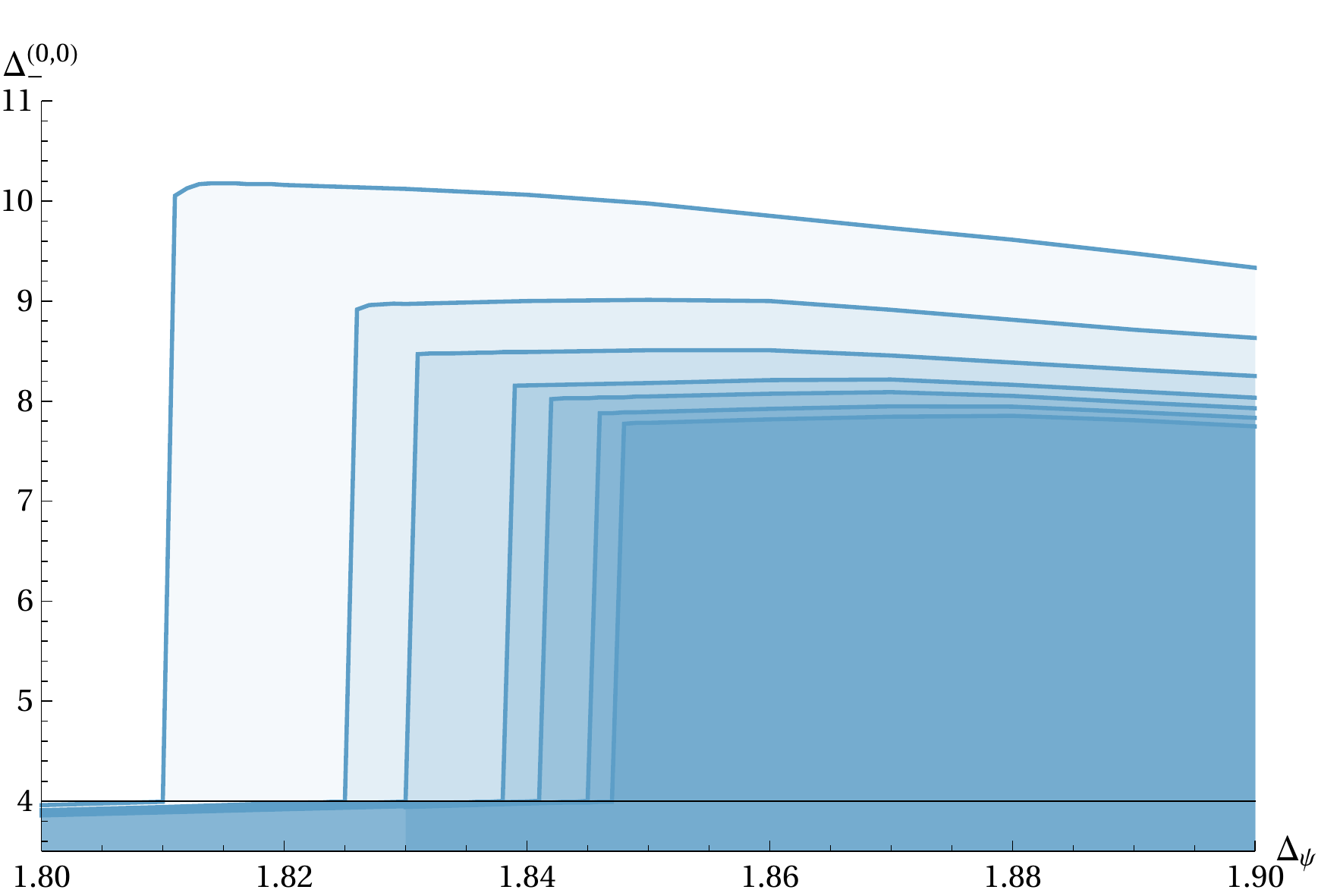}
		\caption{Upper bound on the dimension of the first charged scalar operator in the proximity of the jump as shown in \Figref{fig:charged00}. Different lines corresponds to increasing number of derivatives $\Lambda=14, 16,\ldots, 24, 26$.}
		\label{fig:extrapolationCharged00}
	\end{center}
\end{figure}

We start by presenting the bound on the scaling dimensions $\De^{(0,0)}_{-}$ of the lightest charged scalar as a function of $\De_\psi$ in figure~\ref{fig:charged00}. This plot displays a striking feature that is shared by many other plots presented in this work: the upper bound starts following the GFT line and  then, when it crosses the next integer, 4 in this case, it suddenly jumps to a much higher value.

\begin{figure}[t]
	\begin{center}
		\includegraphics[scale=0.8]{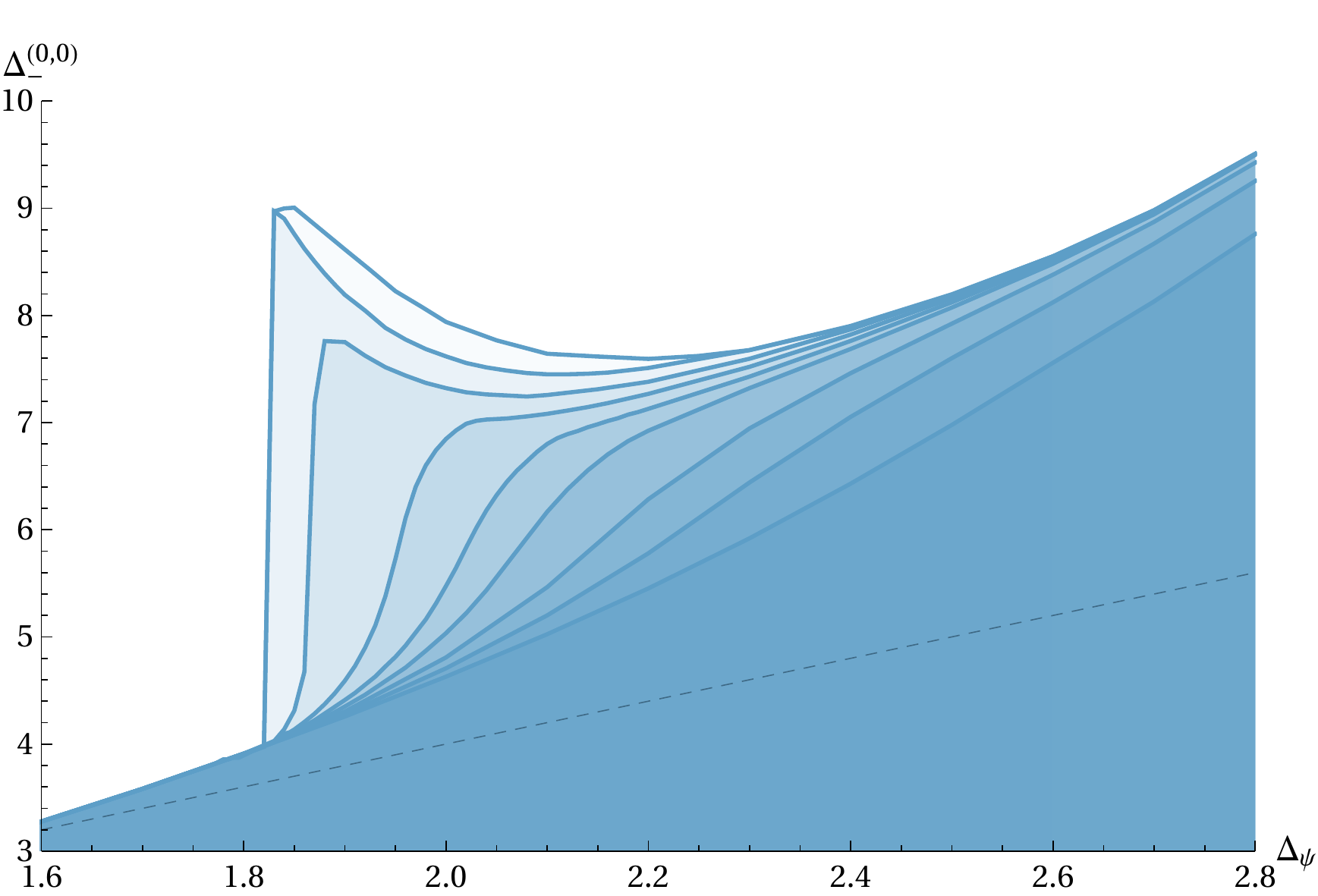}
		\caption{Upper bound on the first charged scalar assuming a gap on the first charged $(1,1)$ operator, namely $\De_-^{(1,1)}\geq 3+\text{gap}$. The values of the gaps 0, 0.1, 0.2, 0.3, 0.4, 0.5, 0.7, 1.0 and 2.0 correspond to the regions from lighter to darker colors respectively.  The bounds are computed at $\Lambda=16$.}
		\label{fig:removingJump}
	\end{center}
\end{figure}

Let us zoom in on the region of $\De_\psi$ where the jump appears and construct the bound for different values of the parameter $\Lambda$ defined in~\eqref{eq:Lambda}. The result is presented in figure~\ref{fig:extrapolationCharged00}.
We observe that the location of the jump in $\De_\psi$ keeps moving as the number of derivatives is increasing. This clearly demonstrates that the jump occurs when the bound crosses the integer value 4: as $\Lambda$ increases the bound gets stronger and the crossing point can only move to the right. Moreover from extrapolation to $\Lambda\rightarrow\infty$ it appears that the presence of jumps remains intact.

This is precisely the jump anticipated in section~\ref{sec:topology_blocks}. Let us reiterate the reasoning. Due to the non-trivial topology of TS blocks in the charged channel, as depicted in the left part of figure~\ref{fig:SigmaTrueTopologyCharged}, the $(1,1)$ block at the unitarity bound $\De=3$ fakes the presence of a scalar operator with the dimension $\De=4$. As long as the bound in figure~\ref{fig:charged00} remains below $\De=4$, it is a bound on the scaling dimension of the first charged scalar operator. However as soon as the bound crosses $\De=4$, we instead get a bound on the dimension of the second charged scalar operator given that the first operator has the scaling dimension 4.

One way to check this statement is to explicitly assume the existence of a scalar charged operator with $\De=4$ and to bound the second one. For $\De_\psi\lesssim 1.84$ no CFTs satisfying it exist due to the bound~\ref{fig:charged00}. For $\De_\psi\gtrsim 1.84$ however this assumption leads to exactly the same upper bound as in figure~\ref{fig:charged00}.

Now let us show how one can remove the jump. According to section~\ref{sec:topology_blocks}, one needs to impose a gap for the charged $(1,1)$ operators above the unitarity bound, namely  $\De_-^{(1,1)}\geq 3+\text{gap}$. The resulting bounds for different values of the gap are shown in figure~\ref{fig:removingJump}. We can observe how the jump transitions into a smooth curve for high enough values of the gap. A finite region of transition from the jump-like behavior into the smooth one is expected, since the vector blocks above the unitarity bound are not exactly equal to the scalar block at $\De=4$, they are still reasonably close to it if the gap is small enough.

Let us mention another interesting feature. In figure~\ref{fig:removingJump} the largest value of the gap is $2$. However, we have also computed the bound for the gap $3$. In the latter case the corresponding bound does not become stronger and coincides precisely with the former one. This can be explained once again by the topology of the charged blocks. Due to the $(2,2)$ TS charged block at the unitarity bound $\Delta=4$ we have always a fake $(1,1)$ charged operator with dimension $\Delta=5$. Thus a gap higher than $2$ is irrelevant since it becomes effectively the gap on the second $(1,1)$ charged operator and not on the first one. To get a stronger bound the gap value should be increased significantly.

\begin{figure}[t]
	\centering
	\subfigure[]{\label{fig:charged11}\includegraphics[scale=0.43]{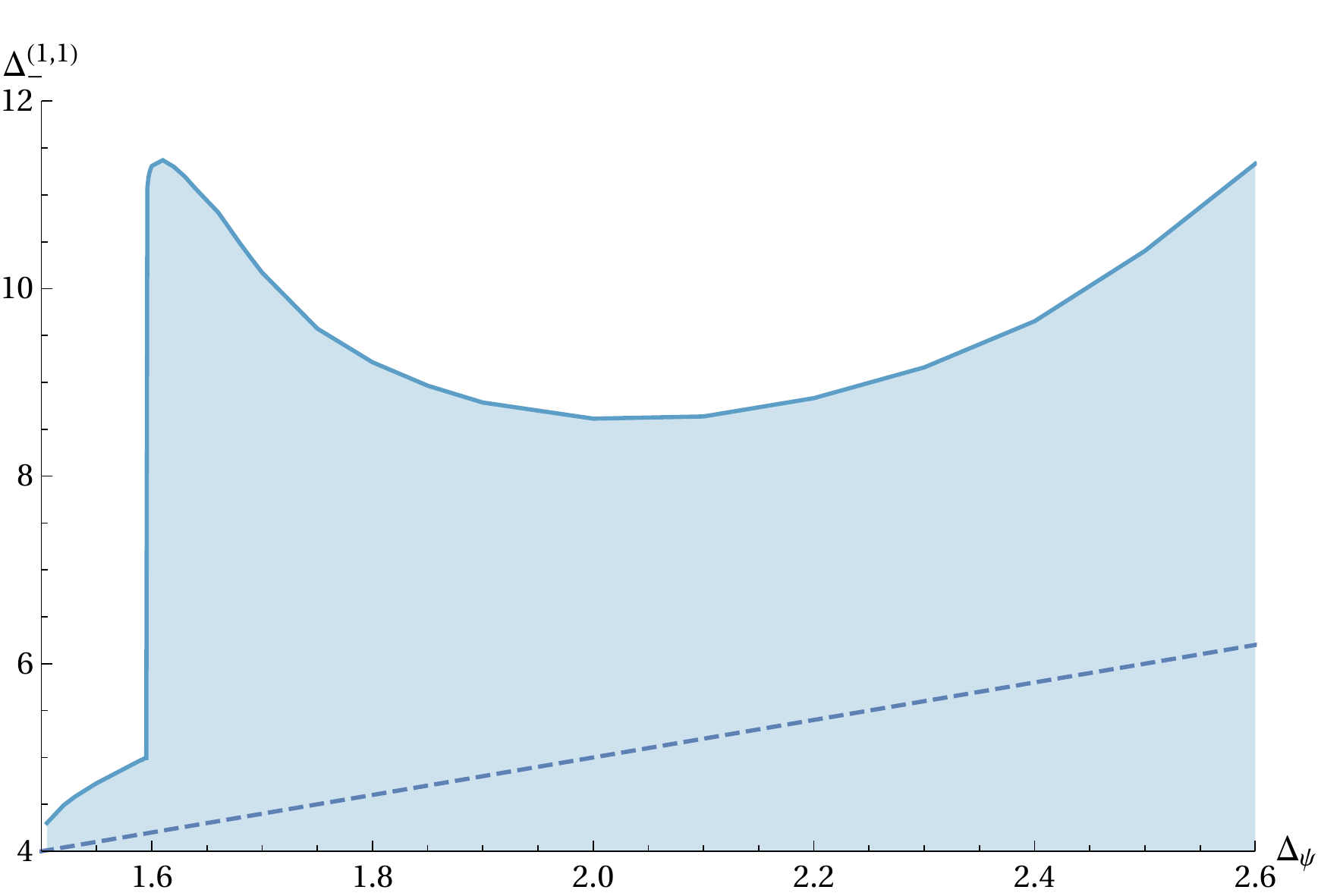}}
	\subfigure[]{\label{fig:charged22}\includegraphics[scale=0.43]{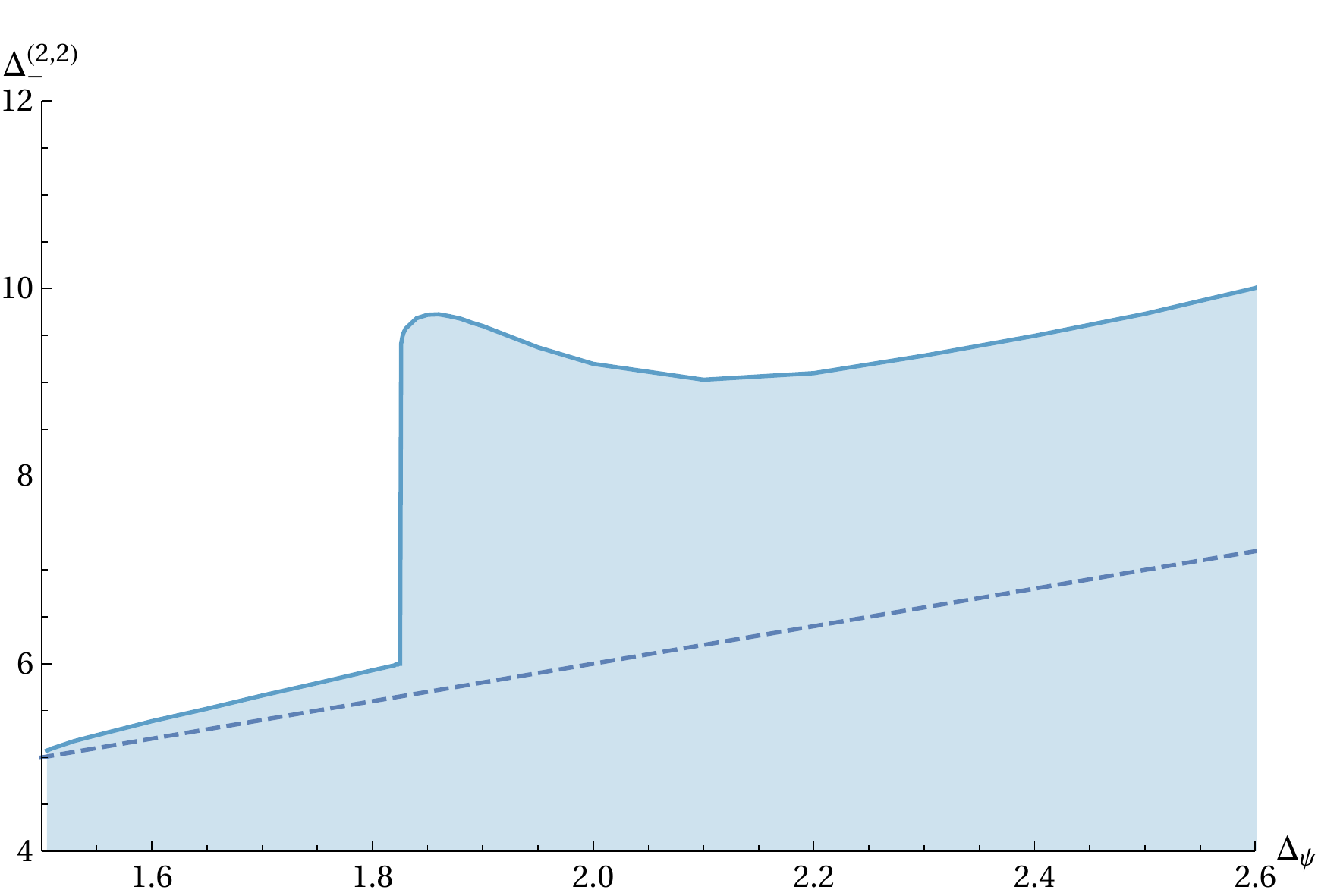}}
	\caption{Left: upper bound on the dimension of the first charged $(1,1)$ operator. Right: upper bound on the dimension of the first charged $(2,2)$ operator. The shaded region is the allowed one. The bounds have been computed at $\Lambda=16$. The dashed lines represent the GFT.}
	\label{fig:charged11-22}
\end{figure}

We now present the bounds on the first $(1,1)$ and $(2,2)$ charged TS operators as a function of $\De_\psi$ in figure~\ref{fig:charged11-22}.
As in the scalar case, jumps occur when the bound hits an integer value 
\be
\Delta_{-\;\text{jump}}^{(\ell,\ell)} = 4+\ell.
\label{eq:Deltajump}
\ee
We have checked explicitly the validity of~\eqref{eq:Deltajump} up to $\ell=5$. As expected this is in precise agreement with the discussion of section~\ref{sec:topology_blocks}.

\begin{figure}[t]
\begin{center}
\subfigure[]{\label{fig:charged13}\includegraphics[scale=0.43]{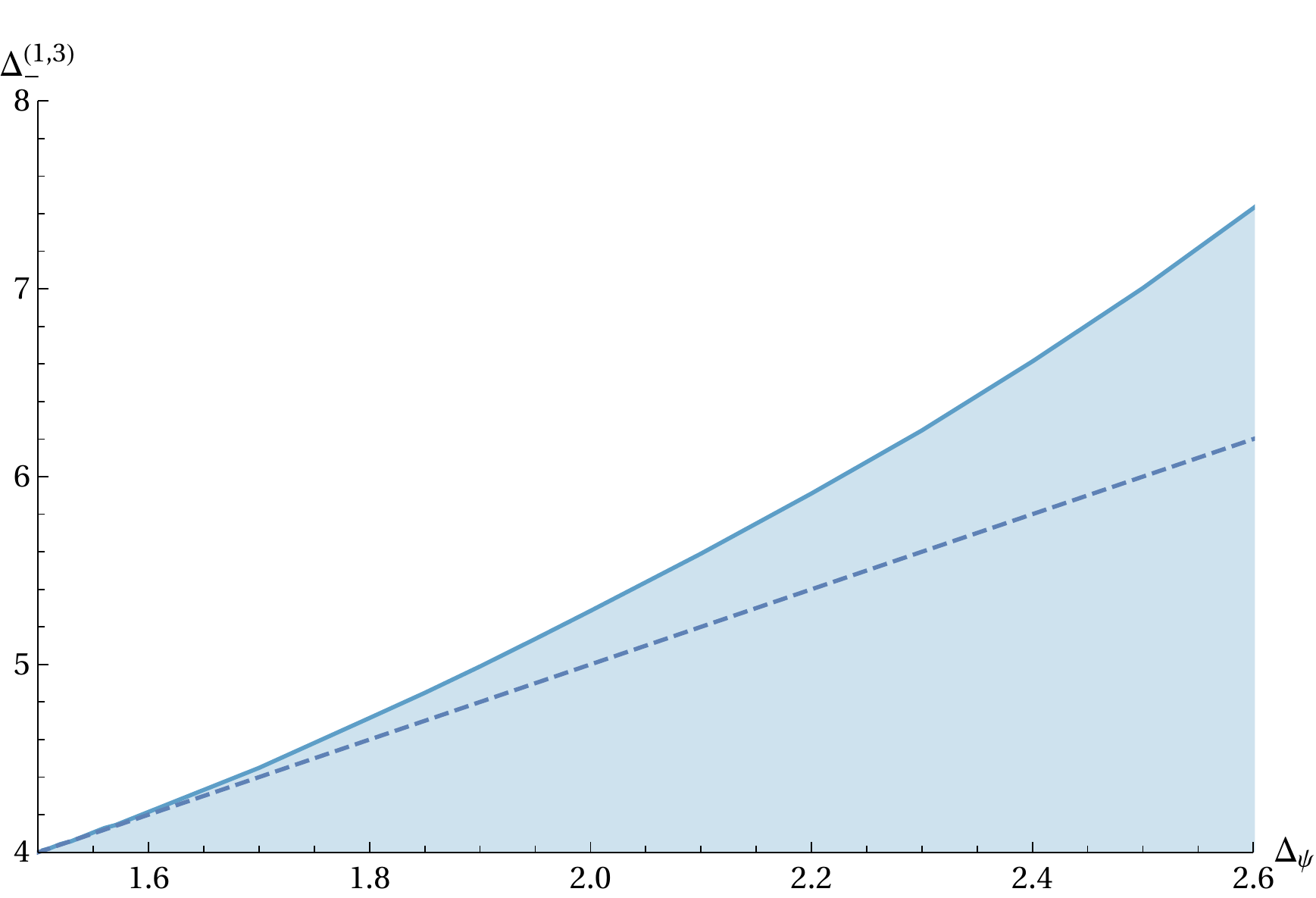}}
\subfigure[]{\label{fig:charged31}\includegraphics[scale=0.43]{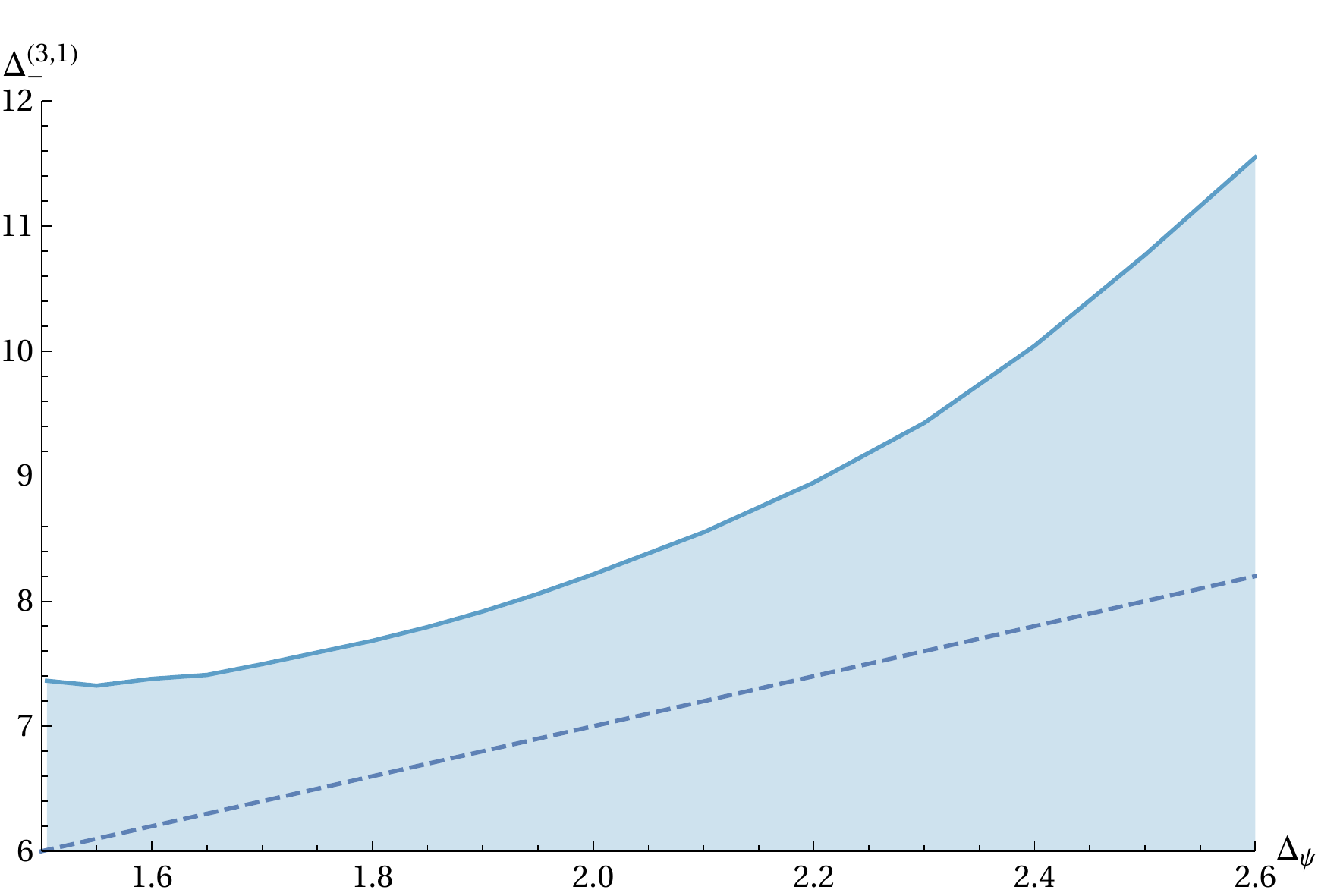}}
\caption{Left: upper bound on the first charged $(1,3)$ NTS operator. Right: upper bound on the first $(3,1)$ NTS operator. The shaded region is the allowed one. The bounds have been computed at $\Lambda=16$. The dashed lines represent the GFT. }
\label{fig:charged13-31}
\end{center}
\end{figure}

\begin{figure}[t]
\begin{center}
\subfigure[]{\label{fig:charged35}\includegraphics[scale=0.43]{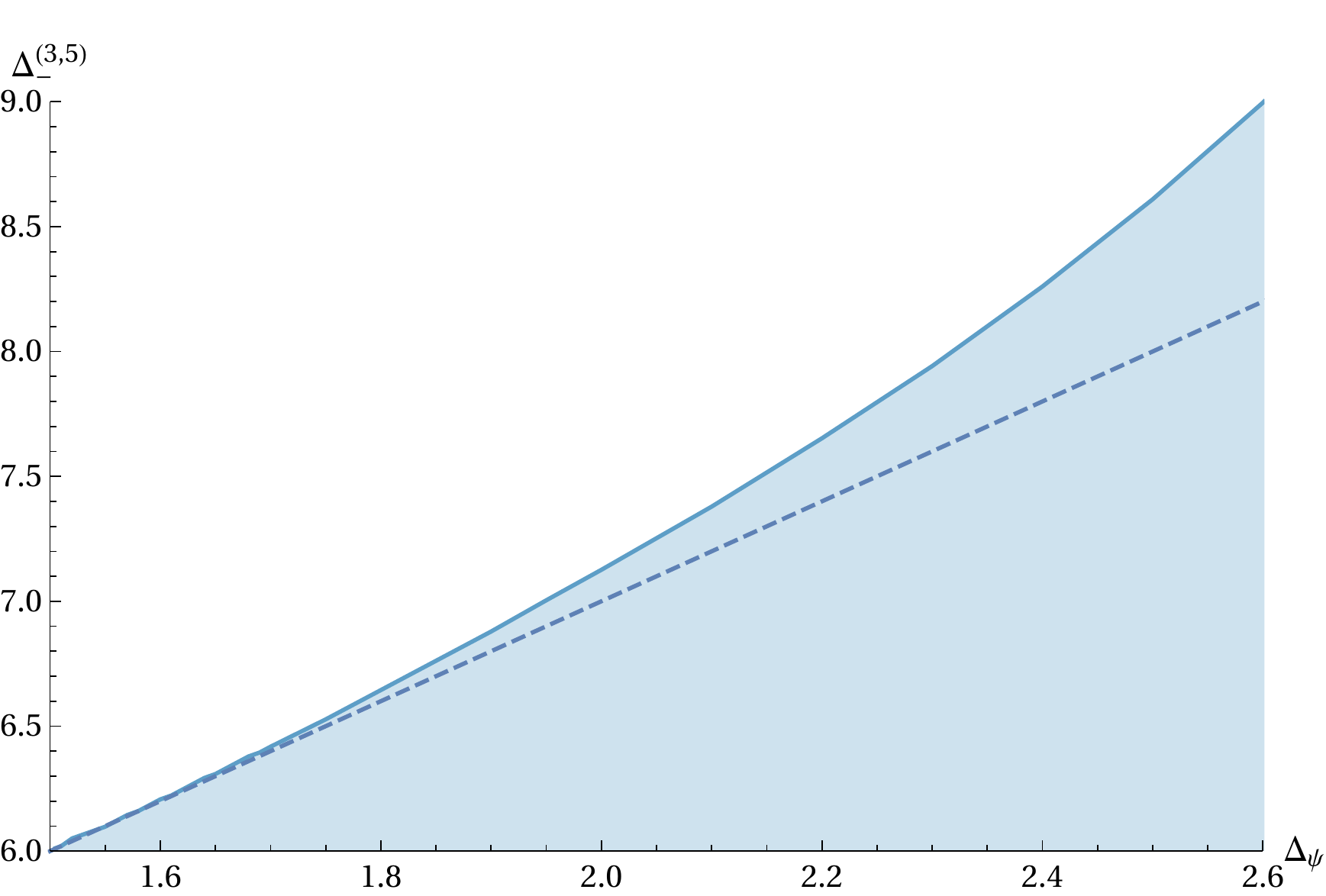}}
\subfigure[]{\label{fig:charged53}\includegraphics[scale=0.43]{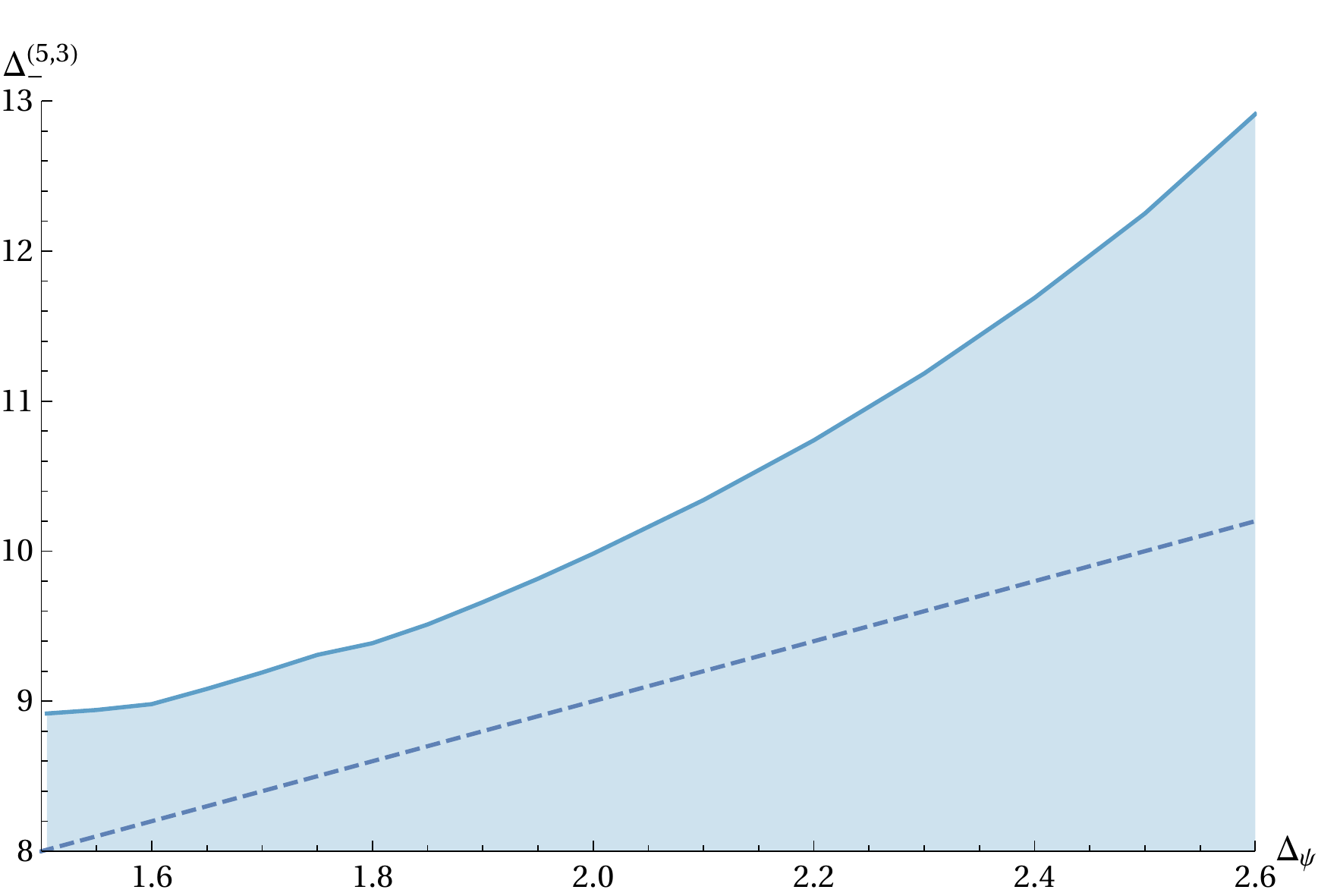}}
\caption{Left: upper bound on the first  charged $(3,5)$ NTS operator. Right: upper bound on the the first $(5,3)$ NTS operator. The shaded region is the allowed one. The bounds have been computed at $\Lambda=16$. The dashed lines represent the GFT. }
\label{fig:charged35-53}
\end{center}
\end{figure}

Finally we present bounds on charged NTS operators as a functions of $\De_\psi$ in figures~\ref{fig:charged13-31} and~\ref{fig:charged35-53}. We stress that this is the first time one is able to get upper bounds on operators that are non-traceless symmetric tensors, although we knew already by analytic bootstrap techniques that at least at large $\ell$ these operators must exist, and their spectrum should  approach the GFT spectrum obtained in \sref{sec:GFT}. This was concretely shown for instance in \cite{Elkhidir:2017iov}.

We remind that in the charged sector, the operators $(\ell,\ell+2)$ and $(\ell+2,\ell)$ are independent of each other and must be treated separately. The spin $\ell$ is constrained to be odd due to presence of identical fermions in the setup.

Note the absence of jumps compared to the charged TS plots, in agreement with the topology of conformal blocks depicted in the right panel of figure~\ref{fig:SigmaTrueTopologyCharged}. Note also
that the bound $(\ell,\ell+2)$ follows closely the GFT line for small value of $\De_\psi$ whereas the bound for $(\ell+2,\ell)$ always stays significantly above it.
An asymmetry between the bounds on $(\ell,\ell+2)$ and $(\ell+2,\ell)$  operators is expected, because the former do and the latter do not exist in the free theory.

\begin{figure}[t]
	\begin{center}
		\includegraphics[scale=0.6]{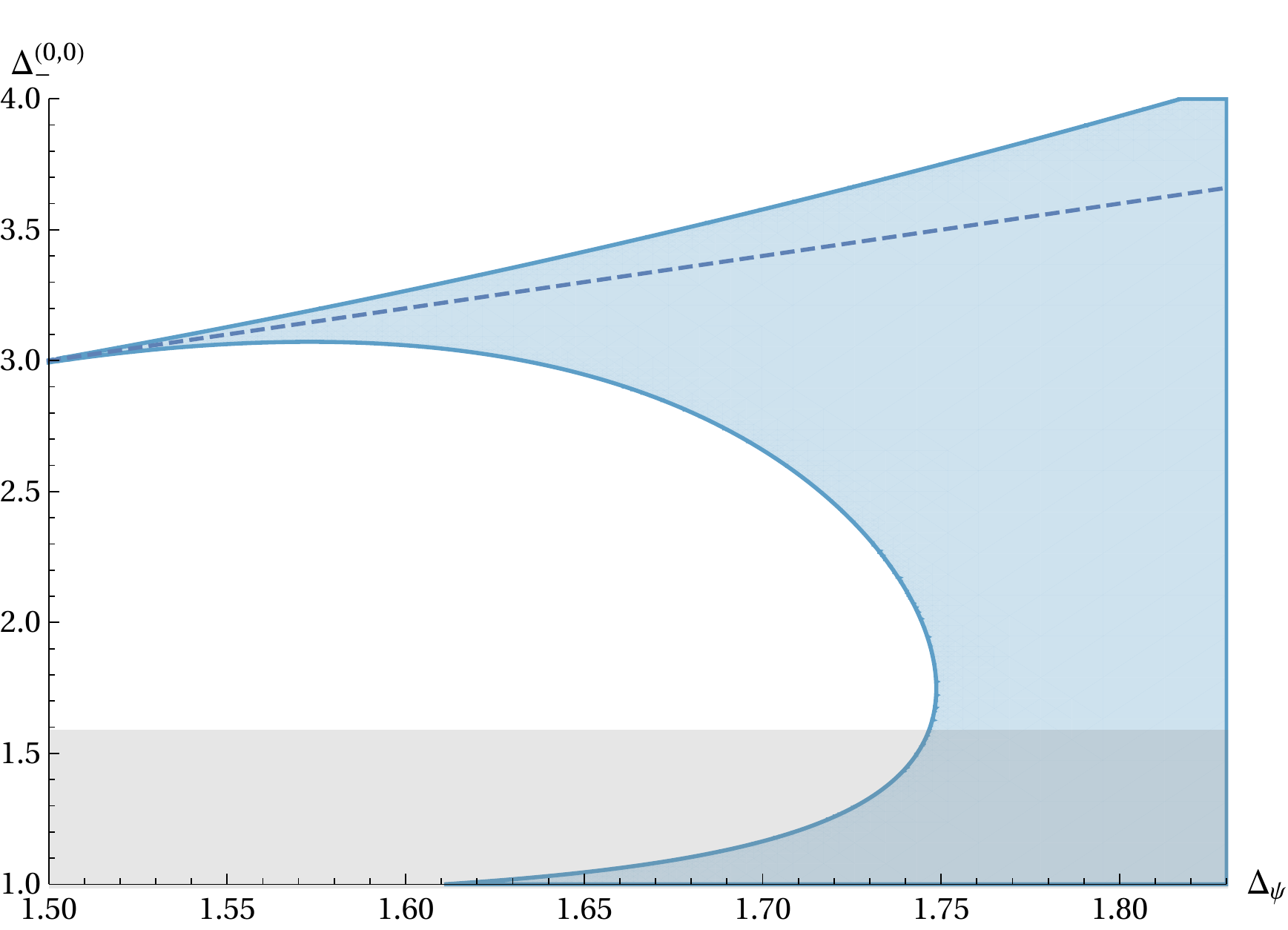}
		\caption{Allowed values of the first scalar charged operator assuming that the second charged scalar is irrelevant. The plot has been done at $\Lambda=16$.}
		\label{fig:peninsula}
	\end{center}
\end{figure}

We conclude the discussion by considering once again the charged scalar $(0,0)$ operator.
According to figure~\ref{fig:charged00}, for $\Delta_\psi\lesssim1.84$ any consistent CFT must contain at least one light relevant ($\De<4$) charged scalar. We can further assume that there is only a single relevant charged operator and all the others are irrelevant ($\De>4$). By imposing this requirement we can construct both an upper and a lower bound on the lightest scalar. The result is presented in figure~\ref{fig:peninsula}. The assumption carves out most of the region leaving only a narrow peninsula surrounding the fermion GFT line. The plot can be compared with figure 6 in~\cite{Iliesiu:2015qra}. Contrary to their case we do not observe any features which might correspond to interesting physical theories.

If one supplements the assumption of a single charged scalar with the complete absence of neutral relevant scalars, figure~\ref{fig:peninsula} is marginally modified: the only effect is to move slightly the lower branch of the allowed region. We do not show this plot here since the region affected by the modification turns out to be unphysical. The reason is that a CFT with a charged scalar must also contain a neutral scalar as dictated by the bootstrap bounds obtained for instance in \cite{Vichi:2011ux,Poland:2011ey}. As it turns out, the absence of neutral relevant scalars is inconsistent with the presence of charged ones below $\De_-^{(0,0)}\lesssim 1.59$. We show this excluded region with a light shading in the plot. Unfortunately the fermion crossing equations alone do not enforce this constraint; on the other hand it would manifest itself in a mixed correlator analysis involving a charged scalar and the fermion.

\begin{figure}[t]
\begin{center}
\includegraphics[scale=0.6]{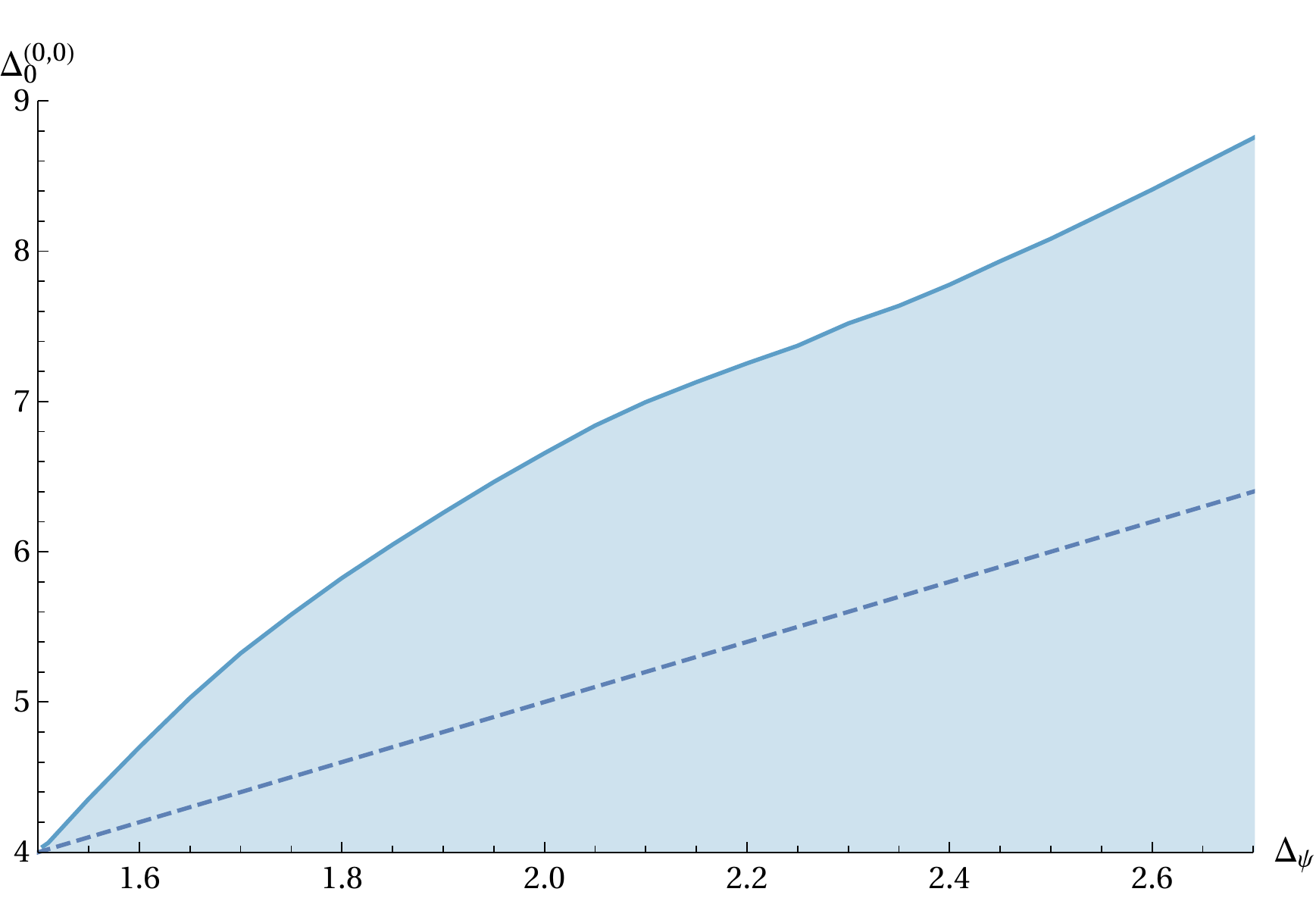}
\caption{Upper bound on the first neutral scalar operator. The shaded region is the allowed one. The bounds has been computed at $\Lambda=20$. The dashed line represents the GFT.}
\label{fig:neutral00}
\end{center}
\end{figure}

\subsection{Bounds on scaling dimensions: neutral channel}
\label{sec:resultsneutral}

We now present our results for the neutral channel. We remind that this channel contains $(1,1)$ conserved current $J$ with $\De=3$ and $(2,2)$ stress tensor $T$ with $\De=4$. In what follows we show bounds on the first neutral $(0,0)$ scalar, second $(1,1)$ operator (after $J$), second $(2,2)$ operator (after $T$), first $(3,3)$ and $(4,4)$ operators. We then show the bounds on the NTS $(0,2)$, $(1,3)$ and $(2,4)$ operators. We remind that the dual operators $(2,0)$, $(3,1)$ and $(4,2)$ are related by hermitian conjugation. As a consequence $(\ell,\ell+2)$ and $(\ell+2,\ell)$ operators enter in the same conformal block~\eqref{eq:1324-symmetric_block} and thus have an identical bound.

As in the charged case, we recall that the values of scaling dimensions of the lightest GFT operators, according to section~\ref{sec:GFTneutral}, are
\begin{equation}
\label{eq:gft_first_neutral_scalings}
\De^{(0,0)}_{0\;\text{GFT}} = 2\De_\psi+1,\quad
\De^{(\ell,\ell)}_{0\;\text{GFT}} = 2\De_\psi+\ell-1,\quad
\De^{(\ell+2,\ell)}_{0\;\text{GFT}} =  2\De_\psi+\ell+1.
\end{equation}
In the second entry $\ell\geq 1$. As before we depict~\eqref{eq:gft_first_neutral_scalings} by dashed lines on all the plots below.

\begin{figure}[t]
\begin{center}
\subfigure[]{\label{fig:neutral11}\includegraphics[scale=0.43]{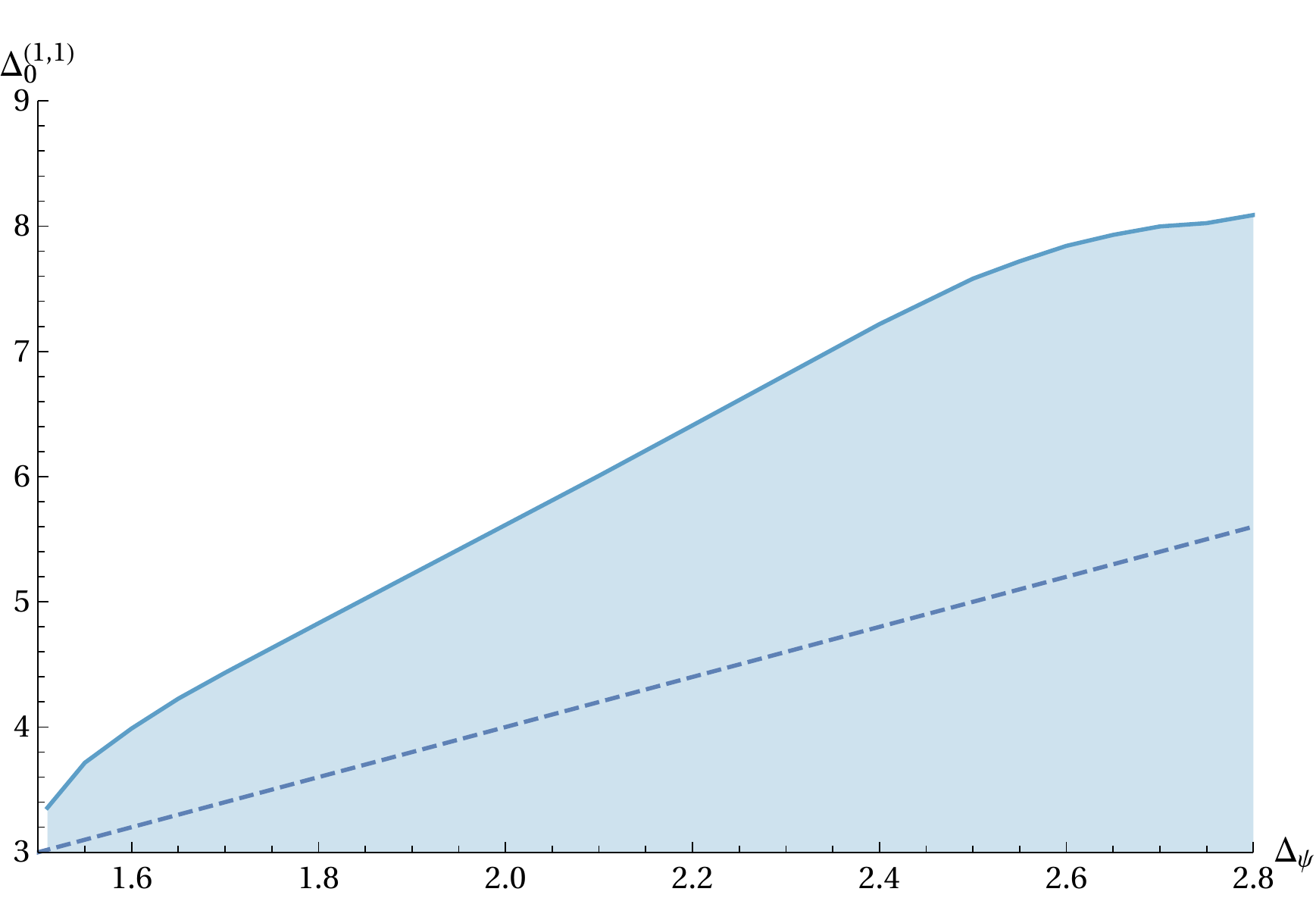}}
\subfigure[]{\label{fig:neutral22}\includegraphics[scale=0.43]{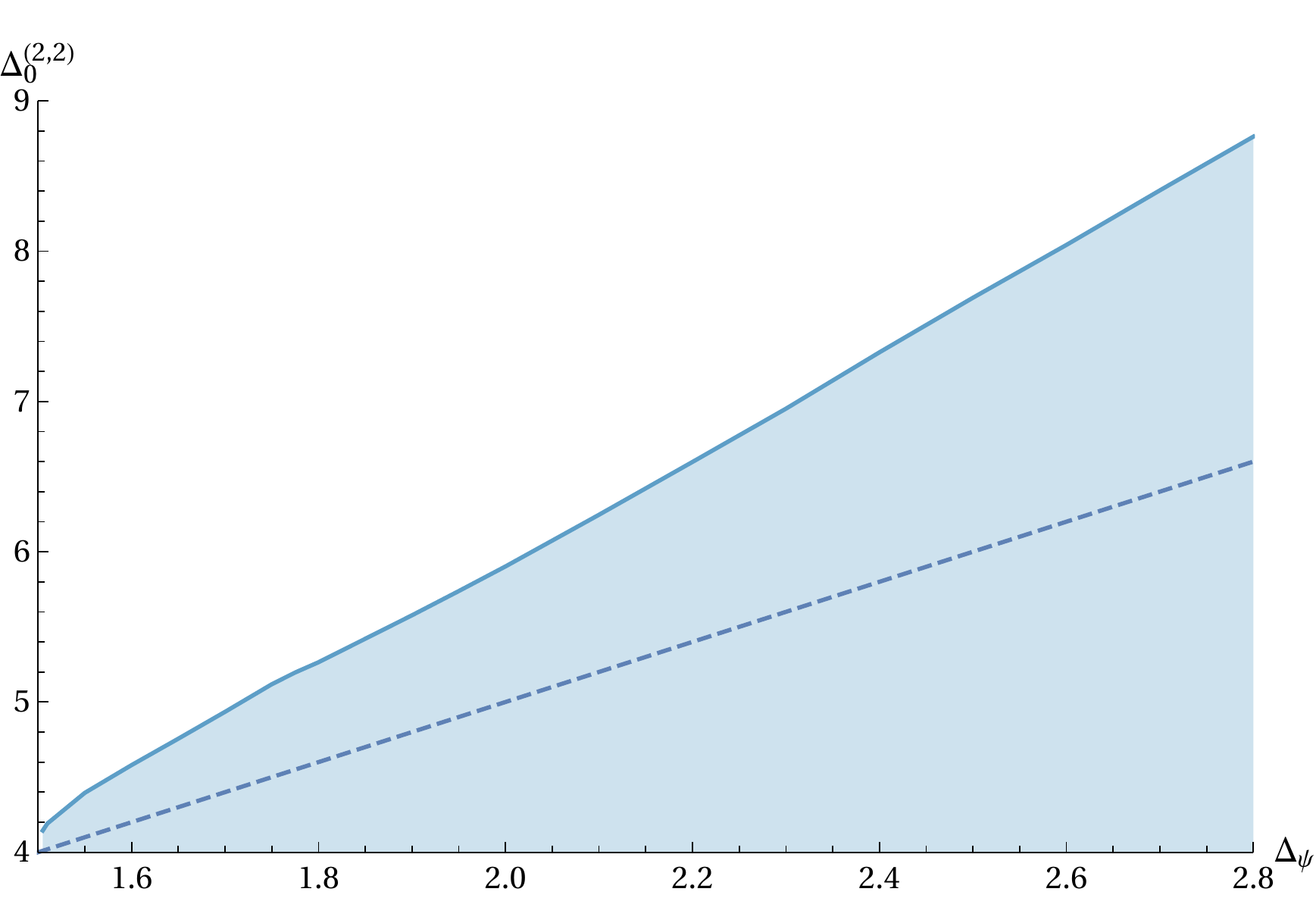}}
\caption{Left: upper bound on the second $(1,1)$ TS operator appearing after the conserved current $J$. Right: upper bound on the second $(2,2)$ TS operator  appearing after the conserved stress tensor  $T$. The bounds have been computed at $\Lambda=16$. The dashed lines represent the GFT lines.}
\label{fig:neutral11-22}
\end{center}
\end{figure}

\begin{figure}[t]
\begin{center}
\subfigure[]{\label{fig:neutral33}\includegraphics[scale=0.43]{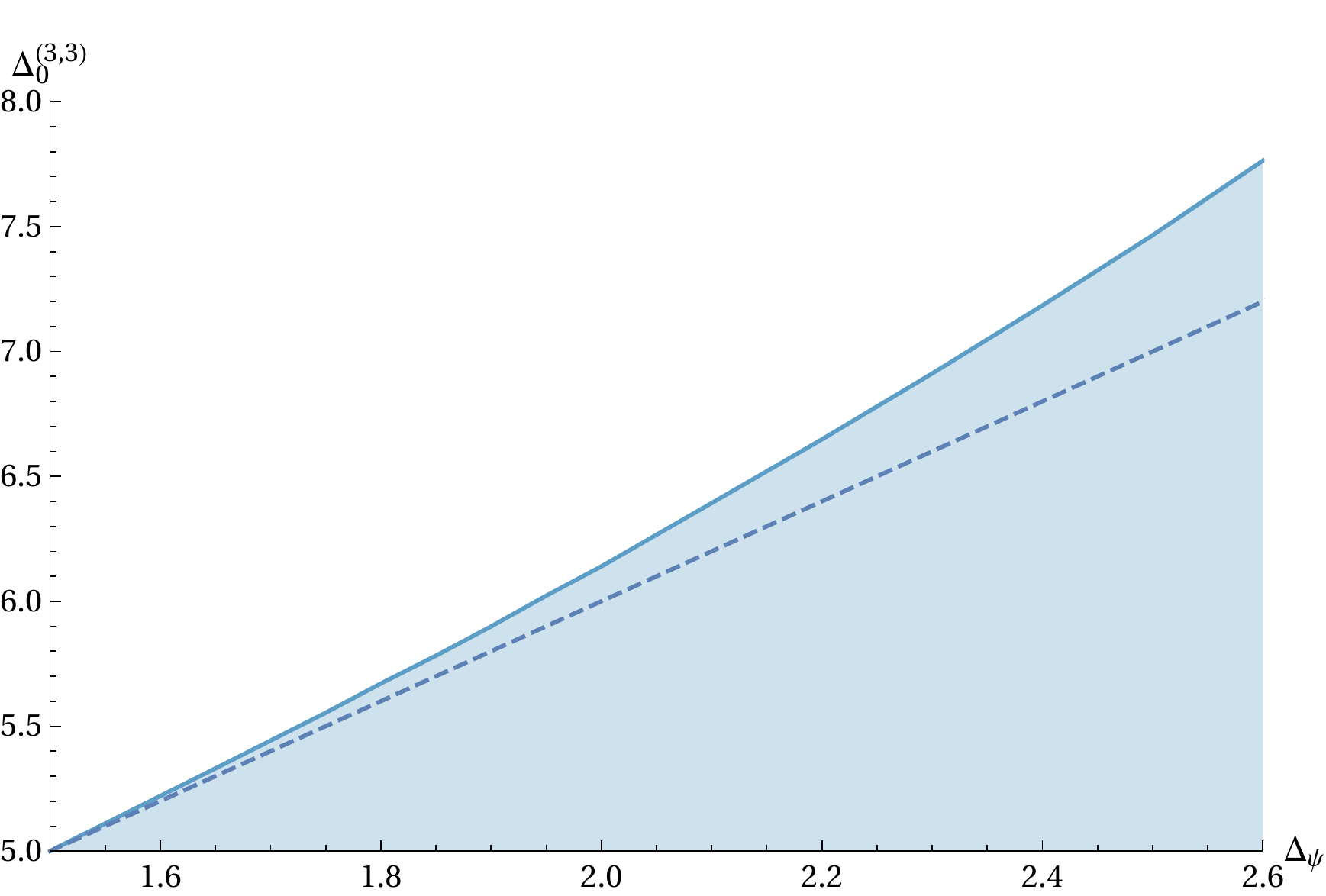}}
\subfigure[]{\label{fig:neutral44}\includegraphics[scale=0.43]{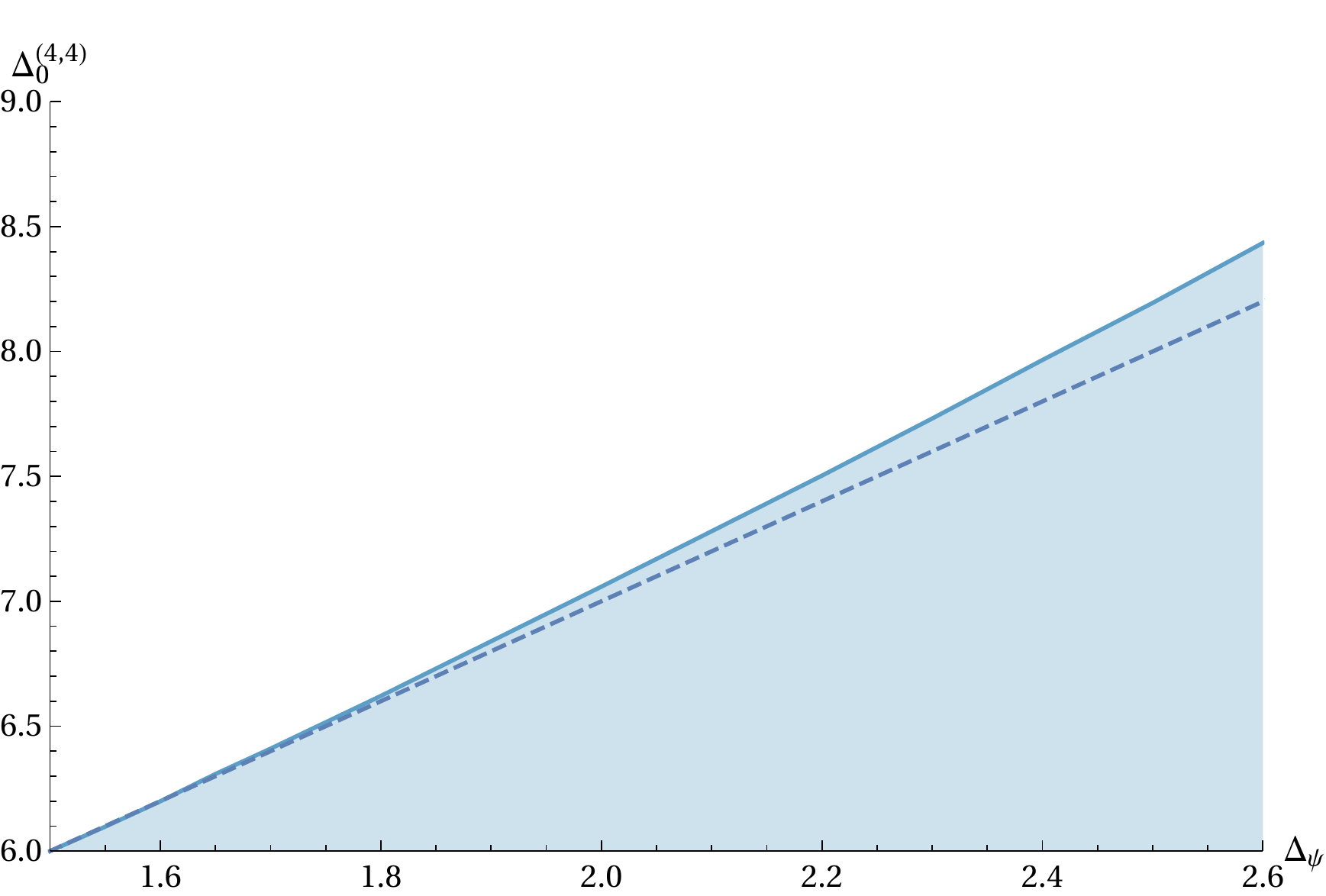}}
\caption{Left: upper bound on the dimension of the neutral $(3,3)$ TS operator. Right: upper bound on the first neutral $(4,4)$ TS operator. The shaded region is the allowed one. The bounds have been computed at $\Lambda=16$. The dashed lines represent the GFT lines. }
\label{fig:neutral33-44}
\end{center}
\end{figure}

We start by considering the bound on the first scalar given in figure~\ref{fig:neutral00}. This bound is usually the principal object of bootstrap investigations since it defines the conditions under which a CFT  allows the absence of relevant perturbations. Unfortunately in our case the bound appears to be very weak, and by construction must allow the GFT solution, which never contains relevant neutral operators. Thus, without  further assumptions the bootstrap does not give any constraint on the stability or naturalness of CFTs containing fermions.

\begin{figure}[t]
	\begin{center}
		\includegraphics[scale=0.6]{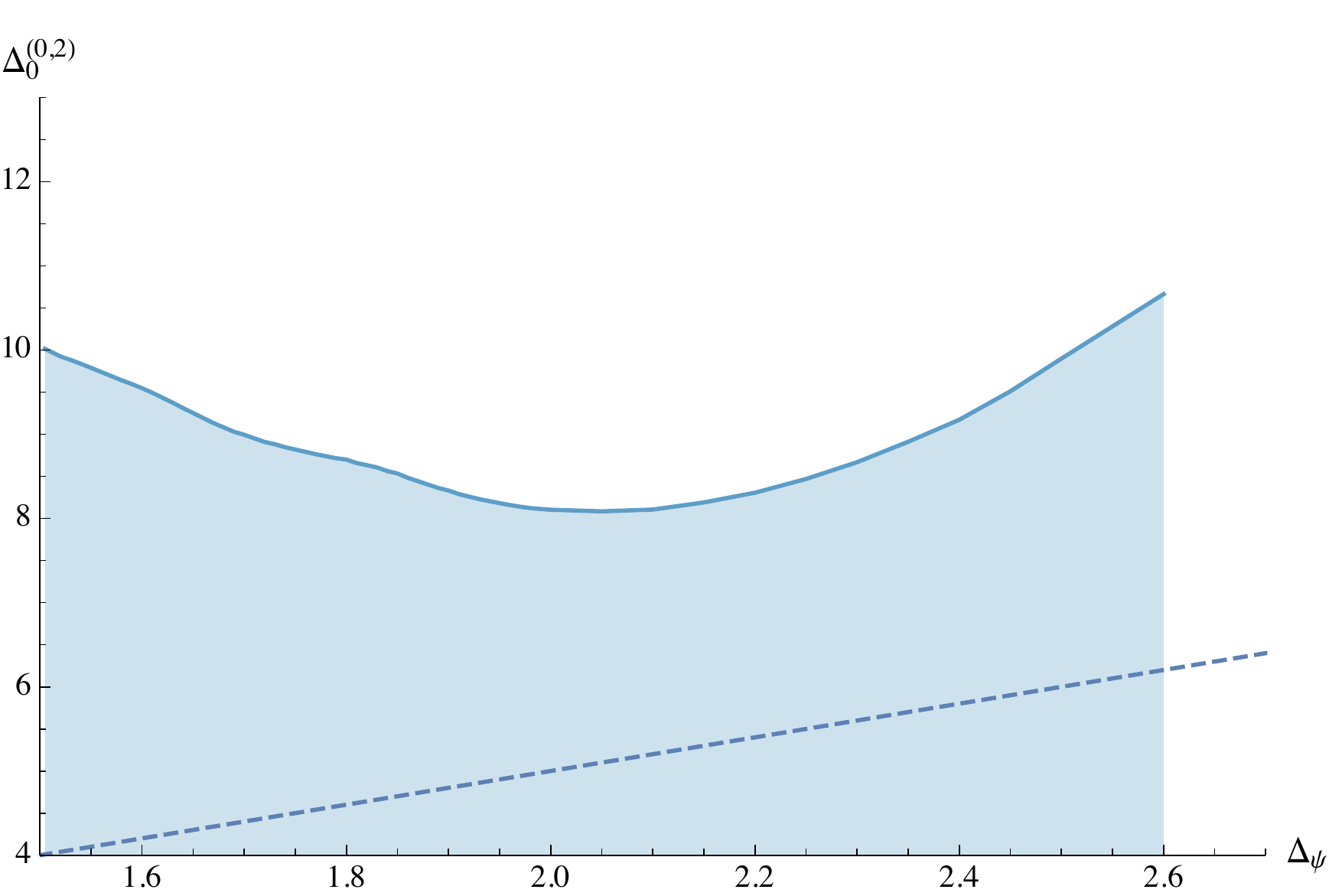}
		\caption{Upper bound on the dimension of the first neutral $(0,2)$ NTS operator. The shaded region is the allowed one. The bounds has been computed at $\Lambda=20$. The dashed line represents the GFT.}
		\label{fig:neutral02}
	\end{center}
\end{figure}

\begin{figure}[t]
\begin{center}
\subfigure[]{\label{fig:neutral13}\includegraphics[scale=0.43]{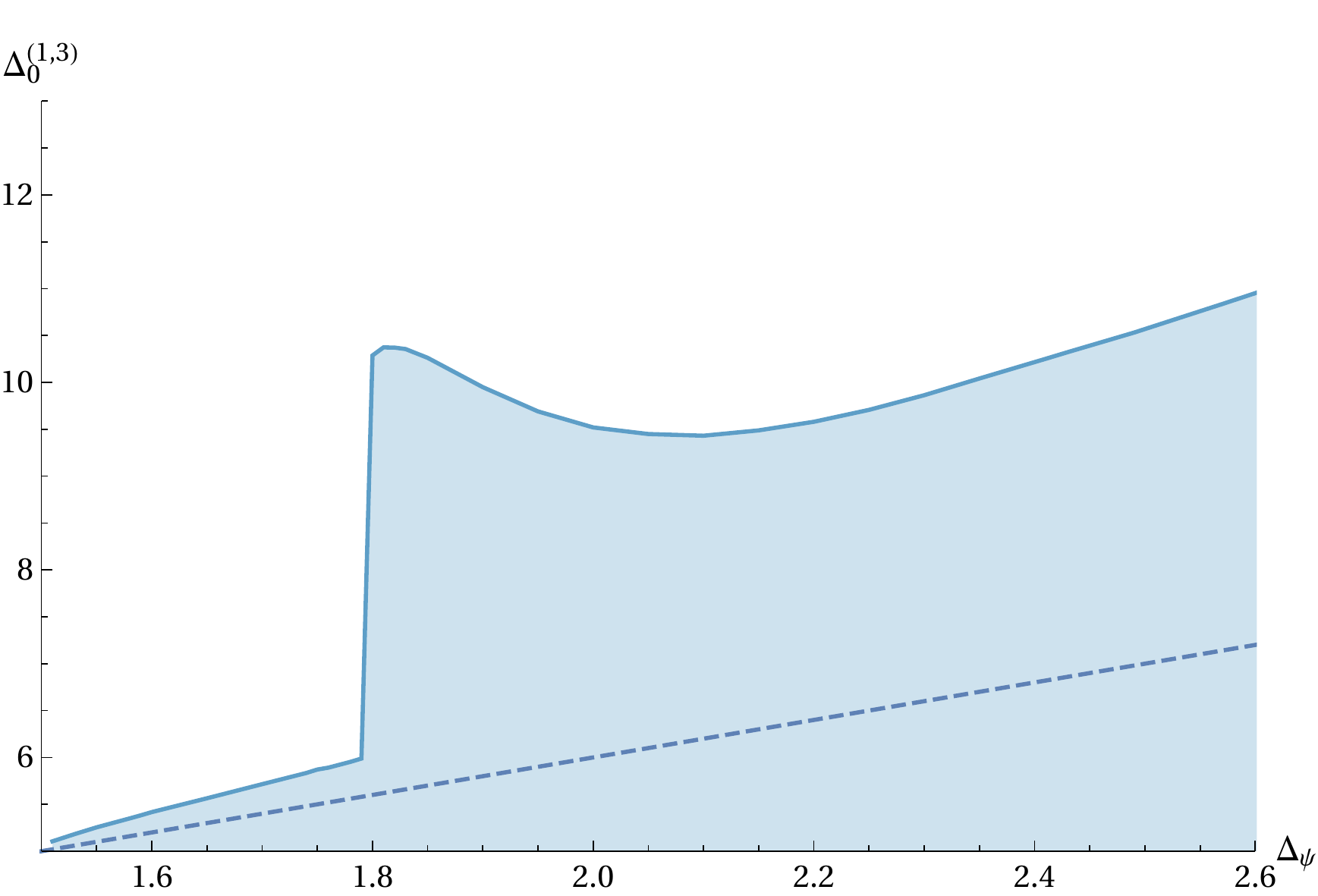}}
\subfigure[]{\label{fig:neutral24}\includegraphics[scale=0.43]{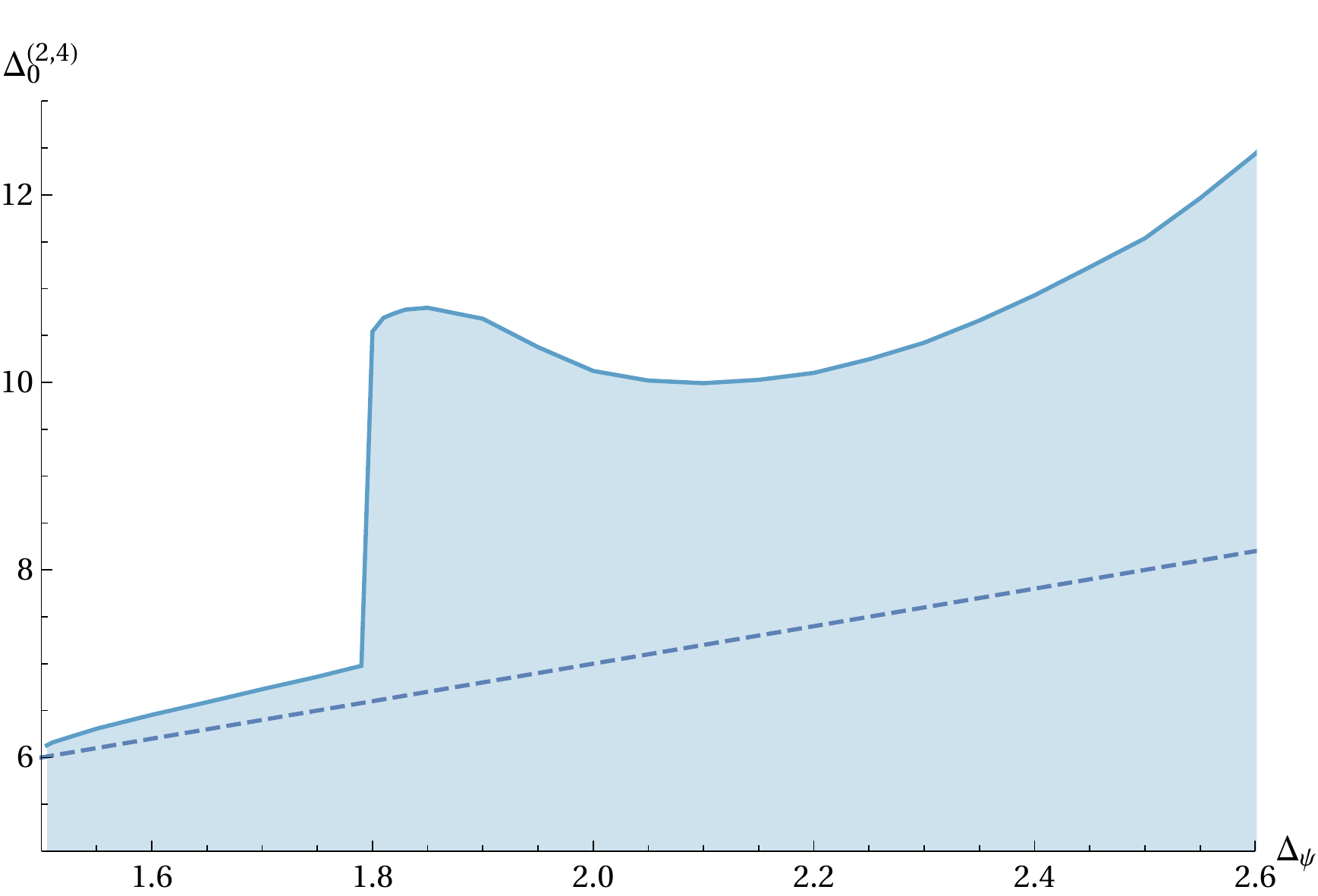}}
\caption{Left: upper bound on the dimension of the first $(1,3)$ neutral NTS operator. Right: upper bound on the first $(2,4)$ neutral NTS operator. The shaded region is the allowed one. The bounds have been computed at $\Lambda=16$. The dashed lines represent the GFT.}
\label{fig:neutral13-24}
\end{center}
\end{figure}

Since we are mostly interested in local CFTs, we assume the presence of the conserved current $J$ and the stress tensor $T$. As a consequence we have not explored the bounds on the very first $(1,1)$ and $(2,2)$ TS operators. Instead we look for the bounds on the second $(1,1)$ and second $(2,2)$  operators. The results are shown in figure~\ref{fig:neutral11-22}. 

The bounds on the first $(3,3)$ and $(4,4)$ operators are given in figure~\ref{fig:neutral33-44}. They have a similar structure: the bound is initially saturated by the GFT line and then eventually smoothly departs from it as $\Delta_\psi$ increases. The larger $\ell$ the closer it stays to the GFT line. This pattern might be related to the results of~\cite{Fitzpatrick:2012yx,Komargodski:2012ek} where GFT operators have been shown to be accumulation points for higher spin operators in CFTs.

We conclude by addressing the $(0,2)$, $(1,3)$ and $(2,4)$ NTS operators. Their bounds are presented in figures~\ref{fig:neutral02}, \ref{fig:neutral13} and~\ref{fig:neutral24} respectively. According to the discussion of section~\ref{sec:topology_blocks} we expect to observe jumps here, similar to the ones in the charged TS sector, at the position
\be
\Delta_{0\;\text{jump}}^{(\ell,\bar{\ell})} =  4 + \frac{\ell+\bar{\ell}}2 = 5+\ell.
\label{eq;DeltaJump}
\ee
For $\ell=1$ and $\ell=2$ this is indeed the case as can be seen from figure~\ref{fig:neutral13-24}. We have also explicitly checked that this is true for $\ell\leq 4$.

\begin{figure}[t]
	\begin{center}
		\includegraphics[scale=0.6]{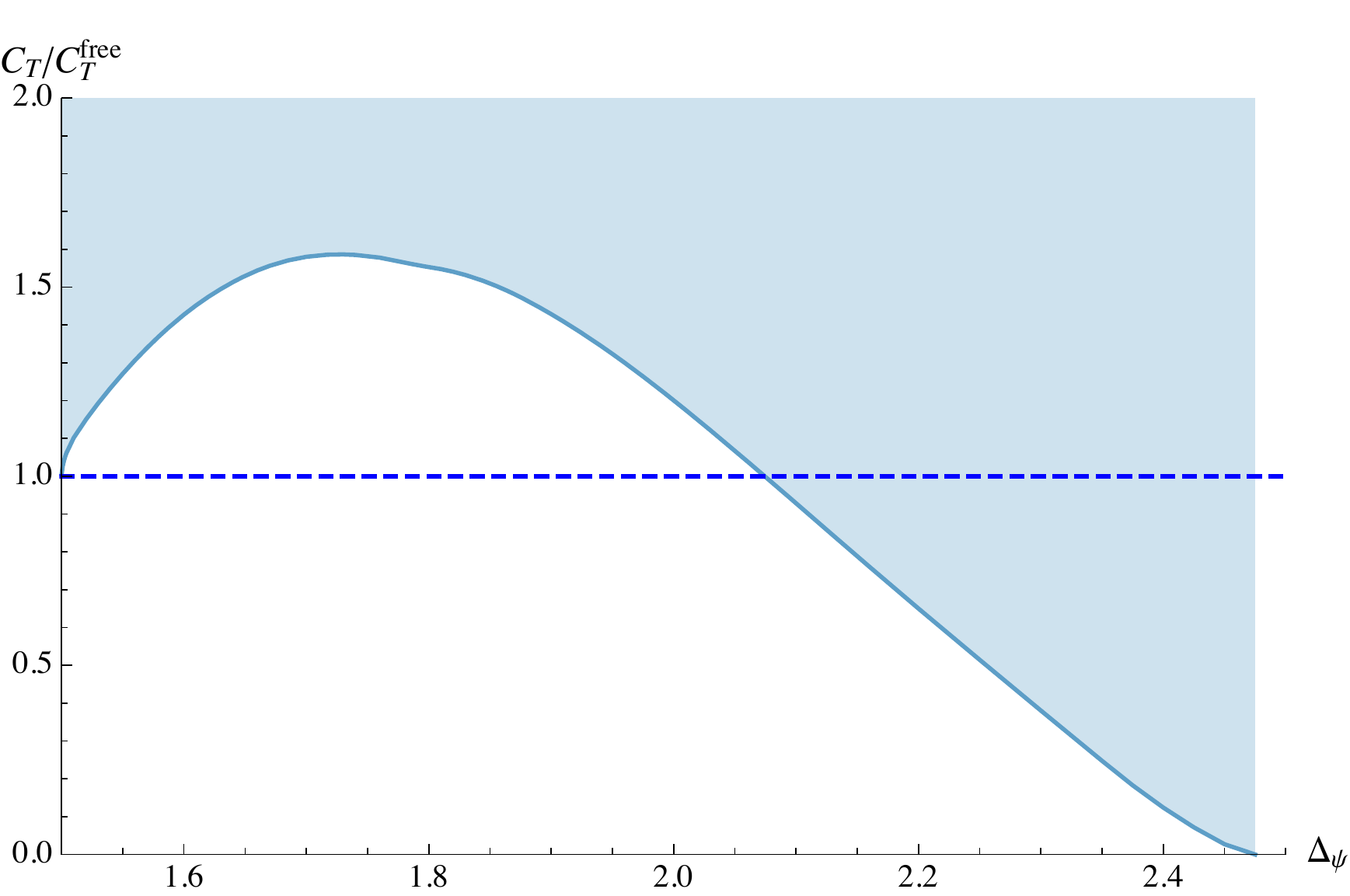} 
		\caption{Lower bound on the central charge normalized to the free fermion value $C_T^\text{free}$. The dashed line indicates the free fermion theory. The bound has been computed at $\Lambda=20$.}
		\label{fig:centralcharge}
	\end{center}
\end{figure}

We do not observe the jump for the $\ell=0$ case because the bound starts above $\De>5$ right from the beginning.\footnote{See discussion at the end of section \ref{sec:implications_numerics} for a possible explanation of why this is the case.}
Thus, in figure~\ref{fig:neutral02} we effectively solve the following problem: given that the first $(0,2)$ operator has the scaling dimension $\De=5$, what is the maximal value of the second lightest $(0,2)$ operator in a consistent CFT? The only reminiscence of the jump is the presence of a little bump at $\De_\psi\approx1.8$. It should also be noticed that this bound is very sensitive to the parameter $\Lambda$, for instance we observed a significant improvement of this bound by increasing $\Lambda$ from 16 to 20. It is then possible
that for high enough values of $\Lambda$ we can push the bound below $5$ for $\De_\psi$ around $1.5$. In that case we would expect to recover the jump.

\subsection{Bounds on central charges}
\label{sec:CentralCharge}

Let us first address the lower bound on the central charge $C_T$. We remind that a generic non-conserved TS operator has two independent OPE coefficients but in the case of the stress tensor they are both fixed by the Ward identities~\eqref{eq:TmnOPEcoeff}. We can then construct an upper bound on the prefactor in~\eqref{eq:PT} or equivalently a lower bound on $C_T$. The result is shown in figure~\ref{fig:centralcharge}, where for convenience we plot the ratio of $C_T$ to the one in the free fermion theory $C_T^{\text{free}}$ derived in~\eqref{eq:CT_free_fermion}. For $\De_\psi\to 3/2$ the bound approaches $C_T^\text{free}$, and the approach is consistent with 
\be
	\frac{C_T}{C_T^\text{free}}\lesssim 1+\a \sqrt{\De_\psi-\frac{3}{2}}
\ee
for some $\a>0$.

\begin{figure}[t]
\begin{center}
\includegraphics[scale=0.6]{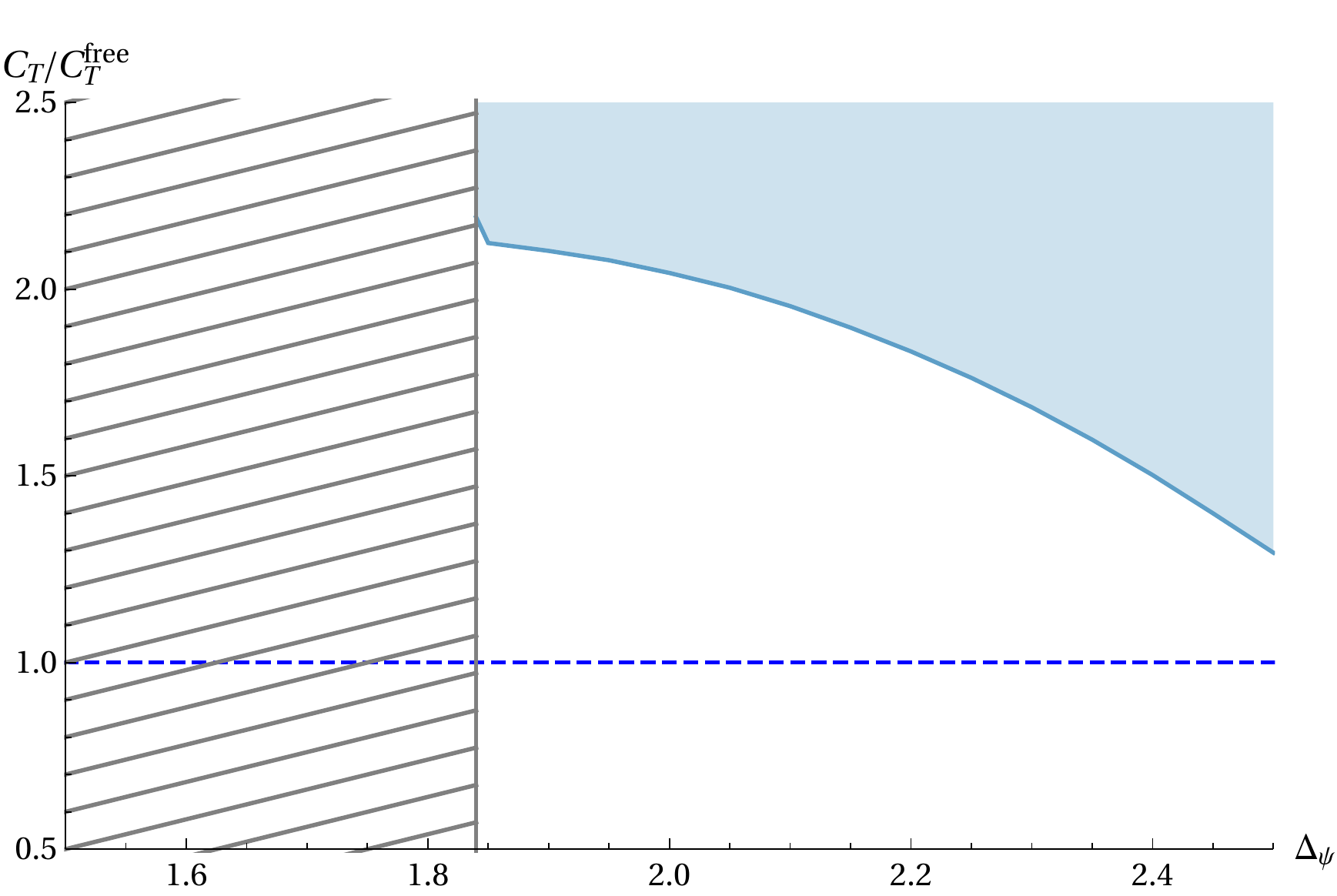} 
\caption{Lower bound on the central charge normalized to the free value under the assumptions that relevant scalars are absent, both in the neutral and charged sector. The shaded region is allowed. The $\Delta_\psi\lesssim1.84$ region is excluded by the condition of no relevant charged scalars, see figure \ref{fig:charged00}. The bound has been computed at $\Lambda=20$.}
\label{fig:centralchargedeadend}
\end{center}
\end{figure}

For the bound in figure~\ref{fig:centralcharge} we assumed nothing besides unitarity and crossing symmetry. One might introduce some assumptions on the spectrum of operators to get a stronger bound. As an example let us focus on CFTs without relevant scalar (charged or neutral) operators known as dead-end CFTs.\footnote{A trivial example of a dead-end CFT is the fermionic GFT studied in section~\ref{sec:GFT}. It does not have relevant neutral scalars and contains the charged ones only for $\Delta_\psi\leq 2$.}
We thus assume that $\Delta_0^{(0,0)}\geq4$ and $\Delta_-^{(0,0)}\geq4$. The result is shown in figure~\ref{fig:centralchargedeadend}. As we can see, we get a stronger, but only slightly, lower bounds for $C_T$. The $C_T$ bound does not exist for $\Delta_\psi\lesssim1.84$, consistently with the bound on $\De_-^{(0,0)}$. More precisely the bound in figure~\ref{fig:charged00} implies that $\Delta_\psi\gtrsim1.84$.

We now address the bound on $C_J$, the central charge associated with the $\mathrm{U}(1)$ conserved current $J$. In this case the two OPE coefficients are related by a single Ward identity~\eqref{eq:J_ope_relation}. As a result we can build an upper bound on the prefactor of~\eqref{eq:PJ} or equivalently a lower bound on $C_J$ as a function of an additional parameter $\theta\in[-\frac{\pi}{4},\;\frac{3\pi}{4}]$ defined in~\eqref{eq:theta_parametrization}.  Note that the Ward identity only fixes a particular linear combination of OPE coefficients, which are still free to be arbitrarily large. In our parametrization this region is mapped to the boundaries of the $\theta$ interval.

\begin{figure}[t]
\begin{center}
\subfigure[]{\label{fig:CJ}\includegraphics[scale=0.43]{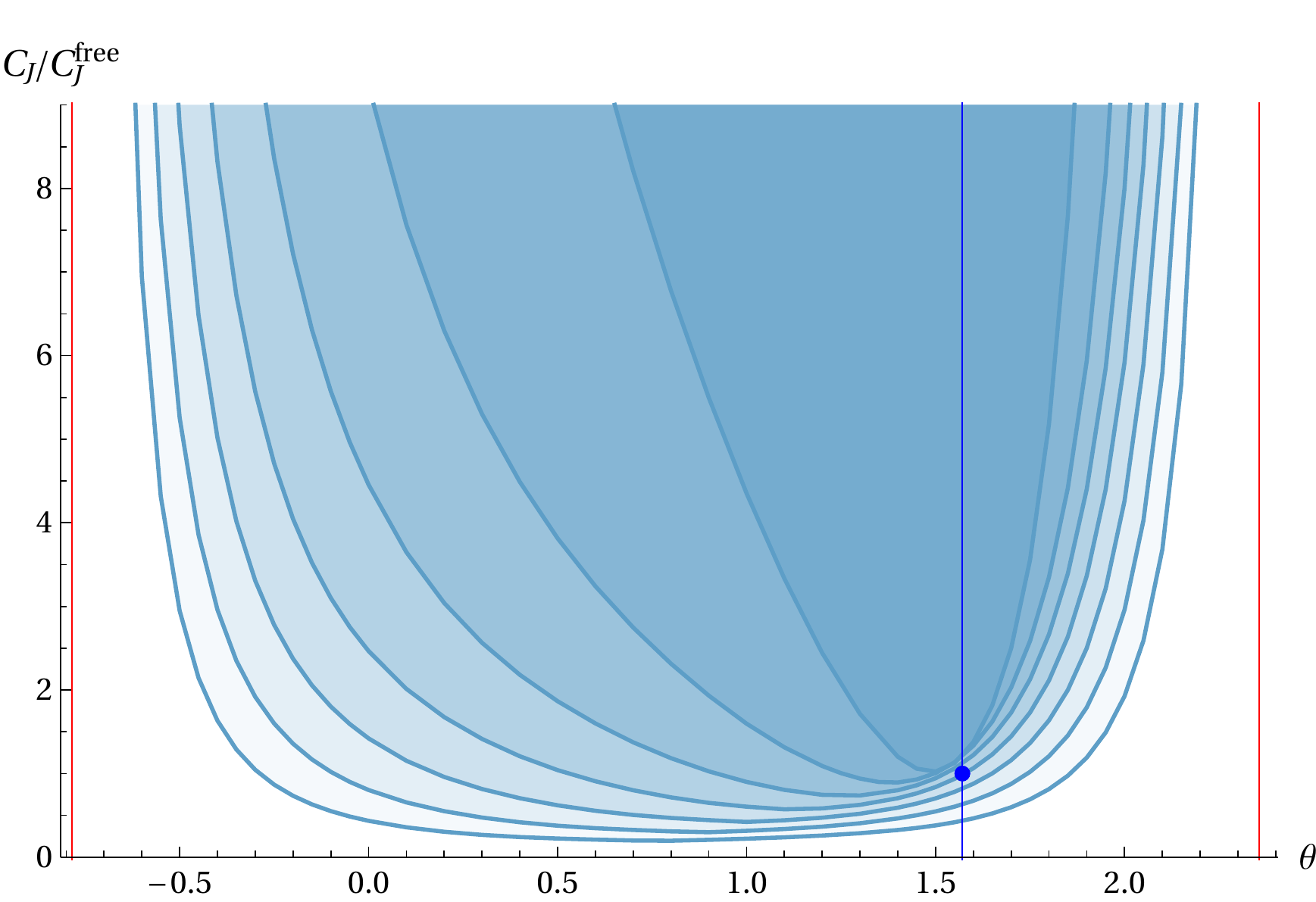} }
\subfigure[]{\label{fig:CJzoom}\includegraphics[scale=0.43]{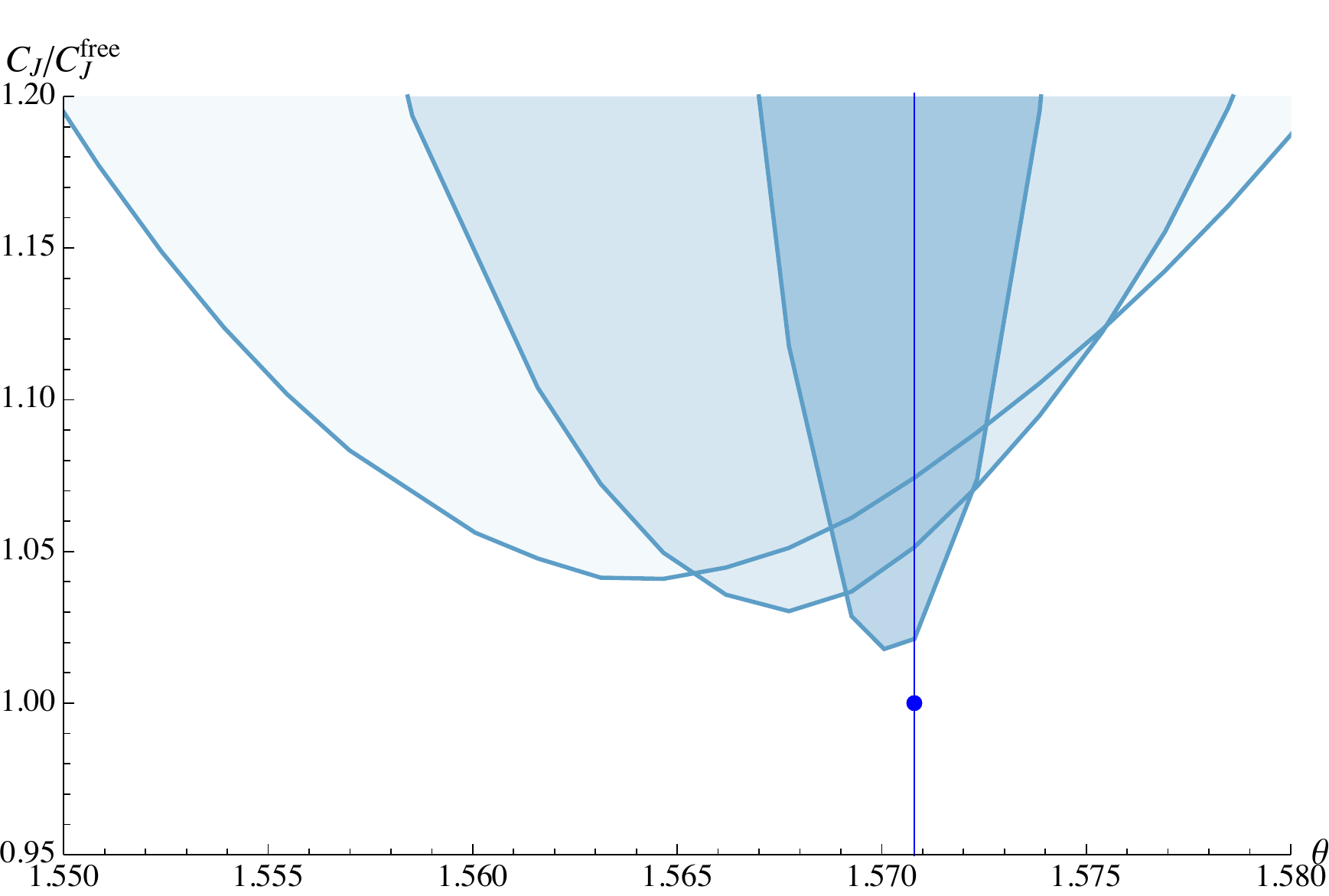}}
\caption{Left: Lower bound on the global symmetry current central charge $C_J$, normalized to the free value, as a function of the parameter $\theta$ defined in~\eqref{eq:theta_parametrization}. The shaded regions are allowed. Different shadings correspond (from darker to lighter) to  $\Delta_\psi=1.6, 1.7, 1.8, 1.9, 2.0, 2.1, 2.2$. Vertical red lines indicate the extremes of the $\theta$ parameter $-\pi/4$ and $3\pi/4$. The blue vertical line indicates the value $\theta=\pi/2$. The blue dot on this line corresponds to the free fermion theory. The bounds have been computed at $\Lambda=16$. Right: same plots zoomed aroung  $\theta=\pi/2$, with different shading corresponding (from darker to lighter) to  $\Delta_\psi=1.501, 1.505, 1.51$.}
\end{center}
\end{figure}

In figure \ref{fig:CJ} we plot a lower bound on $C_J$ as a function of the angle $\theta$ for several values of $\Delta_\psi$. The bounds become stronger when we approach the free fermion theory, $\De_\psi=3/2$, where the CFTs are forced to live in the vicinity of $\theta=\pi/2$, see~\eqref{eq:CJ_free_fermion}. In  figure \ref{fig:CJzoom}, we show that, as $\Delta_\psi\rightarrow1.5$, the bound creates  a sharper and sharper minimum, whose value approaches the free fermion CFT from above. As the fermion dimension increases the bounds get weaker and seem to diverge as $\theta\rightarrow 3\pi/4$ or  $\theta\rightarrow -\pi/4$. As mentioned earlier, the extremes of the $\theta$ interval corresponds to the region of large OPE coefficeints: it is then not surprising that the large central charge compensates the divergent OPE coefficients. One can indeed show that the quantity 
\begin{equation}
\label{eq:finite_ratio_CJ}
\xi_J\equiv (\cos\theta+\sin\theta)^2  \times C_J/ C_J^{\text{free}}
\end{equation}
is finite over the whole $\theta$ interval.

\begin{figure}[t]
\begin{center}
\includegraphics[scale=0.6]{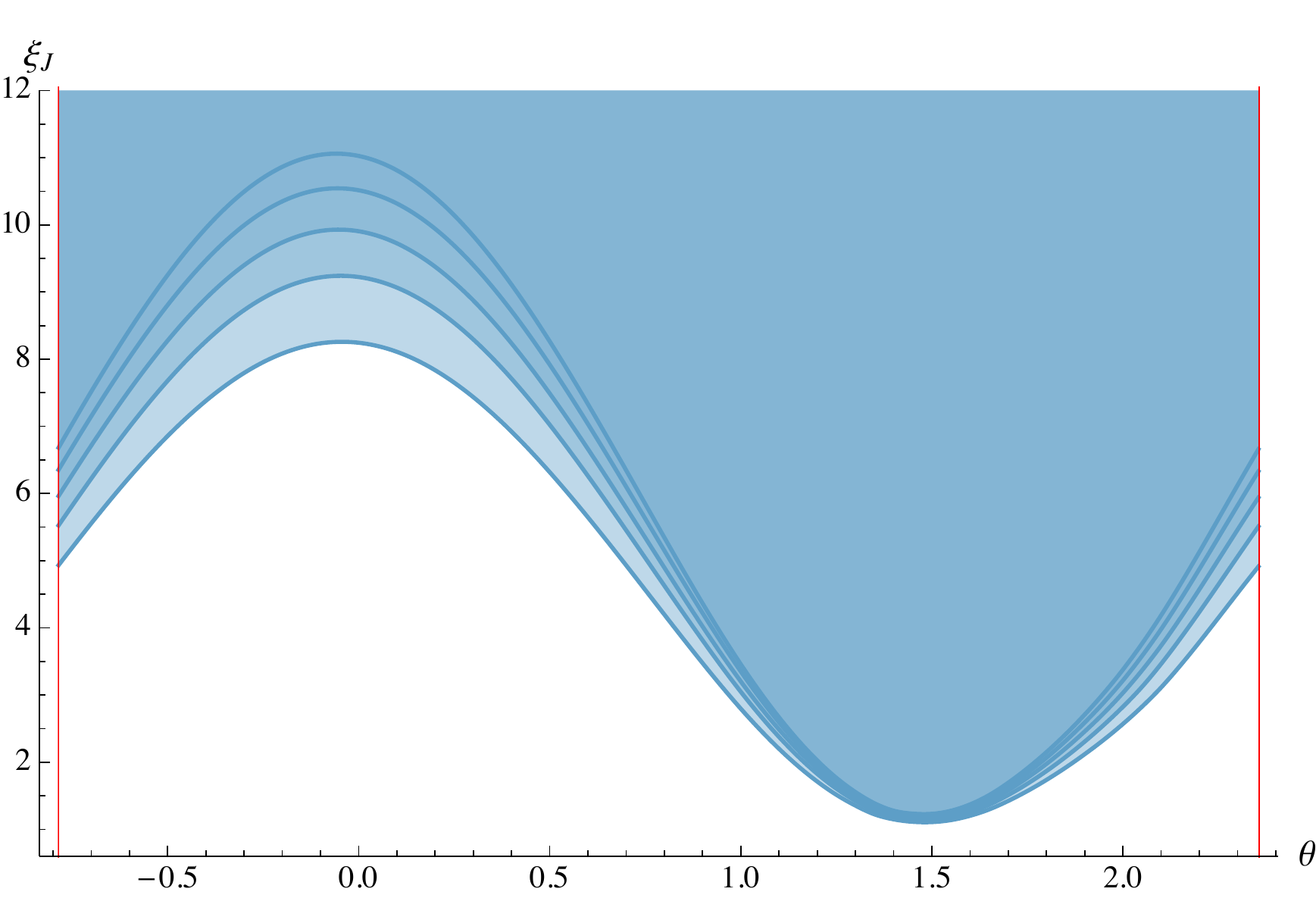} 
\caption{Left: Lower bound on the global symmetry current central charge $C_J$, normalized to the free value, as a function of the parameter $\theta$ defined in~\eqref{eq:theta_parametrization} for $\Delta_\psi = 1.7$. On the vertical axis we plot the rescaled central charge $\xi_J$ defined in~\eqref{eq:finite_ratio_CJ}.  Different shading corresponding (from lighter to darker) to  $\Lambda = 14, 16, 18, 20, 22$. The vertical red lines indicates the values $-\pi/4$ and $3\pi/4$ which are the extremes of the $\theta$ parameter.}
\label{fig:CJhighLambda}
\end{center}
\end{figure}

Finally we want to explore the possibility that by increasing the number of derivatives $\Lambda$ we can exclude part of the $\theta\in[-\frac{\pi}{4},\;\frac{3\pi}{4}]$ region. We thus construct a lower $C_J$ bound as a function of $\theta$ for a fixed value $\De_\psi$ for several values of $\Lambda$. In fact it is convenient to plot the quantity $\xi_J$. The result is shown in figure~\ref{fig:CJhighLambda}.  We do not see evidences that any value of $\theta$ is disallowed in the large $\Lambda$ limit.

\subsection{Bounds on scalar OPE coefficients}
\label{subsec:OPE_scalars}

\begin{figure}[t]
\begin{center}
\subfigure[]{\label{fig:OPEchargedscalar-df1.85}\includegraphics[scale=0.43]{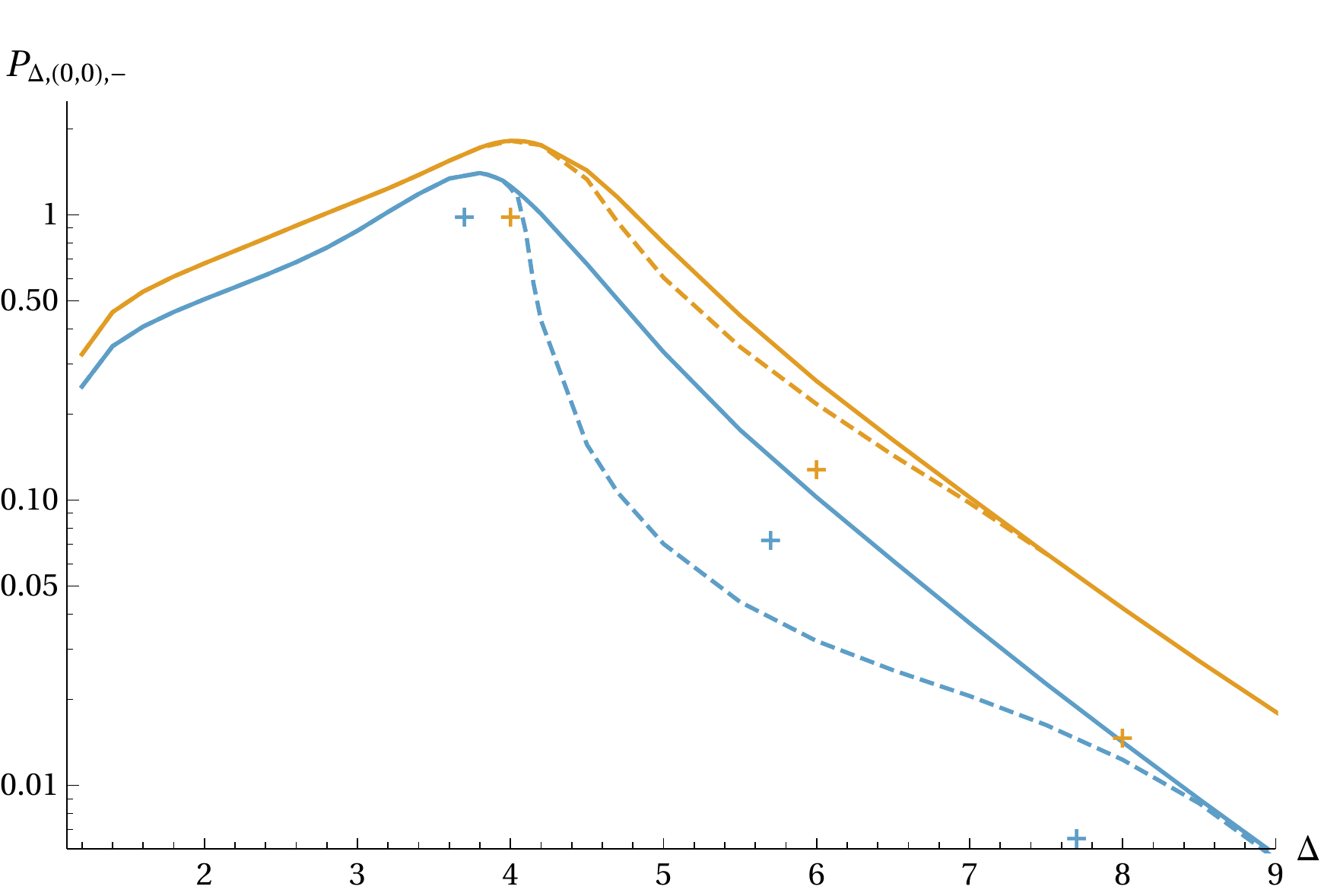} }
\subfigure[]{\label{fig:OPEchargedscalar-df2}\includegraphics[scale=0.43]{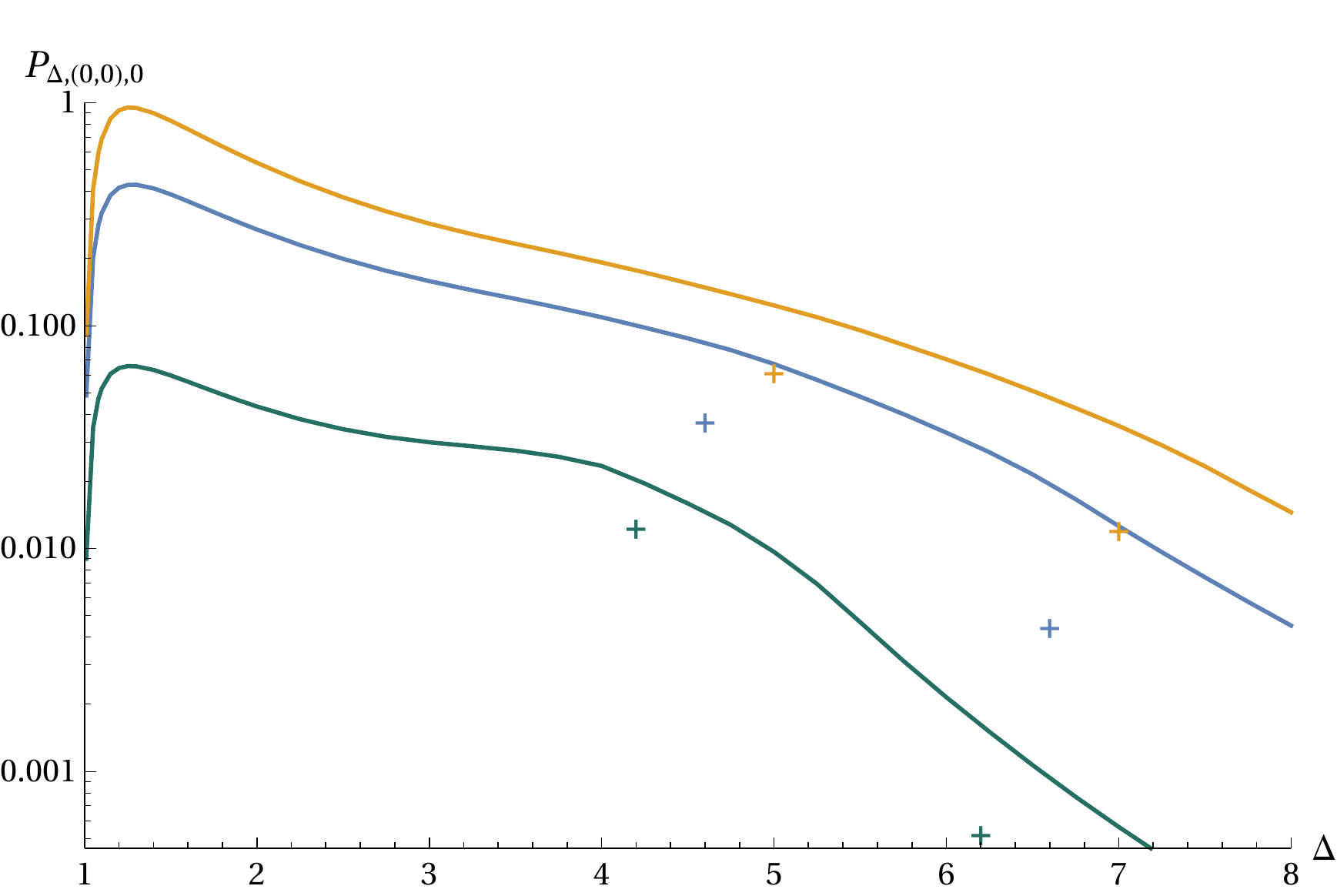}}
\caption{Left: upper bound on the squared OPE coefficient of a charged scalar as a function of its dimension for $\De_\psi=1.85$ (blue line) and $\De_\psi=2.0$ (orange line). Dashed lines are the bounds with the further assumption that the scalar is the lightest in the spectrum. Right: upper bound on the squared OPE coefficient of a neutral scalar as a function of its dimension for $\De_\psi=1.6$ (green line), $\De_\psi=1.85$ (blue line) and $\De_\psi=2.0$ (orange line). The bounds have been computed at $\Lambda=16$. Crosses corresponds to the fermion GFT values. Both plots are given in the log scale.}
\label{fig:OPEbound}
\end{center}
\end{figure}

We conclude the exploration of the parameter space of fermion CFTs by studying upper bounds on products of OPE coefficients for neutral and charged scalar operators as a function of their scaling dimension $\De$. The results are presented in figure~\ref{fig:OPEbound}. Notice that we use a log scale here. Lines with different colors correspond to different values of $\De_\psi$. In figure~\ref{fig:OPEchargedscalar-df1.85} the dashed lines represent the bound under the further assumption that the operator is the lightest in the spectrum. We do not plot the dashed lines on figure~\ref{fig:OPEchargedscalar-df2} because they almost coincide  with the solid ones.

We indicate the values of squared OPE coefficients in the fermion GFT by the little crosses. The consistency of the setup requires the bound to pass above them. In the case when we bound the lightest operator the bound (dashed line) is required to be above only the leftmost cross (the lightest GFT operator).

As $\Delta\rightarrow 1$ the bounds approach zero as expected, due to the presence of a pole at the scalar unitarity bound $\Delta\rightarrow1$. 
The bound becomes weaker as soon as we go to higher values $\De$ and reaches a maximum which is clearly visible. In the non-log scale this feature is much more pronounced.

\begin{figure}[t]
\begin{center}
\includegraphics[scale=0.6]{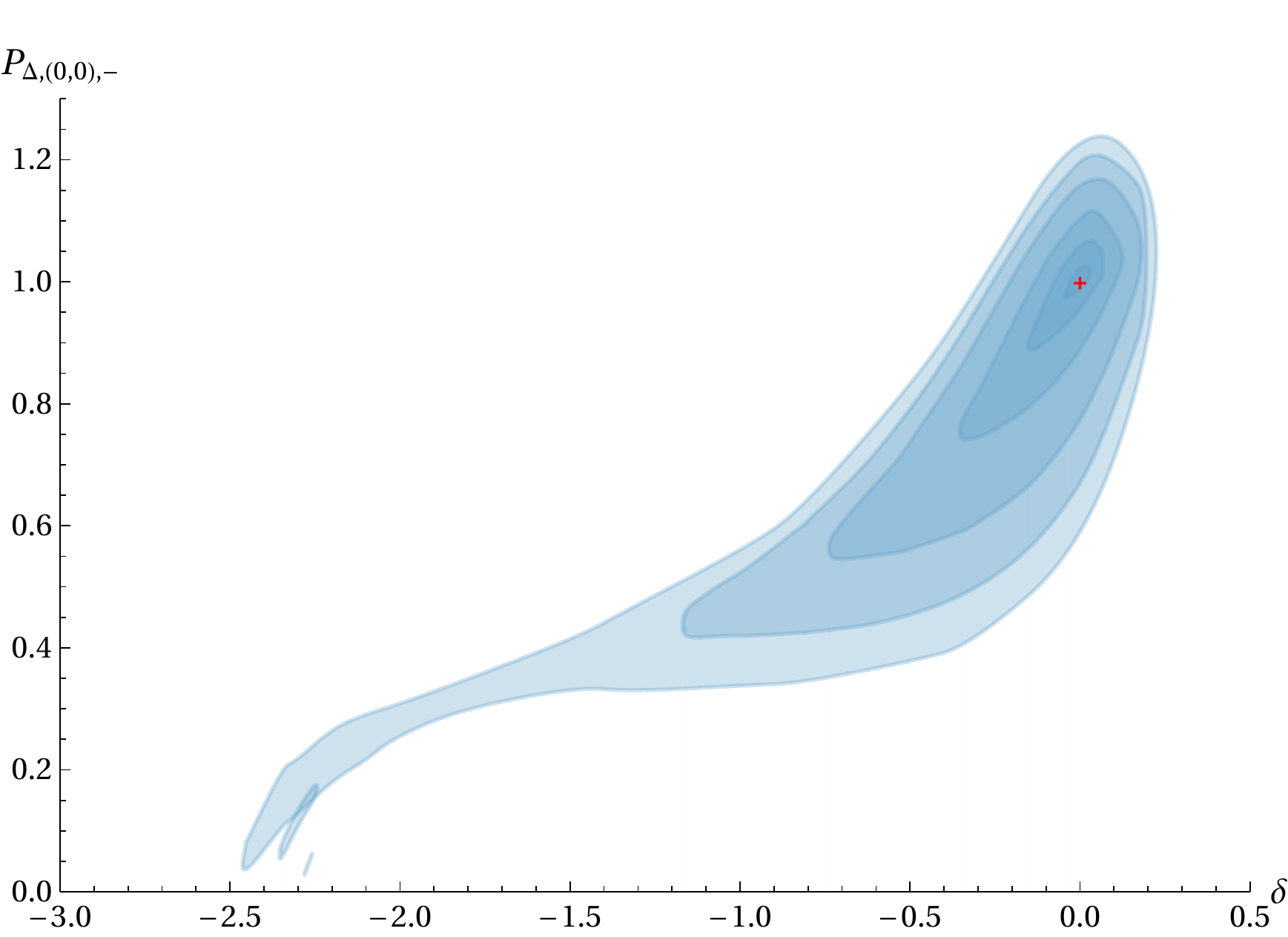} 
\caption{Allowed region in the plane of the dimension and the OPE coefficient of a single relevant scalar charged operator. On the horizontal axis $\delta\equiv \De_-^{(0,0)}-2\De_\psi$. The bounds have been computed at $\Lambda=16$. Shaded regions corresponds to (from darker to ligther) $\Delta_\psi=1.55, 1.6,1.65, 1.7, 1.73,1.75$. The two isolated regions in the lower left corner correspond to $\Delta_\psi=1.7, 1.73$. The red mark shows the value of GFT, see \sref{sec:GFTcharged}. }
\label{fig:OPEboundSingleCharged}
\end{center}
\end{figure}

Finally we address the case of CFTs with a single relevant charged operator. The scaling dimension of such an operator is confined to the region given in figure~\ref{fig:peninsula}. Here we construct in addition the upper and lower bound on its squared OPE coefficient.\footnote{A lower bound exists since this operator is isolated by assumption.} We take several $\De_\psi$ slices of figure~\ref{fig:peninsula} and show the bound in figure~\ref{fig:OPEboundSingleCharged}. For the slices with small $\De_\psi$ we get an island. Going to slices with bigger $\De_\psi$ we see how the island grows. At some point an isolated island appears in the bottom left corner of figure~\ref{fig:OPEboundSingleCharged}, corresponding to the lower part of the allowed region in figure~\ref{fig:peninsula}. For high enough $\De_\psi$
the  two islands merge.

\section{Conclusions}
\label{sec:conclusions}

In this work we studied the constraints imposed by unitarity and crossing symmetry on a generic 4d CFT containing at least one Weyl fermion. By applying numerical bootstrap techniques to the four-fermion correlator~\eqref{eq:the_fermion_four_point_function} we constructed several bounds on operator dimensions, central charges and OPE coefficients, with and without extra assumptions on the operator spectrum. 
The main qualitative advantage of our analysis, compared to previous bootstrap works in four dimensions, is that we are sensitive to operators transforming in the mixed-symmetry representations $(\ell+2,\ell)$ and $(\ell,\ell+2)$, which are invisible in the case of scalar four-point functions. This is also the first time the numerical conformal bootstrap has been applied to a non-scalar correlation function in 4d.

A distinguishing feature of many of our plots are the sharp jumps occurring when the upper bound on charged TS and neutral NTS operators dimensions crosses the value \mbox{$\Delta_\text{jump}=(\ell+\bar{\ell})/2+4$}. While discontinuities on the boundary of the allowed region usually signal the presence of existing CFTs, the integer nature of $\Delta_\text{jump}$ calls for another explanation. Indeed, we tracked down this phenomenon to a peculiar feature of the conformal blocks entering the crossing equations that leads to the appearance of fake primary operators.
 As described in section~\ref{sec:topology_blocks}, the occurrence of simple poles of conformal blocks at the unitarity bound changes the topology of the spectrum of a CFT as probed by a numerical bootstrap analysis. We refer to this phenomenon as the fake primary effect.

This effect is of course not limited to our case, but in general can have an impact for any correlation function with intermediate three-point functions not satisfying equations analogous to~\eqref{eq:vanishing_conditions}, in any d-dimensional CFT.
In particular, we have verified that a jump observed in the study of mixed correlators in the 3d Ising model~\cite{Kos:2014bka} is partly induced by a fake primary effect,  see figure~\ref{fig:isingsigmaprime}. 
As discussed in section~\ref{sec:3dfermion}, the jumps observed in the 3d fermion bootstrap~\cite{Iliesiu:2015qra,Iliesiu:2017nrv} can also be explained in this way.\footnote{Other analysis could be affected by the fake primary effect, for instance~\cite{Nakayama:2016jhq}.}

Note that the fake primary effect should not necessarily be interpreted as a mere artifact of numerical studies. For example, within our four fermion correlator,  if a 4d CFT has an almost conserved charged (1,1) current, 
there could be a solution to crossing with a quite high value of the smallest scaling dimension of the charged primary scalar. CFTs with such a property might then exist and sit on the top of the jump.\footnote{An existence of the solution to a particular set of crossing equations is not enough in general to claim the existence of a CFT.} In order to establish that, one should however check that the effect persists when we increase the precision of the numerics (higher values of $\Lambda$) and  also that it is consistent with bounds coming from other correlators. Figure \ref{fig:extrapolationCharged00} shows that without extra assumptions the effect persists at higher values of $\Lambda$ and a linear extrapolation of the bounds predicts a finite jump value even at $\Lambda\rightarrow \infty$. It is possible, however, that given an arbitrary small gap in the (1,1) charged channel, there exists a sufficiently high value of $\Lambda$ and order of approximation of conformal blocks such that the numerical algorithm is able to distinguish a primary from a fake primary. We have not investigated this possibility in this work.

A second highlight of the present work concerns the development of the rational approximation techniques for spinning conformal blocks in 4d.
More precisely, it was known that generic spinning conformal blocks can be obtained through the action of differential operators on the seed blocks. Despite the latter being known explicitly, their complicated structure makes it extremely hard to efficiently construct their rational approximation, which is ultimately the form needed for numerical studies using~\texttt{SDPB}. To overcome this difficulty, in this work we implemented a recursion relation of~\cite{Karateev:2017jgd} for the seed blocks and used it to express their derivatives in terms of the scalar conformal blocks. The rational approximations then follow from the expansion of the hypergeometric functions appearing in the scalar conformal blocks. 

Fermion four-point functions with a single Weyl fermion do not allow us to put significant constraints on the hypothetical CFTs which are typically invoked in phenomenological considerations (e.g. in composite Higgs models).
A simple enough generalization that might bring us closer to interesting scenarios is the addition of non-abelian global symmetries.\footnote{See e.g. figure 8 of~\cite{Caracciolo:2014cxa} for an example with scalars of how global symmetries can lead to stronger constraints potentially relevant for phenomenological applications.}

However, a severe problem has to be faced: as found in previous numerical studies, the numerical bounds become weaker and weaker as $\De_\psi$ increases from its free field value $3/2$.
As discussed in the introduction, the smallest UV scaling dimensions in gauge theories with fermions in the fundamental or adjoint representation, such as ordinary and adjoint QCD, 
are $\De_\psi^{{\rm UV}}=9/2$ or $\De_\psi^{{\rm UV}}=7/2$.
CFTs with these values appear to be deep in the allowed region of parameter space and cannot manifest themselves as a feature on the boundary even in the presence of extra theory-specific assumptions on the spectrum.\footnote{Global symmetries alone do not seem to be of great help in this context.	We thank Bernardo Zan for sharing with us some unpublished results on studying ``composite'' scalars with global symmetries coming from the $SU(2)$ gauge theory with fundamental fermions.} 
On the other hand, CFTs featuring fermions with $\De_\psi^{{\rm UV}}=5/2$ are more accessible and expected to be not that far from our current bounds. 

The formalism developed in this paper can be straightforwardly extended to superconformal field theories (SCFTs), the only missing ingredients being the precise form of  superconformal blocks. The simplest possibility is to identify the external fermion operator $\psi$ with the unique spinor present in a chiral scalar supermultiplet in $\mathcal N=1$ 4d SCFTs. However, past boostrap works already considered this problem by studying the correlation function of the superprimary operator sitting in the same supermultiplet \cite{Poland:2010wg, Vichi:2011ux, Poland:2011ey,Poland:2015mta,Li:2017ddj} and we do not expect the fermion bootstrap to lead to new results.\footnote{The correlation function in super-space  of chiral operators is indeed fully determined by the lowest component four-point function \cite{Fitzpatrick:2014oza}.}
A second more interesting possibility is to interpret $\psi$ as the lowest component of a chiral spinor superfield $W_\alpha$  which would also contain a two-form $F_{\mu\nu}$.   

Finally, the knowledge of 4d spinning conformal blocks, together with their rational approximations developed in this work, allows us to investigate other spinning correlators. Among these, the four-point function of conserved currents is one of the simplest correlators that should be considered next.

\section*{Acknowledgments}

We thank the organizers of the Bootstrap 2017 conference held in San Paulo and the Bootstrap 2018 conference held in Calthech for the hospitality and all the participants for interesting discussions. We especially thank Luca Iliesiu, Madalena Lemos, Miguel Paulos, Jo\~ao Penedones, David Poland, David Simmons-Duffin, Riccardo Rattazzi,  Slava Rychkov, Andreas Stergiou and Ning Su, for useful comments. 
We also thank Andrea Manenti for making a few supersymmetric checks.
We thank Tom DeGrand and Viljami Leino for discussing recent CFT data results from lattice techniques.

PK is supported by DOE grant No. DE-SC0009988. DK is supported by Simons Foundation grant 488649 (Simons Collaboration on the Nonperturbative Bootstrap) and by the National Centre of Competence in Research SwissMAP funded by the Swiss National Science Foundation. AV is supported by the Swiss National Science Foundation under grant no. PP00P2-163670 and by the European Research Council Starting Grant  under grant no. 758903. The computations in this paper were run on the EPFL SCITAS cluster and the IAS Hyperion cluster.

\appendix

\section{Connection between tensor and spinor formalisms}
\label{app:vectorspinor}
The spinor formalism used in this paper allows to work with operators in arbitrary spin representation $(\ell, \bar \ell)$. In case of traceless symmetric operators $(\ell,\ell)$ one can use the tensor formalism~\cite{Osborn:1993cr,Costa:2011mg} instead. Many results in the literature were obtained using the latter (for example the values of $C_J$ and $C_T$ in free theories), it is thus important to establish a precise connection between them.

The index-free operators in both formalism have the following form
\begin{align}
\label{eq:tensor_operators}
\cO^{(\ell,\ell)}_{\text{tensor}}(x,z) &\equiv \cO_{\mu_1\ldots\mu_\ell}(x)z^{\mu_1}\ldots z^{\mu_\ell},\quad z^2=0,\\
\label{eq:spinor_operators}
\cO^{(\ell,\bar\ell)}_{\text{spinor}}(x,s,\bar s) &\equiv \cO_{\alpha_1\ldots\alpha_\ell}^{\dot\beta_1\ldots \dot\beta_{\bar\ell}}(x)s^{\alpha_1}\ldots s^{\alpha_\ell}
\bar s_{\dot\beta_1}\ldots \bar s_{\dot\beta_{\bar\ell}}.
\end{align}
Here $z^\mu$ are constant null vector polarizations and $s^\alpha$ and $\bar s_{\dot\beta}$ are spinor polarization. In the $\ell=\bar \ell$ traceless symmetric case one can relate~\eqref{eq:tensor_operators} and~\eqref{eq:spinor_operators} by requiring
\begin{equation}
\label{eq:tensor_spinor_connection}
z^\mu = c\times(\bar s \, \bar\sigma^\mu  s),
\end{equation}
where $c$ is an arbitrary constant which is a matter of convention. Requiring than that~\eqref{eq:tensor_operators} and~\eqref{eq:spinor_operators} are equal then fixes also the relation between $\cO$ with spinor and vector indices.

Tensor structures of $n$-point functions are constructed as products of basic invariants. In the tensor (parity invariant) formalism there are two of them~\cite{Costa:2011mg}, they read as
\begin{align}
\label{eq:invariant_H}
H_{ij} &= x_{ij}^2\times\left((z_1\cdot z_2)-2\,\frac{(z_1\cdot x_{12})(z_2\cdot x_{12})}{x_{12}^2}\right),\\
\label{eq:invariant_V}
V_{k,ij} &= \frac{x_{ki}^2x_{kj}^2}{x_{ij}^2}\times
\left(\frac{(z_k\cdot x_{ki})}{x_{ki}^2}-\frac{(z_k\cdot x_{kj})}{x_{kj}^2}\right).
\end{align}
The tensor invariants in spinor formalism are summarized in appendix D of~\cite{Cuomo:2017wme}. Given the connection~\eqref{eq:tensor_spinor_connection} one can express the tensor invariants~\eqref{eq:invariant_H} and~\eqref{eq:invariant_V} in terms of the spinor ones as follows
\begin{align}
\label{eq:tensor_invariants_connection}
H_{ij}   = 2c^2\times \hat\II^{ij}\hat\II^{ji},\quad
V_{k,ij} = -c\times\hat\JJ^k_{ij}.
\end{align}

Two-point correlation functions are uniquely determined and are given in two formalisms by 
\begin{align}
\label{eq:2ptTS_tensor}
\<
\bar\cO^{(\ell,\ell)}_{\text{tensor}}(x_1,z_1)
\cO^{(\ell,\ell)}_{\text{tensor}}(x_2,z_2)
\> &= \text{const}\;\times\frac{H_{12}^\ell}{x_{12}^{2\,(\De+\ell)}},\\
\label{eq:2ptTS_spinor}
\<
\bar\cO^{(\ell,\ell)}_{\text{spinor}}(x_1,s_1,\bar s_1)
\cO^{(\ell,\ell)}_{\text{spinor}}(x_2,s_2,\bar s_2)
\> &= \text{const}'\times\frac{\left(\hat\II^{12}\hat\II^{21}\right)^{\ell}}{x_{12}^{2\,(\De+\ell)}}.
\end{align}
Here $\text{const}$ and $\text{const}'$ are positive real numbers which specify normalization of the CFT in two formalisms. The two-point function~\eqref{eq:2ptTS_spinor} is a special case of~\eqref{eq:two-point functions} with $\ell = \bar \ell$ and arbitrary normalization. It is very convenient to require
\begin{equation}
\text{const} = \text{const}'.
\end{equation}
Due to ~\eqref{eq:tensor_invariants_connection} this requirement leads to
\begin{equation}
c = \pm 2^{-1/2}.
\end{equation}
The same relation was found in (C.107) in~\cite{Karateev:2018oml} by equivalently requiring that the conformal two-point pairing is the same in two formalisms. Note that this is not the convention used in Wess-Bagger~\cite{Wess:1992cp}. In addition we can also remove an inconvenient minus sign in the second equality of~\eqref{eq:tensor_invariants_connection}. This leads to our final convention
\begin{equation}
c = - 2^{-1/2}
\quad\Rightarrow\quad
H_{ij}   = \hat\II^{ij}\hat\II^{ji},\quad
V_{k,ij} = 2^{-1/2}\times\hat\JJ^k_{ij}.
\end{equation}

\section{Ward identities}
\label{app:ward_identities}

In this section we show how to use weight shifting operators to compute $T_{\mu\nu}$ and $J_\mu$ Ward identities starting from the more familiar scalar ones. Let us start with the former. We recall that starting from the stress tensor we can construct the set of conserved charged associated to various conformal symmetry generators by
\be
\mathcal Q_\epsilon = \int_S dS^\mu \epsilon^\nu(x) T_{\mu\nu}(x)
\ee
where the integral is taken over any complete spacelike surface $S$, which can be taken to be $x^0=0$. For example, we have the Hamiltonian
\be
	H=\int d^3 x\, T_{00}(x),
\ee 
which corresponds to $\e^\mu=(1,0,0,0)$. We can get all the conformal charges $L_{AB}$\footnote{Here $A$ and $B$ are vector indices in $\R^{d,2}$, and $L_{AB}=-L_{BA}$ are the generators of the conformal algebra $\SO(d,2)$.} by using appropriate Killing vectors $\e^\mu_{AB}(x)$,
\be
	L_{AB}\propto\int_S dS^\mu \epsilon^\nu_{AB}(x) T_{\mu\nu}(x).
\ee
Here the proportionality coefficient depends on the normalization convention for $\e_{AB}$ and $L_{AB}$; we will not need it. The expressions for $\e_{AB}$ are the simplest in the embedding formalism. For example, in the embedding formalism of~\cite{Costa:2011mg} these Killing tensors are given by
\be
	\e_{AB}(X,Z)=X_{[A}Z_{B]}.
\ee
In the 6d embedding formalism used in this paper they are given by
\be
\e^{a}{}_b(\point) = S^a \bar S_b.
\ee
Here we used an equivalent set of indices for the adjoint of $\SO(4,2)$ by using the isomorphism with $\SU(2,2)$. It will be convenient to work with a formal primary vector operator $\cQ^\mu(y)$ of dimension $-1$ defined as
\be
	\cQ^\mu(y)=\e^\mu_{AB}(y)L^{AB}.
\ee
The fact that it transforms as a primary follows from transformation properties of $\e^\mu_{AB}$ and $L^{AB}$. Note that despite the fact that $\cQ^\mu(y)$ is labeled by a point $y$, it is by no means a local operator. Using the fact that\footnote{This follows straightforwardly from the explicit expressions for $\e_{AB}$. Alternatively, it simply suffices to check the conformal transformation properties on both sides.}
\be
	\e^\mu_{AB}(y)\e^{\nu,AB}(x)\propto \<\e^\mu(y)\e^\nu(x)\>,
\ee
where in the right hand side we mean the standard two-point function for vector operators of dimension $-1$, we can write
\be\label{eq:RTdefn}
	\cQ^\mu(y)=\mathbf{R}[T]^\mu(y)\equiv \cN \int_S dS^\s \<\e^\mu(y)\e^\nu(x)\>T_{\s\nu}(x),
\ee
for some normalization factor $\cN$. The notation $\mathbf{R}[T]$ stresses that we have an object which is obtained from $T$ by applying an integral transform. This transform is conformally-invariant, since it yields a vector primary $\cQ^\mu$ of dimension $-1$, and is a special case of the integral transforms considered in~\cite{Kravchuk:2018htv}. We will see that $\mathbf{R}[T]$ encodes the Ward identities in a convenient form.

The Ward identities express the fact that if we use $T$ in a three-point function to construct the charges, these charges should act appropriately on the other two primary fields. In particular,
\be
	\<0|\cO_1(\point_1) \mathbf{R}[T](\point_3) \cO_2(\point_2)|0\>=&\<0|\cO_1(\point_1) \cQ(\point_3) \cO_3(\point_2)|0\>\nn\\
	=&\e_{AB}(\point_3)\<0|\cO_1(\point_1) L^{AB} \cO_2(\point_2)|0\>\nn\\
	=&\e_{AB}(\point_3)\<0|\cO_1(\point_1) (\cL^{AB} \cO_2)(\point_2)|0\>,
\ee
where $\cL$ are the differential operators which implement the action of the conformal group generators on primaries,
\be
	[L^{AB},\cO(\point)]=(\cL^{AB}\cO)(\point).
\ee
The right-hand side contains a two-point function and thus is only non-trivial if $\cO_2=\bar \cO_1$, which we assume in what follows.\footnote{Even though the right hand side vanishes, the left hand side can still be non-zero and give a non-trivial condition on the three-point function coefficients. This will not be relevant for our discussion.} The general Ward identity is then
\be\label{eq:generalWardIdentity}
	\<0|\bar\cO(\point_1) \mathbf{R}[T](\point_3) \cO(\point_2)|0\>=\e_{AB}(\point_3)\<0|\bar\cO(\point_1) (\cL^{AB} \cO)(\point_2)|0\>.
\ee

Let us analyze the general features of the equation~\eqref{eq:generalWardIdentity}. On both sides we have natural conformally-invariant objects. For example, on the left hand side we have a conformally-invariant integral transform applied to a conformally-invariant three-point function. As we discussed above, $\mathbf{R}[T]=\cQ$ transforms as a primary vector field of dimension $-1$. Furthermore, it satisfies the conformal Killing equation. This is simply by definition~\eqref{eq:RTdefn}, since $\e^\mu(x)$ satisfies it in the two-point function. The same is true of the right hand side. Therefore, we can expand both sides in the appropriate basis of three-point tensor structures $\mathbb{T}_{3}^a$. The tensor structures $\mathbb{T}_{3}^a$ are the conformally-invariant three-point tensor structures for operators $\bar\cO$, $\cO$, and a vector primary of dimension $-1$. Furthermore, these structures should satisfy the conformal Killing equation for the vector primary. After the expansion~\eqref{eq:generalWardIdentity} takes form 
\be
	\sum_a l_a \mathbb{T}_{3}^a = \sum_a r_a \mathbb{T}_{3}^a,
\ee
where $l_a$ and $r_a$ are the coefficients of the left- and right-hand side expressions. The Ward identity simply requires that $l_a=r_a$. 

As we can see, the Ward identity for a given three-point function can in general include more then one equation if there is more that one $\mathbb{T}_3^a$. The counting of such three-point functions is a bit complicated due to the conformal Killing equation that they must solve. Fortunately, precisely such structures have been considered in~\cite{Karateev:2017jgd}. Their result is simply that the number of $\mathbb{T}_3^a$ is the number of Lorenz invariants\footnote{This rule works if there are no differential equations imposed on $\cO$.}
\be
	(\rho_\cO\otimes \rho^*_\cO\otimes(\bullet\oplus\text{adj}))^{\SO(d-1,1)},
\ee
where, $\rho_\cO$ is the Lorentz irrep of $\cO$, $\rho^*_\cO$ is the dual irrep, and $\text{adj}$ is the adjoint irrep of $\SO(d-1,1)$. In our case of $d=4$ we have $\text{adj}=(2,0)\oplus (0,2)$, and we are interested in $\cO=\psi$, which has irrep $\rho=(1,0)$. We find
\be
	(1,0)\otimes (1,0) \otimes (\bullet\oplus (2,0)\oplus (0,2))=2\bullet\oplus\ldots
\ee
where $\ldots$ represent non-scalar irreps. This means that there are two possible tensor structures and thus $2$ constraints from the Ward identity.

To find these constraints it is useful to employ weight-shifting operators and use the scalar case $\cO=\phi$ as the seed. In the case $\cO=\phi$ the three-point function 
\be\label{eq:correctffT}
	\<\phi(\point_1)\phi(\point_2)T(\point_3)\> = -\frac{\De_\f}{3\pi^2}\p{\hat{\mathbb{J}}^3_{1,2}}^2\cK_3,
\ee
where $\cK_3$ is the appropriate three-point kinematic factor, and the two-point function 
\be
	\<\f(\point_1)\f(\point_2)\> = \frac{1}{X_{12}^{\De_\f}}
\ee
satisfy the Ward identity~\eqref{eq:generalWardIdentity}~\cite{Osborn:1993cr}. We can compute the right-hand side of~\eqref{eq:generalWardIdentity} straightforwardly by using \texttt{CFTs4d} package. Using \texttt{opL} function of $\texttt{CFTs4D}$ we find
\be
	\e^{b}{}_a(\point_3)\<0|\f(\point_1) (\cL^a{}_b \f)(\point_2)|0\>=2\De_\f \hat{\mathbb{J}}^3_{1,2}\cK_3.
\ee
This means that a correctly-normalized $\mathbf{R}$ operation should take the standard structure $\<\f\f T\>^{(\uniq)}$ (which is the same as~\eqref{eq:correctffT} but without $-\De_\f/3\pi^2$ prefactor) to 
\be
	\<0|\f_1(\point_1)\mathbf{R}[T](\point_3)\f(\point_2)|0\>=-6\pi^2 \hat{\mathbb{J}}^3_{1,2}\cK_3.
\ee

Since the transform $\mathbf{R}$ acts only on $\point_3$, it commutes with weight-shifting operators acting on $\point_1$ and $\point_2$, so we can use the above equation as a seed to compute action of $\mathbf{R}$ on other three-point functions involving $T$. In particular, using the differential operators defined in section~\ref{sec:fermion_block_computation} we find
\be
	\<0|\bar\psi(\point_1)\mathbf{R}[T](\point_3)\psi(\point_2)|0\>\small{=}2\pi^2
	\left(-(3\l^1_{\<\bar\psi\psi T\>}+2\l^2_{\<\bar\psi\psi T\>})\hat{\mathbb{I}}^{1,2}\hat{\mathbb{J}}^3_{1,2}+\l^2_{\<\bar\psi\psi T\>}\hat{\mathbb{I}}^{1,3}\hat{\mathbb{I}}^{3,2}\right)\cK_3,
\ee
while the right hand side computed by \texttt{CFTs4D} is
\be
	\e^{b}{}_a(\point_3)\<0|\bar\psi(\point_1) (\cL^a{}_b \psi)(\point_2)|0\>=\p{-i(2\De_\psi+1)\hat{\mathbb{I}}^{1,2}\hat{\mathbb{J}}^3_{1,2}+2i\hat{\mathbb{I}}^{1,3}\hat{\mathbb{I}}^{3,2}}\cK_3.
\ee
This leads precisely to the relations~\eqref{eq:TmnOPEcoeff}. 

The discussion above applies almost identically to the case of spin-1 conserved current, except that everywhere the adjoint irrep of $\SO(2,d)$ must be replaced with the trivial irrep, which makes matters much simpler. In particular, $\mathbf{R}[J]$ is given by
\be
	\mathbf{R}[J]=\int d^3 x J^0(x)
\ee 
and produces a constant function simply equal to the $\mathrm{U}(1)$ charge $Q$. Accordingly, the right-hand side of~\eqref{eq:generalWardIdentity} is replaced with 
\be
	Q_\cO\<\bar\cO(\point_1)\cO(\point_2)\>. 
\ee
Due to this, we effectively get that the quantities equated in the $J$ analog of~\eqref{eq:generalWardIdentity} are two-point functions and we always get a single condition. In the case of scalars it is straightforward to check that the standard three-point function $\<\bar\phi\phi J\>^{(\uniq)}=\hat{\mathbb{J}}^3_{12}K_3$ gets sent by $\mathbf{R}$ to
\be
	2\sqrt{2}\pi^2 i\<\bar\phi(\point_1)\f(\point_2)\>.
\ee
Applying weight-shifting operators we find
\be
	\<0|\bar\psi(\point_1)\mathbf{R}[J](\point_3)\psi(\point_2)|0\>=i\sqrt{2}{\pi^2}\p{2\l^1_{\<\bar\psi\psi J\>}+\l^2_{\<\bar\psi\psi J\>}}\hat{\mathbb{I}}^{12}X_{12}^{-\De_\psi-\half},
\ee
which has to be equal to 
\be
	q\<\bar\psi(\point_1)\psi(\point_2)\>=iq \hat{\mathbb{I}}^{12}X_{12}^{-\De_\psi-\half},
\ee
leading to~\eqref{eq:J_ope_relation}.

\section{Parity constraints}
\label{app:parity}
If the CFT under consideration preserves space parity, there is a unitary operator $\mathcal{P}$ which relates various local operators in the spectrum.\footnote{By combining $\cP$ with $\cC\cP\cT$ symmetry, one can construct $\cC\cT$ symmetry. In principle, $\cC\cT$ is a valid anti-unitary time-reversal symmetry, so we may say that if theory preserves parity, then it also preserves time-reversal. If we demand additional properties from time-reversal (such as particular commutation rules with global charges, which is traditional in some contexts) which are not satisfied by $\cC\cT$, it then may be meaningful to say that there is no time-reversal.} Given an operator $\cO$ in the $(\ell,\bar\ell)$ spin representation, the generic action of space parity is, according to equation (A.26) in~\cite{Cuomo:2017wme},
\begin{equation}
\label{eq:parity_transformation}
\mathcal{P}\cO^{(\ell,\bar\ell)}_{\De,\rho}(\point) \mathcal{P}^{\dag} = \widetilde\cO^{(\bar\ell,\ell)}_{\De,\rho'}(\mathcal{P}\point),\quad
\mathcal{P}\point\equiv(\mathcal{P}x, \mathcal{P}s,\mathcal{P}\bar s),
\end{equation}
where the arguments in the right-hand side are given by
\begin{equation}
\label{eq:parity_transformation_arguments}
(\mathcal{P}x)^\mu=(x^0,-x^i),\quad
(\mathcal{P}s)^\alpha = i\bar s_{\dot\alpha},\quad
(\mathcal{P}\bar s)_{\dot\alpha} = i s^{\alpha},
\end{equation}
and $\widetilde\cO$ is some local operator in the $(\bar\ell,\ell)$ representation. In equation~\eqref{eq:parity_transformation} we added an explicit index $\rho$ specifying the representation of local operators under the global symmetry. If parity commutes with the global symmetry we have $\rho=\rho'$. At the level of correlation functions parity implies
\begin{equation}
\label{eq:parity_constraint_correlation_functions}
\<\cO(\point_1)\ldots\cO(\point_n)\> =
\<\widetilde\cO(\mathcal{P}\point_1)\ldots\widetilde\cO(\mathcal{P}\point_n)\>.
\end{equation}

In our setup we have a $\mathrm{U}(1)$ symmetry, thus the charge under this $\mathrm{U}(1)$ plays the role of $\rho$. We can distinguish between vectorial and axial $\mathrm{U}(1)$ symmetry by the commutation rule of $\cP$ with the charge $Q$,\footnote{If the full global symmetry group is larger than $\mathrm{U}(1)$ then there can be more general options, which we do not consider here for the sake of simplicity.}${}^,$\footnote{One might worry that axial $\mathrm{U}(1)$ symmetries can be broken by the ABJ anomaly. However, the charge $Q$ can still be a Cartan of a non-abelian symmetry. Furthermore, even if $\mathrm{U}(1)$ is broken by the ABJ anomaly, it is generically only broken down to a sufficiently large $\Z_n$ subgroup, and most of the analysis still applies.}
\be\label{eq:axivector}
	\cP Q\cP^\dagger = \pm Q,
\ee 
where $(+)$ sign is for vectorial and $(-)$ for axial.
According to~\eqref{eq:parity_transformation} we can write the parity transformation property for the Weyl fermion~\eqref{eq:weyl_fermion} as
\begin{equation}
\label{eq:parity_operator_1}
\mathcal{P}\psi_{\De_\psi,q}(\point) \mathcal{P}^{\dag} = \widetilde\psi^{(0,1)}_{\De, \pm q}(\mathcal{P}\point),
\end{equation}
where in the right-hand side the sign is the same as in~\eqref{eq:axivector}. The operator $\widetilde{\psi}$ can either be related to the hermitian conjugate of the same Weyl fermion $\bar\psi$ or to a hermitian conjugate of a different Weyl fermion which we denote by $\bar\chi$,
\begin{align}
\label{eq:parity_cases}
\widetilde\psi^{(0,1)}_{\De_\psi, \pm q}(\point) = \eta_\psi \bar\psi_{\De_\psi,\pm q}(\point)
\quad
\text{or}
\quad
\widetilde\psi^{(0,1)}_{\De_\psi, \pm q}(\point) = \eta_\psi \bar\chi_{\De_\psi,\pm q}(\point).
\end{align}

We first address the second case in~\eqref{eq:parity_cases}. The two Weyl fermions $\psi$ and $\bar\chi$ can be combined into a four-component Dirac fermion $\Psi_D\equiv (\psi_\alpha, \bar\chi^{\dot\alpha})$, and this option is consistent with both axial and vectorial symmetry. In this work we do not consider this situation since we deal only with one Weyl fermion. In order to bootstrap such theories we would need to add extra mixed correlators with both $\psi$ and $\chi$, which would complicate the setup significantly. The results of this paper still apply to theories with Dirac fermions but are not optimal.

Now let us address the first case in~\eqref{eq:parity_cases}. Clearly, for the vectorial $\mathrm{U}(1)$ symmetry $q=0$ and thus we drop the charge label everywhere below. As a consequence we can construct a four-component Majorana fermion $\Psi_M=(\psi_{\alpha},\psi^{\dag\dot\alpha})$. The absence of charges imply that there is no distinction between the ``charged'' and ``neutral'' sectors discussed in section~\ref{sec:three_point_functions}, in other words the exchanged operators appearing in both channels are actually the same. In practice when constructing the bounds if the gap is imposed on an operators in neutral channel, the same gap should be imposed on the operator with the same spin in the charged channel and vice versa. For an axial symmetry the charge $q$ can be non-zero, and the analysis is not modified.

The parity transformation rules can be summarized as
\begin{align}
\label{eq:Majorana_fermion}
\mathcal{P}\psi(\point) \mathcal{P}^{\dag} &=
\eta_\psi \bar\psi(\mathcal{P}\point),\\
\label{eq:Majorana_fermion_bar}
\mathcal{P}\bar\psi(\point) \mathcal{P}^{\dag} &=
\eta_{\bar\psi} \psi(\mathcal{P}\point).
\end{align}
Applying hermitian conjugation to~\eqref{eq:Majorana_fermion} and comparing it to~\eqref{eq:Majorana_fermion_bar} we deduce that 
\begin{equation}
\eta_{\bar\psi} = \eta_\psi^*.
\end{equation}
Applying parity transformation to~\eqref{eq:Majorana_fermion} we should get back the original operator, thus
\begin{equation}
\mathcal{P}^2\psi(\point) \mathcal{P}^{\dag2} =
-|\eta_\psi|^2 \psi(\point),
\end{equation}
where the $(-)$ sign comes from $(\cP^2 s)^\a=i(\cP\bar s)_{\dot\a}=i^2 s^\a=-s^\a$.
This implies that states $\int d\point f(\point)\psi(\point)|0\>$ are eigenstates of $\cP^2$ with eigenvalue $-|\eta_\psi|^2$. Since $\cP$ is unitary we find
\be\label{eq:intrinsic_parity}
	|\eta_\psi|^2=1,\quad \cP^{2}\psi(\point)\cP^{\dagger 2}=-\psi(\point).
\ee
If all operators in the theory can be obtained by repeated OPE of $\psi$, this implies that $\cP^2=(-1)^F$, but we won't be needing this conclusion.

Let us see what are the implications of~\eqref{eq:Majorana_fermion} and~\eqref{eq:Majorana_fermion_bar} for our setup. Applying~\eqref{eq:parity_cases} to the four-point function~\eqref{eq:masterfourpt} we get
\be
\label{eq:parity_constraint}
\<\bar\psi(\point_1)\psi(\point_2)\bar\psi(\point_3)\psi(\point_4)\> = 
|\eta_{\psi}|^4
\<\psi(\mathcal{P}\point_1)\bar\psi(\mathcal{P}\point_2)\psi(\mathcal{P}\point_3)\bar\psi(\mathcal{P}\point_4)\>.
\ee
Using the expansion~\eqref{eq:ordering_1_tensor_structure_decomposition} of the four-point correlation function into tensor structures, the anti-commutation properties of Weyl fermions and~\eqref{eq:intrinsic_parity}, we can rewrite the parity constraint~\eqref{eq:parity_constraint} as
\be
\label{eq:parity_constraint_final}
\sum_{i=1,\pm}^3 g_{i,\pm}(z,\bar z) \mathbb{T}_{i,\pm} = 
\sum_{i=1,\pm}^3 g_{i,\pm}(z,\bar z) \pi_{12}\pi_{34}\mathcal{P}\mathbb{T}_{i,\pm},
\ee
where $\mathcal{P}\mathbb{T}$ are the tensor structures obtained from $\mathbb{T}$ by applying~\eqref{eq:parity_transformation_arguments}. Notice that $(z,\bar z)$ are invariant under parity which is clear from their definition~\eqref{eq:definition_(z,zb)}. We have
\begin{equation}
\mathcal{P}\mathbb{T}_{i,\pm} = \mathbb{T}_{i,\pm},\quad i=1,2;\quad
\mathcal{P}\mathbb{T}_{2,+} = \mathbb{T}_{2,+};\quad
\mathcal{P}\mathbb{T}_{2,-} = -\mathbb{T}_{2,-}.
\end{equation}
As a result the only constraint we get from~\eqref{eq:parity_constraint_final} is the requirement that
\begin{equation}
g_{2,-}(z,\bar z) = 0.
\end{equation}
We see that this constraint is automatically satisfied by~\eqref{eq:permutation_constraints} coming from permutation symmetry. Thus, parity requirement does not bring extra constraints on the four-point function~\eqref{eq:masterfourpt}.

In the same way one can study parity constraints on the three-point functions defined in section~\ref{sec:three_point_functions} assuming that there are no new operators in the spectrum, so parity relates existing operators among themselves. They will enforce some extra reality properties on the OPE coefficients $\lambda$. Our setup is insensitive to such constraints. Indeed, the only way that $\l's$ enter into our equations is, schematically, through
\be\label{eq:PGproduct}
	\mathrm{tr}\p{PG},
\ee
where $P$ are the Hermitian matrices
\be
	P^{ba}=\sum_\l \l^a (\l^b)^*
\ee
and $G$ are various conformal blocks. If phase of $\l$'s is fixed, then $P$ is restricted to be real symmetric. However, the conformal blocks $G$ turn out to be themselves real symmetric matrices. This implies that only the real part of $P$ contributes to~\eqref{eq:PGproduct}, and in practice there is no difference whether $P$ is Hermitian or real symmetric. 

Since our setup is insensitive to the reality properties of $\l$'s, we will not discuss them further.

\section{Smoothness constraints}
\label{app:analyticity}
In this section we derive the constraints in \eqref{eq:gzzbregularity}, which follow by imposing smoothness properties of the four-point function in conformal frame. Constraints of this kind have been  described in appendix A of \cite{Kravchuk:2016qvl}, but we repeat the logic here, adding a few details for the reader convenience. We focus on the ordering $\<\bar\psi(\point_1)\psi(\point_2)\bar\psi(\point_3)\psi(\point_4)\> $, but the logic is similar for other orderings.

As discussed in \cite{Kravchuk:2016qvl}, at generic $z,\bar{z}$, four-point tensor structures must be invariant under the conformal frame stabilizer group, in our case $\SO(2)$.  This means that the six structures $\T^0_i$ defined in \eqref{eq:gchan4structures0} are singlets under $\SO(2)$. On the other hand, at the special configuration $z=\bar z$,  the stabilizer group enhances from $\SO(2)$ to $\SO(3)$,\footnote{In our case $\SO(3)$ is actually $\SO(2,1)$ and stabilizes the third spatial axis. We will keep using $\SO(3)$ since it does not alter the discussion.} and the structures $\T^0_i$ can be recast in $SO(3)$ representations. Since the external fermions transform in the $j=1/2$ representation of $\SO(3)$, we have the tensor product
\be \label{eq:so3vsso2}
	\half\otimes\half\otimes\half\otimes\half=2\oplus 1\oplus 1\oplus 1\oplus 0\oplus 0.
\ee
We then see that the six structures  $\T^0_i$ can be seen as the neutral components of 1 quintuplet, 3 triplets and 2 singlets under $\SO(3)$. Let us denote by $\hat \T^0_i$ the $\SO(3)$ diagonal structures defined by~\eqref{eq:so3vsso2}. Instead of~\eqref{eq:g0}, we could alternatively  expand the four-point function in this basis:
\be\label{eq:widetildeg0}
	\<\bar\psi(\point_1)\psi(\point_2)\bar\psi(\point_3)\psi(\point_4)\>=\sum_{i=1}^6 \hat{\mathbb{T}}^{0}_i \hat g_i^0(z,\bar z).
\ee
We see from~\eqref{eq:so3vsso2} that at $z=\bar z$ there are only two $\SO(3)$ invariant structures, which implies that the four functions $\hat g_i(z,\bar z)$ associated to non-$\SO(3)$ invariant structures must vanish in the limit $\bar z\rightarrow z$. We can determine the way in which they vanish by matching the $\SO(2)$ and $\SO(3)$ descriptions close to the $z=\bar z$ line.
Indeed, in the conformal frame~(\ref{eq:cf}), we can expand each $\hat g_i^0(z,\bar z)$ in the variable $y^p= (\bar{z}-z,0,0)$, which is a vector under the $\SO(3)$ stabilizer group:
\be
\hat g_i^0(z,\bar z) = \hat g_i^{p_1\ldots p_j}(z+\bar z) y_{p_1}\ldots y_{p_j} +{\cal O}(y^2) \,,
\ee
where $j=0,1,2$ depending on the corresponding structure defined by~\eqref{eq:so3vsso2}.
We conclude that, in the limit $z\rightarrow\bar{z}$, the function $\hat g_i^0$ vanishes as $(z-\bar{z})^{j}$, where $j$ is the $\SO(3)$ representation of the associated structure $\hat \T^0_i$.
Furthermore, these combinations have to be even (odd) under $z\leftrightarrow \bar z$ for even (odd) $j$. 

The relation between the structures $\T_i^0$ and  $\hat \T^0_i$ is easily found using the tabulated Clebsch-Gordan coefficients up to spin 2.
We get 
\be
	\T^0&=R\,\hat\T^0,
\ee
where $R$ is the orthogonal matrix
\be\label{eq:regularitymatrix}
	R =\left(
	\begin{array}{cccccc}
		\frac{1}{\sqrt{6}} & \frac{1}{\sqrt{6}} & \frac{1}{\sqrt{6}} & \frac{1}{\sqrt{6}} & \frac{1}{\sqrt{6}} & \frac{1}{\sqrt{6}} \\
		0 & 0 & 0 & 0 & -\frac{1}{\sqrt{2}} & \frac{1}{\sqrt{2}} \\
		0 & 0 & -\frac{1}{\sqrt{2}} & \frac{1}{\sqrt{2}} & 0 & 0 \\
		-\frac{1}{\sqrt{2}} & \frac{1}{\sqrt{2}} & 0 & 0 & 0 & 0 \\
		-\frac{1}{2} & -\frac{1}{2} & 0 & 0 & \frac{1}{2} & \frac{1}{2} \\
		-\frac{1}{2 \sqrt{3}} & -\frac{1}{2 \sqrt{3}} & \frac{1}{\sqrt{3}} & \frac{1}{\sqrt{3}} & -\frac{1}{2 \sqrt{3}} & -\frac{1}{2 \sqrt{3}} \\
	\end{array}
	\right)
\ee
and the structures $\hat\T_i^0$ are ordered as eigenstates corresponding to spins $\{2,1,1,1,0,0\}$, respectively. 
Alternatively, the decomposition in $\SO(3)$ irreducible representations can be obtained by solving the eigenproblem for the $\SO(3)$ quadratic Casimir operator in the space of the $\T_i^0$'s. We remind that the Casimir reads
\be
	C_{\SO(3)}=M_{01}M_{01}+M_{02}M_{02}-M_{12}M_{12}.
\ee
Using \texttt{CFTs4D} we compute the action of the Casimir\footnote{The action of the $\SO(3)$ Casimir operator is implemented in a \texttt{Mathematica} notebook attached to this work.} to be
\be
	C_{\SO(3)}\T_i^0=M_{i}{}^j\T_j^0,
\ee
where
\be
	M=\begin{pmatrix}
		2 & 0 & 1 & 1 & 1 & 1\\
		0 & 2 & 1 & 1 & 1 & 1\\
		1 & 1 & 2 & 0 & 1 & 1\\
		1 & 1 & 0 & 2 & 1 & 1\\
		1 & 1 & 1 & 1 & 2 & 0\\
		1 & 1 & 1 & 1 & 0 & 2
	\end{pmatrix}.
\ee
Diagonalizing the matrix $M$ gives back the rotation matrix~\eqref{eq:regularitymatrix}.

We thus find that the structures $\hat\T_1,\hat\T_5,\hat\T_6$ are even under $z\leftrightarrow \bar z$, while the structures $\hat\T_2,\hat\T_3,\hat\T_4$ are odd under $z\leftrightarrow \bar z$, 
$\hat g_1^0$ should vanish as $(z-\bar z)^2$, while $\hat g_{2,3,4}^0$ should vanish as $z-\bar z$. The last requirement is in fact trivial since these functions are odd anyway, while the first 
reads
\be\label{eq:regularity}
	\hat g_1^0= R_1{}^i g_i^0(z,\bar z)=\frac{1}{\sqrt 6}\sum_{I=1}^6 g_i^0(z,\bar z)=O((z-\bar z)^2).
\ee
Rephrasing the above condition in terms of $g_{i,\pm}$ one obtains precisely \eqref{eq:gzzbregularity}.

Finally let us understand $z\leftrightarrow \bar z$ symmetry.  Note that it is implemented by a boost by $i\pi$ in the plane 0-1 or 0-2 \cite{Kravchuk:2016qvl}. It does not matter which one to use because of $\SO(2)$-invariance. We will use 0-2.  In the notation of~\cite{Cuomo:2017wme}, it sends
\be
	\eta\to\xi,\,\xi\to-\eta,\,\bar\eta\to-\bar\xi,\,\bar\xi\to\bar\eta,
\ee 
and thus, in the notation of~\cite{Cuomo:2017wme},
\be
	\structgeneral\to(-1)^{(\ell-\bar\ell)/2} 
	\struct{-q_1}{-q_2}{-q_3}{-q_4}
	       {-\bar q_1}{-\bar q_2}{-\bar q_3}{-\bar q_4}.
\ee
In our case we have $(-1)^{(\ell-\bar\ell)/2}=1$. By looking at our definitions of tensor structures, we see that
\be
	&\T_1^0-\T_2^0,\qquad  \T_3^0-\T_4^0, \qquad\T_5^0-\T_6^0
\ee
are odd under $z\leftrightarrow \bar z$, while
\be
	&\T_1^0+\T_2^0,\qquad  \T_3^0+\T_4^0, \qquad \T_5^0+\T_6^0
\ee
which implies the relation in \eqref{eq:gzzbparity}. 

We do not repeat the analysis here, but similar arguments lead to the constraints listed in \eqref{eq:gpzzbparity_and_hzzbparity}
and \eqref{eq:gpzzbregularity}.

\section{Rational approximations of scalar blocks}
\label{app:rationalscalar}

In even dimensions the standard Zamolodchikov-like recursion relations~\cite{Zamolodchikov:1987,Kos:2013tga,Kos:2014bka} traditionally used for rational approximations of scalar blocks becomes more complicated,\footnote{Closer to the completion of this work we have implemented the scalar Zamolodchikov-like recursion relations directly in 4d. This approach was used in this work only for the upper bound on $C_T$. It will be described elsewhere.} and in this work we instead use the exact Dolan-Osborn expression~\cite{DO1,DO2},
\be\label{eq:dolanosborn}
G^{a,b}_{\Delta,\ell}(z,\bar z)&=(-1)^\ell\frac{z\bar z}{z-\bar z}\left[k_{\Delta-\ell-2}(a,b;z)k_{\Delta+\ell}(a,b;\bar z)-k_{\Delta-\ell-2}(a,b;\bar z)k_{\Delta+\ell}(a,b;z)\right],\\
k_\beta(a,b;z)&=z^{\beta/2}{}_2F_1(a+\beta/2,b+\beta/2,\beta,z).
\ee
The parameters $a$ and $b$ are given by
\be
	a=-\frac{\De_1-\De_2}{2},\quad b=\frac{\De_3-\De_4}{2},
\ee
where $\De_i$ are the scaling dimensions of the external scalars. The notation above slightly differs from the main text and we use it here for convenience.
Let us start by analyzing the $k$-function, which in large-$\beta$ limit behaves as
\be
k_\beta(a,b;z)&=(4\rho)^{\beta/2}\left[\frac{1}{\sqrt{1-\rho^2}}\left(\frac{1+\rho}{1-\rho}\right)^{a+b}+O(\beta^{-1})\right],\\
\rho&=\frac{1-\sqrt{1-z}}{1+\sqrt{1+z}}=\frac{z}{(1+\sqrt{1-z})^2},\\
z&=\frac{4\rho}{(1+\rho)^2},
\ee
which can be derived from an integral representation or directly from the hypergeometric equation. This function has poles at $\beta=-m$, $m\in Z_{\geq 0}$,
\be
k_\beta(a,b;z)&=\frac{R^{a,b}_m}{\beta+m}k_{m+2}(a,b;z)+O(1),\quad \beta\to -m,\\
R^{a,b}_m&=\frac{(-1)^{m}(a-m/2)_{m+1}(b-m/2)_{m+1}}{m!(m+1)!}.
\ee
Note that for (half-)integral $a$ or $b$, half of the residues vanish starting from sufficiently large $m$. In practice this is useful for simplifying the denominator of the rational approximation when there are relations between external scalar dimensions, which is the case for the scalar blocks we need.

We can therefore as usual define
\be
h_\beta(a,b;z)=(4\rho)^{-\beta/2}k_\beta(a,b;z),
\ee
which has the poles
\be
h_\beta(a,b;z)=(4\rho)^{m+1}\frac{R^{a,b}_m}{\beta+m}h_{m+2}(a,b;z)+O(1),\quad \beta \to -m. 
\ee
At the same time we know the behavior of $h_\beta$ at $\beta\to\infty$, so that we can conclude
\be\label{eq:appDhexpansion}
h_\beta(a,b;z)=\frac{1}{\sqrt{1-\rho^2}}\left(\frac{1+\rho}{1-\rho}\right)^{a+b}+\sum_{m=0}^\infty (4\rho)^{m+1}\frac{R^{a,b}_m}{\beta+m}h_{m+2}(a,b;z).
\ee
Naively this appears to be an expansion in powers of $4\rho$, since $h_{m+2}$ approaches a constant value at large $m$, but in fact it is in powers of $\rho$ because there is a $4^{-m}$ in the asymptotic behavior of $R^{a,b}_m$,
\be
R^{a,b}_m=4^{-m}\frac{(-1)^m}{2\pi}\sin(\pi a-\pi m/2)\sin(\pi b-\pi m/2)(1+O(m^{-1})).
\ee
This expression also reproduces the selection rule for poles at (half-)integral $a$ or $b$.

This structure of $k$-functions implies that there are poles in $G_{\Delta,\ell}^{a,b}$ at
\be
\Delta=\ell+2-m, \quad m\in \Z_{\geq 0},
\ee
and the poles at $m> 2\ell+1$ are in general double poles.

To compute rational approximations of the blocks $G_{\De,\ell}^{a,b}$ first note that we can set $z=\thalf+x+y,\, \bar z=\thalf+x-y$ in~\eqref{eq:dolanosborn}, and then straightforwardly Taylor-expand. This essentially gives us
\be
\partial_x^n\partial_y^m G_{\Delta,\ell}^{a,b}(\thalf,\thalf)=\sum_{n',m'}c^{n,m}_{n',m'}\partial^{n'} k_{\Delta-\ell-2}(a,b;\thalf)\partial^{m'}k_{\Delta+\ell}(a,b;\thalf)
\ee
It is easy to check that $n+m-2\leq n'+m'\leq n+m+1$. This gives us an expression for derivatives of $G_{\Delta,\ell}^{a,b}$ in terms of products of derivatives of $k$-functions. We can furthermore reduce it to products of derivatives of $h$-functions with polynomial (in $\Delta$) coefficients, so we can simply approximate these $h$-functions.

We thus consider approximating 
\be\label{eq:hproduct}
h_{\Delta-\ell-2}(a,b;z)h_{\Delta+\ell}(a,b;\bar z)
\ee
to a fixed order in $\rho$-expansion. In other words, we will simply substitute expansions~\eqref{eq:appDhexpansion} and truncate them so that the highest-order terms in the sum are proportional to $\rho^n\bar\rho^m$ with $n+m=$\texttt{keptPoleOrder}, a parameter to our approximation. Note that this is as good as keeping first \texttt{keptPoleOrder} terms in both expansions (and sometimes even better), but produces fewer terms.

Let us comment a bit on the structure of the poles in~\eqref{eq:hproduct}. Consider first-order poles in~\eqref{eq:hproduct}, $\Delta=\ell+2-m$, $0\leq m\leq 2\ell+1$. The behavior near these poles is given by
\be
\frac{(4\rho)^{m+1}R^{a,b}_m}{\Delta-(\ell+2-m)}h_{m+2}(a,b;z)h_{2\ell+2-m}(a,b;\bar z)
\ee
and thus they contribute at $\rho$-order $m+1$. Now consider the second order poles at $\Delta=-\ell-m$, $m\geq 0$. The pole behavior is
\be\label{eq:secondorderpole}
&\frac{(4\rho)^{2\ell+3+m}(4\bar\rho)^{m+1}R^{a,b}_{2\ell+2+m}R^{a,b}_m}{(\Delta-(-\ell-m))^2}h_{2\ell+4+m}(a,b;z)h_{m+2}(a,b;\bar z)+\nn\\
&+\frac{(4\rho)^{2\ell+3+m}R^{a,b}_{2\ell+2+m}}{\Delta-(-\ell-m)}h_{2\ell+4+m}(a,b;z)h^\mathrm{reg}_{-m}(a,b;\bar z)+\nn\\
&+\frac{(4\rho)^{m+1}R^{a,b}_{m}}{\Delta-(-\ell-m)}h^\mathrm{reg}_{-2\ell-2-m}(a,b;z)h_{m+2}(a,b;\bar z).
\ee
Here $h^\mathrm{reg}_{\beta}(a,b;z)$ is defined by
\be
h_\beta(a,b;z)=\frac{(4\rho)^{m+1} R^{a,n}_m}{\beta+m}h_{m+2}(a,b;z)+h^\mathrm{reg}_{-m}(a,b;z)+O(\beta+m),
\ee
and can be computed from~\eqref{eq:appDhexpansion}. We have a second order pole at $\rho$-order $2m+2\ell+4$, a first order pole at $\rho$-order $2\ell+3+m$ and a first order pole at $\rho$-order $m+1$. Since these contribute at different $\rho$-orders, they are cut off at different values of $m$. Therefore, effectively in our ansatz we will have first-order poles which come from second-order poles with second-order piece neglected due to its high $\rho$-order. Similarly, when computing $h^\mathrm{reg}_\beta$ from~\eqref{eq:appDhexpansion}, we truncate~\eqref{eq:appDhexpansion} at the $\rho$-order dictated by the power of $\rho$ or $\bar\rho$ multiplying $h^\mathrm{reg}_\beta$ in~\eqref{eq:secondorderpole}. In practice we do not care to do the same for $h_\beta$ themselves since they, together with their derivatives, are efficiently calculated by the Mathematica built-in support for ${}_2F_1$.

\section{Parameters of numerical searches}

\begin{table}[!h]
	\centering
	\begin{tabular}{|l|c|c|}
		\hline
		problem & (A) & (B) \\
		\hline
		\texttt{precision}                     & 400 & 400\\
		\texttt{findPrimalFeasible}            & True & False\\
		\texttt{findDualFeasible}              & True & False\\
		\texttt{detectPrimalFeasibleJump}      & True & False\\
		\texttt{detectDualFeasibleJump}        & True & False\\
		\texttt{dualityGapThreshold}           & $10^{-10}$ & $10^{-10}$\\
		\texttt{primalErrorThreshold}          & $10^{-30}$ & $10^{-30}$\\
		\texttt{dualErrorThreshold}            & $10^{-20}$ & $10^{-20}$ \\
		\texttt{initialMatrixScalePrimal}      & $10^{20}$ & $10^{20}$\\
		\texttt{initialMatrixScaleDual}        & $10^{20}$ & $10^{20}$\\
		\texttt{feasibleCenteringParameter}    & 0.1 & 0.1\\
		\texttt{infeasibleCenteringParameter}  & 0.3 & 0.3\\
		\texttt{stepLengthReduction}           & 0.7 & 0.7 \\
		\texttt{choleskyStabilizeThreshold}    & $10^{-120}$ & $10^{-120}$ \\
		\texttt{maxComplementarity}            & $10^{100}$ & $10^{100}$\\
		\hline
	\end{tabular}
	\caption{SDPB parameters for bounds on scaling dimensions (A) and OPE coefficients (B).}
	\label{tab:sdpb_parameters}
\end{table}

There are two separate instances where various parameters needed for numerical searches should be chosen appropriately. First, we address parameters governing truncation of the bootstrap equations of section~\ref{subsec:crossing_equations} and rational approximations of conformal blocks in section~\ref{sec:rational_approximation}. Second, we address SDPB parameters~\cite{Simmons-Duffin:2015qma} used for solving the semidefinite problems of section~\ref{sec:sdp}.

The strength of numerical bounds depends on the number of total derivatives $\Lambda$ (which in principle should be $\infty$). In the majority of plots we use $\Lambda=16$ or $\Lambda=20$. Another truncation parameter is the maximal spin of ``exchanged'' operators denoted by \verb|maxSpin|. For a given $\Lambda$ any bound should be independent of \verb|maxSpin|, we find that for our setup one can choose
\begin{equation}
\verb|maxSpin| = \Lambda + 15.
\end{equation}
Failure of taking \verb|maxSpin| big enough results in stronger but incorrect bounds. The parameter characterizing the precision of the conformal block approximation is denoted by \verb|keptPoleOrder|. We have discovered that the choice of this parameter depends on the scaling dimension of the Weyl fermion $\De_\psi$. We divide all the values $\De_\psi$ into two parts: close and far from the fermion unitarity bound. We then use 
\begin{eqnarray}
&&\verb|keptPoleOrder| = \Lambda + 15,\quad \De_\psi\geq 1.6;\\
&&\verb|keptPoleOrder| = \Lambda + 25,\quad \De_\psi\in[1.505,1.6].\nonumber
\end{eqnarray}
Better ways of approximating the scalar blocks than the one described in appendix~\ref{app:rationalscalar} will solve this inconvenience.
In obtaining the final bootstrap equations we use the following number of digits after the comma: \verb|prec|=200.

In general we have two distinct types of semi-definite problems: bounds on scaling dimensions (A) and bounds on OPE coefficients (B). We thus use two different sets of SDPB parameters for (A) and (B). We make a choice independent on $\Lambda$ which is summarized in table~\ref{tab:sdpb_parameters}.

\bibliographystyle{JHEP}
\bibliography{refs}

\end{document}